\documentclass[12pt]{article}
\pdfoutput=1
\usepackage{amsmath,amssymb,amsthm,bm,graphicx,float,array,multirow,multicol,rotfloat,caption,subcaption,hyperref,cleveref,enumerate,geometry,mathdots,adjustbox,booktabs,parskip,mathtools,tikz,tikz-cd,pdflscape,csquotes,lscape,rotating,empheq,mathrsfs}
\usepackage[para]{threeparttable}
\usepackage[all]{xy}
\usepackage{wasysym}
\usepackage[normalem]{ulem}
\usepackage[numbers,sort&compress]{natbib}
\usepackage{longtable}
\usepackage{pdflscape}
\numberwithin{equation}{section}
\numberwithin{figure}{section}
\allowdisplaybreaks
\usetikzlibrary{arrows,shapes.misc,positioning,decorations.pathmorphing,decorations.markings,matrix,patterns,backgrounds}
\makeatletter

\theoremstyle{plain}
\newtheorem{alg}{Algorithm}

\makeatother

\begin{document}

\begin{titlepage}
\vspace*{-3cm} 
\begin{flushright}
{\tt DESY-23-204}\\
{\tt ZMP-HH/23-20}

\end{flushright}
\begin{center}
\vspace{2cm}
{\LARGE\bfseries The Bestiary of 6d \boldmath{$(1,0)$} SCFTs: \\[5mm]{\Large Nilpotent Orbits and Anomalies}}

\vspace{1.2cm}

{\large
Florent Baume$^{1}$ and Craig Lawrie$^{2}$\\} 
\vspace{.7cm}
{ $^1$ II. Institut für Theoretische Physik, Universität Hamburg,}\par
{Luruper Chaussee 149, 22607 Hamburg, Germany}\par
\vspace{.2cm}
{ $^2$ Deutsches Elektronen-Synchrotron DESY,}\par
{Notkestr.~85, 22607 Hamburg, Germany}\par
\vspace{.2cm}

\vspace{.3cm}

\scalebox{1.0}{\tt florent.baume@desy.de, craig.lawrie1729@gmail.com}\par
\vspace{1.2cm}
\textbf{Abstract}
\end{center}
\noindent 
Many six-dimensional $(1,0)$ SCFTs are known to fall into families labelled by nilpotent orbits of certain simple Lie algebras. For each of the three infinite series of such families, we show that the anomalies for the continuous zero-form global symmetries of a theory labelled by a nilpotent orbit $O$ of $\mathfrak{g}$ can be determined from the anomalies of the theory associated to the trivial nilpotent orbit (the parent theory), together with the data of $O$. In particular, knowledge of the tensor branch field theory is bypassed completely. We show that the known anomalies, previously determined from the geometric/atomic construction, are reproduced by analyzing the Nambu--Goldstone modes inside of the moment map associated to the $\mathfrak{g}$ flavor symmetry of the parent SCFT. This provides further evidence for the physics underlying the labelling of the SCFTs by nilpotent orbits. We remark on some consequences, such as the reinterpretation of the 6d $a$-theorem for such SCFTs in terms of group theory.

\vfill 
\end{titlepage}

\tableofcontents
\newpage

\section{Introduction}\label{sec:intro}

In recent years, the study of six-dimensional superconformal field theories
(SCFTs) has undergone rapid progress. While such theories were conjectured to
exist from an analysis of superconformal algebras \cite{Nahm:1977tg}, the
absence of any concrete bottom-up construction led to the widespread doubt of
their existence. Such SCFTs can be shown to lack any relevant or marginal
(supersymmetry-preserving) parameters
\cite{Cordova:2015fha,Louis:2015mka,Cordova:2016xhm,Cordova:2016emh, Buican:2016hpb},
precluding the existence of a straightforward Lagrangian approach, and
rendering the usual weakly-coupled perturbative techniques impotent. Instead, it is necessary to develop alternative techniques, often string-theoretic, to both understand the existence of such inherently strongly-coupled quantum field theories, and to extract their physical behavior. 

It was not until the heyday of string theory that a construction giving rise to
an interacting 6d SCFT with maximal supersymmetry was developed. Consider Type
IIB string theory compactified on an orbifold singularity:
\begin{equation}\label{eqn:orbifold}
    \mathbb{C}^2/\Lambda \,,
\end{equation}
where $\Lambda$ is a finite subgroup of $SU(2)$. As the orbifold is a
non-compact Calabi--Yau twofold, the compactification does not break
supersymmetry entirely, but preserves half the supersymmetry in the resulting
effective six-dimensional theory. The fundamental degrees of freedom of the
compactified theory are tensionless self-dual strings, which arise from the
Type IIB perspective from D3-branes wrapping the collapsed, zero-volume cycles
associated to the orbifold singularity. However, it was shown in
\cite{Witten:1995zh,Strominger:1995ac,Seiberg:1996qx}, that these are, in fact,
local superconformal field theories, and each tensionless string couples to a
self-dual two-form potential which belongs to a tensor multiplet. There is a moduli space of supersymmetric vacua parametrized by the vacuum expectation values of the scalar fields inside of these tensor multiplets; this is the \emph{tensor branch} of the SCFT. At the generic point of the tensor branch all of the self-dual strings are tensionful. The
classification of finite subgroups of $SU(2)$ is an ADE-classification; thus,
there are two infinite series of 6d $(2,0)$ SCFTs and three sporadic SCFTs,
corresponding to:
\begin{equation}\label{eqn:20theories}
    A_{N \geq 1} \,, \quad D_{N \geq 4} \,, \quad E_6 \,, \quad E_7 \,, \quad E_8 \,.
\end{equation}
Typically, we use this ADE-classification to label each 6d $(2,0)$ SCFT by a
simple and simply-laced Lie algebra. The top-down construction of
strongly-coupled superconformal field theories, either from string theory or
from higher-dimensional field theory, has been extremely powerful over the last
thirty years; see \cite{Argyres:2022mnu} for a recent review.

The Type IIB string theoretic construction just described yields 6d SCFTs with
maximal supersymmetry. It is natural to consider an analogous construction for
6d SCFTs with minimal supersymmetry. Instead of the orbifold in equation
\eqref{eqn:orbifold}, we can consider the compactification of Type IIB string
theory on a non-compact K\"ahler surface which is \emph{not} Calabi--Yau.
Naively, this breaks all of the supersymmetry in the effective six dimensional
theory, however, it is possible to simultaneously turn on a non-trivial
axio-dilaton profile such that one quarter of the supersymmetry is preserved.
In this way, we should replace the non-compact Calabi--Yau twofold of equation
\eqref{eqn:orbifold} with a non-compact elliptically-fibered Calabi--Yau
threefold, where the elliptic fibration captures the axio-dilaton profile; this
puts us squarely in the realm of F-theory
\cite{Vafa:1996xn,Morrison:1996na,Morrison:1996pp}. 

The generalization to such Calabi--Yau threefold compactifications has been
worked out in \cite{Heckman:2013pva,Heckman:2015bfa}. Consider a non-compact
elliptically-fibered Calabi--Yau threefold
\begin{equation}
    \pi : Y \rightarrow B \,,
\end{equation}
where the base of the fibration, $B$, contains no complex curves of finite
volume. The base may be singular, in which case we assume that it has at most
one singular point, which we label as $b_0$. Further, we assume that
$\pi^{-1}(b)$ is an irreducible, possibly degenerate, genus-one curve, for all
points $b \in B$; thus \cite{MR977771}, we can write the elliptic fibration $Y$
as a Weierstrass model over $B$. Assume that the Weierstrass model has at most
one non-minimal fiber,\footnote{The definition of non-minimal is somewhat
technical and not particularly illuminating, so we suppress it here. We refer
the reader to \cite{Heckman:2018jxk} for a full review of the construction of
6d $(1,0)$ SCFTs from F-theory.} supported over the point $b_0$ in $B$. Then
F-theory compactified on $Y$ leads to an interacting 6d $(1,0)$ SCFT with a
single energy-momentum tensor.\footnote{The edge-case, where there is neither a
singular point in the base, nor a non-minimal singular fiber, leads to a
trivial SCFT.} Thus, to construct 6d $(1,0)$ SCFTs, it is necessary to know how
to construct Calabi--Yau threefolds satisfying the requisite properties. 

We know that for the construction of the 6d $(2,0)$ SCFTs, the base $B$ takes
the form of an orbifold singularity:
\begin{equation}\label{eqn:orbifold2}
    B = \mathbb{C}^2/\Lambda \,.
\end{equation}
In \cite{Heckman:2013pva}, it was shown that for any $Y$ that engineers a 6d
SCFT, the base must instead be a ``generalized orbifold''. These take the same
form as in equation \eqref{eqn:orbifold2}, however $\Lambda$ is now allowed to
be particular finite subgroups of $U(2)$, instead of $SU(2)$. We refer the reader to \cite{Heckman:2013pva} for a review of the generalized orbifolds, in particular the action
of the finite group $\Lambda$ on the coordinates of $\mathbb{C}^2$. In the end,
there are two families of generalized orbifolds, known as generalized $A$-type
and generalized $D$-type orbifolds, each parameterized by a pair of integers
$p$, $q$, and which we write as
\begin{equation}
    \mathcal{A}_{(p, q)} \,, \qquad \mathcal{D}_{(p+q, q)} \,.
\end{equation}
For particular combinations of the parameters, the generalized orbifolds reduce
to the standard orbifold singularities. 
There is no generalization of the standard $E$-type orbifolds.

Once a generalized orbifold $\mathbb{C}^2/\Lambda$ has been specified, it is necessary to provide information which captures the nature of the non-minimal singular fiber supported over the orbifold point. In fact, as we review in Section \ref{sec:bestiary}, it is sufficient to encode this data in a choice of (possibly trivial) ADE Lie algebra, $\mathfrak{g}$. In the end, then, one can obtain the following families of 6d $(1,0)$ SCFTs:
\begin{equation}
    \mathcal{A}_{(p,q)}^{\mathfrak{g}}  \,, \quad \mathcal{D}_{(p+q,q)}^{\mathfrak{g}} \,, \qquad E_6^{\mathfrak{g}} \,, \qquad E_7^{\mathfrak{g}} \,, \qquad E_8^{\mathfrak{g}} \,.
\end{equation}
This constitutes a natural generalization of the families of 6d $(2,0)$ SCFTs
in equation \eqref{eqn:20theories}. Here the possibilities for $\mathfrak{g}$ are constrained by
the surface singularity. Similarly, the values of $p$ and $q$ that can appear
are constrained as only certain combinations
 correspond to 6d $(1,0)$ SCFTs, see \cite{Heckman:2013pva}. 

In this paper, we focus on families of theories where the number of tensor
multiplets can be taken to be arbitrarily large. As such, we do not consider
the E-type bases further here, however, see \cite{Lawrie:2024zon} for an analysis of those
SCFTs. Similarly, when the theories $\mathcal{A}_{(p,q)}^\mathfrak{g}$ and
$\mathcal{D}_{(p+q,q)}^\mathfrak{g}$ admit a large $N$ limit in the number of
tensors, the combinations $(p, q, \mathfrak{g})$ are further constrained. In
particular, for $\mathcal{A}_{(p,q)}^\mathfrak{g}$, they are specified by an
integer $N \geq 1$, and two fractions $f_L$ and $f_R$ which belong to the set
\begin{equation}\label{eqn:possfractions}
    \left\{ \frac{1}{6},\, \frac{1}{5},\, \frac{1}{4},\, \frac{1}{3},\, \frac{2}{5},\, \frac{1}{2},\, \frac{3}{5},\, \frac{2}{3},\, \frac{3}{4},\, \frac{4}{5},\, \frac{5}{6},\, 1 \right\} \,.
\end{equation}
The $f_L$ and $f_R$ are related to the numbers of
fractional M5-branes in the M-theory dual descriptions of these 6d $(1,0)$
SCFTs, as we discuss momentarily. For a choice of $(p, q)$, specified by
fractions $f_L$ and $f_R$, the choice of algebra $\mathfrak{g}$ must satisfy
that
\begin{equation}\label{eqn:fractionconditions}
    \operatorname{max}(\operatorname{denom}(f_L), \operatorname{denom}(f_R)) \leq \frac{n_\mathfrak{g}}{2} \,,
\end{equation}
where $\operatorname{denom}(\cdot)$ is the denominator of the fraction and
$n_\mathfrak{g} = 2, 4, 6, 8, 12$ for $\mathfrak{g} = \mathfrak{su}(K)$,
$\mathfrak{so}(2K)$, $\mathfrak{e}_6$, $\mathfrak{e}_7$, $\mathfrak{e}_8$,
respectively. Similarly, the $\mathcal{D}_{(p+q, q)}^\mathfrak{g}$ theories
that exist for large numbers of tensors can be specified by an integer $N \geq
4$, a fraction $f$ belonging to the set in equation \eqref{eqn:possfractions},
and an ADE Lie algebra $\mathfrak{g}$ which satisfies the analogous condition
to equation \eqref{eqn:fractionconditions}. We write these theories as
\begin{equation}\label{eqn:someparents}
    \mathcal{A}_{N-1;f_L, f_R}^\mathfrak{g} \,, \qquad \text{ and } \qquad \mathcal{D}_{N; f}^\mathfrak{g} \,.
\end{equation}
for convenience. These are the families of 6d $(1,0)$ SCFTs that generalize the
infinite series of AD-type 6d $(2,0)$ SCFTs.\footnote{When $f_L = f_R = 1$ or $f = 1$, then the generalized orbifold is just the standard orbifold. We will use the shorthand
$A_{N-1}^\mathfrak{g} = \mathcal{A}_{N-1;1,1}^\mathfrak{g}$ and $D_{N}^\mathfrak{g} = \mathcal{D}_{N;1}^\mathfrak{g}$ for convenience.}

Here, we have constructed the theories in equation \eqref{eqn:someparents} from
Type IIB string theory (or its non-perturbative avatar, F-theory). However,
each member of these families of theories can also be realized in M-theory. The
description of $A_{N-1}^\mathfrak{g}$ in terms of M5-branes is straightforward.
It is the 6d $(1,0)$ SCFT that lives on the worldvolume of a stack of $N$
M5-branes probing a $\mathbb{C}^2/\Gamma$ orbifold singularity, where $\Gamma$
is the finite subgroup of $SU(2)$ with the same ADE-type as $\mathfrak{g}$.
Notice that both the M-theory and F-theory descriptions involve an orbifold,
but in the former it is associated to the ``fiber data'', $\mathfrak{g}$, and
in the later it is instead associated to the base of the elliptic fibration:
$A_{N-1}$. As we discuss later, when the fractional numbers are different than $1$, there are ``fractional'' M5-branes, and when the base is of generalized D-type, there are orientifold 5-branes in the M-theory description.

The theory $\mathcal{A}_{N-1; f_L, f_R}^\mathfrak{g}$, which is known as
(fractional) conformal matter \cite{DelZotto:2014hpa}, typically has a
non-Abelian flavor algebra which is
\begin{equation}
    \mathfrak{f} = \mathfrak{g}_{f_L} \oplus \mathfrak{g}_{f_R} \,,
\end{equation}
where $\mathfrak{g}_{f_L}$ and $\mathfrak{g}_{f_R}$ are simple Lie
algebras fixed by $f_L$ and $f_R$. When $\mathcal{A}_{N-1; f_L, f_R} =
A_{N-1}$,
this theory is simply rank $N$ $(\mathfrak{g}, \mathfrak{g})$ conformal matter,
and $\mathfrak{g}_{f_L=1} = \mathfrak{g}_{f_R=1} = \mathfrak{g}$. It
has been proposed that giving a nilpotent vacuum expectation value to the
moment map associated to the $\mathfrak{g}_{f_L} \oplus
\mathfrak{g}_{f_R}$ flavor symmetry triggers a Higgs branch
renormalization group flow to a new interacting 6d $(1,0)$ SCFT
\cite{Heckman:2016ssk,Hassler:2019eso,Heckman:2018pqx}, for $N$ sufficiently
large.\footnote{The Higgs branch renormalization group flows triggered by such
vacuum expectation values do not exhaust the interacting fixed points on the
Higgs branch, in general. For example, there are also flows triggered by giving
vevs to so-called end-to-end operators, such as have been studied in
\cite{Baume:2020ure,Heckman:2020otd,Razamat:2019mdt,Bergman:2020bvi,Baume:2022cot,DKL}.
We do not study the effects of this ``end-to-end Higgsing'' in this paper.}
Such vacuum expectation values depend only on a choice of nilpotent
orbit, rather than the nilpotent element itself, and thus we can consider a
family of 6d $(1,0)$ SCFTs 
\begin{equation}\label{eqn:Afamily}
    \mathcal{A}_{N-1; f_L, f_R}^\mathfrak{g}(O_L, O_R)  \,,
\end{equation}
where $O_L$ and $O_R$ are nilpotent orbits of $\mathfrak{g}_\text{fr}^L$ and
$\mathfrak{g}_\text{fr}^R$, respectively. Similarly, the theories
$\mathcal{D}_{N; f}^\mathfrak{g}$ have only a single non-Abelian flavor factor
\begin{equation}
    \mathfrak{f} = \mathfrak{g}_{f} \,,
\end{equation}
and again new interacting 6d SCFTs can be obtained via nilpotent Higgsing of
that flavor symmetry. Picking $O$ as a nilpotent orbit of
$\mathfrak{g}_{f}$, we then have the family of theories\footnote{The
Higgs branch of the $\mathcal{D}_{f, N}^\mathfrak{g}(O)$ family of SCFTs is
analyzed extensively in \cite{Lawrie:2024zon}.}
\begin{equation}\label{eqn:Dfamily}
    \mathcal{D}_{N; f}^\mathfrak{g}(O) \,.
\end{equation}

Another interesting family of 6d $(1,0)$ SCFTs that are realized in the
geometric F-theory construction are the so-called rank $N$ $(\mathfrak{e}_8,
\mathfrak{g})$ orbi-instanton theories \cite{DelZotto:2014hpa}. From the
M-theory perspective, these SCFTs live on the worldvolume of $N$ M5-branes
probing a $\mathbb{C}^2/\Gamma$ orbifold, where $\Gamma$ is the finite ADE
group corresponding to the simple Lie algebra $\mathfrak{g}$, and contained
inside of an end-of-the-world M9-brane. Furthermore, we can choose an $f$
belonging to the set in equation \eqref{eqn:possfractions}, such that $f$ and
$\mathfrak{g}$ satisfy the analogous condition to that in equation
\eqref{eqn:fractionconditions}. Let $N \geq 1$, then we can denote this family
of theories as
\begin{equation}
    \mathcal{O}_{N; f}^\mathfrak{g} \,.
\end{equation}
These theories possess a flavor symmetry which is
\begin{equation}
    \mathfrak{f} = \mathfrak{e}_8 \oplus \mathfrak{g}_f \,,
\end{equation}
where $\mathfrak{g}_{f}$ is again fixed by the fraction $f$. The
$\mathfrak{g}_{f}$ arises from the orbifold singularity in M-theory, and
again it is expected that giving a vacuum expectation value to the associated
moment map generically triggers a Higgs branch renormalization group flow that
leads to a new interacting 6d $(1,0)$ SCFT. On the other hand, the
$\mathfrak{e}_8$ flavor symmetry does not arise from the orbifold singularity,
but instead from the M9-brane; in particular, it is necessary to pick a choice
of boundary conditions, on the $S^3/\Gamma$ boundary of the
$\mathbb{C}^2/\Gamma$ orbifold, for the $E_8$-bundle associated to the
M9-brane. Such a boundary condition corresponds to a choice of homomorphism 
\begin{equation}
    \sigma : \Gamma \rightarrow E_8 \,.
\end{equation}
The $\mathfrak{e}_8$ flavor symmetry factor is realized when $\sigma$ is the
trivial homomorphism, and the symmetry is broken to a subalgebra for any other
choice of $\rho$. It is widely believed that there exists a Higgs branch
renormalization group flow from the theory with $\sigma$ trivial to any theory
where the homomorphism is non-trivial \cite{Frey:2018vpw}. That is, there is a
family of 6d $(1,0)$ SCFTs
\begin{equation}\label{eqn:OIfamily}
    \mathcal{O}_{N; f}^\mathfrak{g}(\sigma, O) \,,
\end{equation}
where $\sigma$ belongs to $\operatorname{Hom}(\Gamma, E_8)$ and $O$ is a
nilpotent orbit of $\mathfrak{g}_{f}$, which arise via Higgs branch
renormalization group flows from $\mathcal{O}_{N; f}^\mathfrak{g}$.

It turns out that, once the effective field theory at the generic point of the
tensor branch involves a sufficiently large number of tensor
multiplets,\footnote{The required number of tensor multiplets is not
particularly large.} every known 6d SCFT belongs to one of the families in
equations \eqref{eqn:Afamily}, \eqref{eqn:Dfamily} and
\eqref{eqn:OIfamily}.\footnote{There are two subtleties here. First, we are
		considering only the local operator spectrum of the 6d SCFT; when
		considering extended operators, each member of the families given here
		may correspond to multiple SCFTs that differ only in their spectrum of
		extended operators. Second, there are discretely-gauged versions of
		some members of these families; these are distinct as local SCFTs,
		however, for the purposes of the anomalies that we consider in this
		paper, they can be treated as equivalent. For more details on the
discretely-gauged 6d $(1,0)$ SCFTs, and especially their Higgs branches, see
\cite{LLM}.} 

We have now discussed, in some detail, a mechanism for constructing 6d $(1,0)$
SCFTs via the compactification of F-theory on certain non-compact
elliptically-fibered Calabi--Yau threefolds. However, since these theories are
inherently strongly-coupled, and cannot be written down in a Lagrangian
formulation, it is generally hard to extract the physical properties. Anomalies
are by nature topological, and thus it should be possible to determine them
without detailed access to the microscopics of the SCFT.

Consider first the 6d $(2,0)$ SCFTs that are engineered via Type IIB string
theory compactified on an orbifold singularity, $\mathbb{C}^2/\Lambda$. When
$\Lambda = \mathbb{Z}_K$, the SCFT has an alternative interpretation as the
worldvolume theory on a stack of $K$ M5-branes in M-theory. The anomalies of
the $(2,0)$ SCFT can then be determined by considering anomaly inflow from the
M-theory bulk. The dependence of the anomalies of all 6d $(2,0)$ SCFTs on the
finite group $\Lambda$ has been determined
\cite{Duff:1995wd,Witten:1996hc,Freed:1998tg,Harvey:1998bx,Intriligator:2000eq,Yi:2001bz,Ohmori:2014kda}:
\begin{equation}\label{eqn:AP20}
    I_8 = \frac{h_\mathfrak{g}^\vee d_\mathfrak{g}}{24} p_2(N) + \frac{r_\mathfrak{g}}{48} \big( p_2(N) - p_2(T) + \frac{1}{4}(p_1(T) - p_1(N))^2 \big) \,.
\end{equation}
This is a formal eight-form polynomial in the characteristic classes of the
global symmetries of the SCFT. The $p_i(T)$ are the Pontryagin classes of the
tangent bundle to the 6d spacetime, this captures the $\mathfrak{so}(1,5)$
Lorentz group; $p_i(N)$ are the Pontryagin classes of the bundle associated to
the $\mathfrak{so}(5)_R$ R-symmetry. The coefficients $h_\mathfrak{g}^\vee$,
$d_\mathfrak{g}$, and $r_\mathfrak{g}$ are, respectively, the dual Coxeter
number, the dimension, and the rank of the ADE Lie algebra $\mathfrak{g}$ of
the same ADE-type as the finite group $\Lambda$.

Similarly, the anomaly polynomials for the 6d $(1,0)$ SCFTs can be worked out
from the geometric description of the effective tensor branch field theory.
Generically, the anomaly polynomial can be written as
follows:\footnote{Throughout this paper, we typically ignore Abelian flavor
symmetries; it is straightforward to generalize the analysis to include the
anomalies for such symmetries. Abelian symmetries require additional care due
to the presence of ABJ anomalies which does not occur with non-Abelian Lie algebras as their generators are traceless \cite{Lee:2018ihr,Apruzzi:2020eqi}.}
\begin{equation}\label{eqn:AP}
  \begin{aligned}
		  I_8 &=  \frac{\alpha}{24} c_2(R)^2+ \frac{\beta}{24}  c_2(R) p_1(T) + \frac{\gamma}{24}  p_1(T)^2 + \frac{\delta}{24} p_2(T) \cr &\quad + \sum_a \text{Tr}F_a^2 \left(\kappa_a p_1(T) + \nu_a c_2(R) + \sum_b \rho_{ab} \text{Tr}  F_b^2\right) + \sum_a \mu_a \text{Tr}F_a^4  \,.
  \end{aligned}
\end{equation}
Now, $c_2(R)$ is the second Chern class of the bundle associated to the
$\mathfrak{su}(2)_R$ R-symmetry; and, $\text{Tr}F_a^2$ and $\text{Tr}F_a^4$ are
the one-instanton normalized traces of the curvature, $F_a$, for each simple
non-Abelian factor in the flavor algebra.

On the tensor branch of the SCFT, where the strings become tensionful, the
superconformal symmetry is broken, however, the Lorentz symmetry, the
R-symmetry, and any flavor symmetry remains unbroken. As the coefficients in
the anomaly polynomial in equation \eqref{eqn:AP} are coefficients of
characteristic classes of unbroken symmetry, they are unchanged under the
movement onto the tensor branch. Thus, one can determine the anomaly polynomial
of the effective field theory at the generic point of the tensor branch, and
then use a variant of 't Hooft anomaly matching \cite{tHooft:1979rat}, to
determine the anomaly polynomial of the SCFT at the origin of the tensor branch
\cite{Ohmori:2014pca,Ohmori:2014kda,Intriligator:2014eaa,Baume:2021qho}.\footnote{Alternatively,
one can attempt to determine certain combinations of the 't Hooft anomaly
coefficients from the conformal bootstrap; this orthogonal approach has been
shown, in certain cases, to recover the anomaly coefficients determined from
the F-theory geometry \cite{Chang:2017xmr,Baume:2019aid,Baume:2021chx}.} In
particular, see Algorithm 1 of \cite{Baume:2021qho} for a concise and
comprehensive algorithm to determine the SCFT anomaly polynomial from any
tensor branch configuration in the F-theory construction.

For the 6d $(2,0)$ SCFTs, we could see precisely how the anomaly coefficients
in equation \eqref{eqn:AP20} depended on the data of the F-theory
compactification; in this case, the non-compact elliptically-fibered
Calabi--Yau threefold is the trivial elliptic fibration over the orbifold
singularity in equation \eqref{eqn:orbifold}, and we see directly how the coefficients in equation \eqref{eqn:AP20}
depend on $\Lambda$. Of course, using the tensor branch effective
field theory, it is straightforward to apply 't Hooft anomaly matching to
determine the anomaly coefficients \emph{in terms of the data of the tensor
branch theory}. In particular, the dependence of the anomaly coefficients on
the tensor branch data, such as the number of vector multiplets,
hypermultiplets, the Green--Schwarz couplings, etc., has been determined for
many of theories studied here, see, for example, \cite{Ohmori:2014kda,Mekareeya:2016yal,Mekareeya:2017sqh,Mekareeya:2017jgc,Kim:2018lfo,DelZotto:2018tcj,Chen:2019njf,Baume:2021qho}. 

In this paper, we take an orthogonal approach: we would like to know how the
anomalies of the theories 
\begin{equation}\label{eqn:thetheories}
    \mathcal{A}_{N-1;f_L,f_R}^\mathfrak{g}(O_L, O_R) \,, \qquad \mathcal{D}_{N;f}^\mathfrak{g}(O) \,, \qquad \mathcal{O}_{N;f}^\mathfrak{g}(\varnothing, O) \,,
\end{equation}
where $\varnothing$ represents the trivial homomorphism $\sigma : \Gamma_\mathfrak{g}
\rightarrow E_8$, can be determined from a bottom-up SCFT perspective. We determine the anomalies of the,
so-called, parent, or ultraviolet, theories
\begin{equation}\label{eqn:parents}
    \mathcal{A}_{N-1;f_L,f_R}^\mathfrak{g} \,, \qquad \mathcal{D}_{N;f}^\mathfrak{g} \,, \qquad \mathcal{O}_{N;f}^\mathfrak{g} \,,
\end{equation}
from the tensor branch effective field theory. Then, we argue that the
anomalies of the infrared theories in equation \eqref{eqn:thetheories} can be
written in terms of the anomalies of the parent theories in equation
\eqref{eqn:parents} and the nilpotent orbits, without further recourse to the
effective field theory on the tensor branch. In particular, we determine the anomalies from the tensor branch, and we show
that the resulting anomalies are exactly what one would expect from the
bottom-up nilpotent Higgsing of the moment map of an SCFT, where the
only modes to decouple in the infrared are the Nambu--Goldstone modes inside of
the moment map. 

The structure of this paper is as follows. In Section \ref{sec:bestiary}, we review the atomic construction of 6d $(1,0)$ SCFTs in F-theory, and detail the three infinite series of SCFTs whose anomalies we explore in this paper. We determine the anomaly polynomials for the three infinite series of 6d $(1,0)$ SCFTs, written in terms of the nilpotent orbit data, that we consider in this paper in Section \ref{sec:I8}. In Section \ref{sec:NG}, we compute the contribution to the anomaly polynomial from the Nambu--Goldstone modes inside of the moment map upon Higgsing, and show that the bottom-up approach, captured in Algorithm \ref{alg:I8IR}, reproduces the anomaly polynomial known from the tensor branch description. We discuss some consequences and future directions in Section \ref{sec:discussion}. Finally, in Appendices \ref{app:representations}, \ref{app:nilps}, \ref{app:hasse}, and \ref{app:nilp2TB}, we provide a comprehensive review of the necessary data for nilpotent orbits, and enumerate how nilpotent orbits are related to 6d $(1,0)$ tensor branch geometries.

\section{The Bestiary of Long 6d SCFTs}\label{sec:bestiary}

In Section \ref{sec:intro}, we have explained how a non-compact elliptically-fibered Calabi--Yau threefold, subject to certain conditions, can give rise to a 6d $(1,0)$ SCFT via the medium of F-theory. Unfortunately, these elliptic fibrations involve non-minimal fibers supported over points of the base of the fibration, which, a priori, renders them challenging to work with directly. Luckily, a method is known through which such elliptic fibrations can be obtained \cite{Heckman:2013pva,Heckman:2015bfa}.

The general strategy to obtain a 6d $(1,0)$ SCFT via F-theory is to
construct an elliptically-fibered Calabi--Yau threefold, $\widetilde{Y}$, with a smooth base
$\widetilde{B}$ containing a set of curves $\Sigma^i\subset \widetilde{B}$, such that the elliptic fibration is minimal. F-theory compactified on this elliptic fibration in fact gives a description of the theory on the generic point of the tensor
branch of the SCFT. The conformal fixed point is reached by shrinking all curves
in the base to zero volume. The possible $\widetilde{Y}$ such that the contraction map leads to a $Y$ which engineers a 6d $(1,0)$ SCFT are highly constrained. The curves must then have self-intersection
$\Sigma^i\cdot\Sigma^i = -n$ with $12\leq n\leq1$, and to be able to contract them 
simultaneously for all curves, their adjacency matrix must furthermore be negative definite:
\begin{equation}
		A^{ij} = \Sigma^i\cdot \Sigma^j\prec 0\,.
\end{equation}

One can enumerate every elliptically-fibered Calabi--Yau threefold
$\widetilde{Y}$ satisfying the necessary conditions. Collapsing all curves to zero size one can then reach the geometries $Y$ describing any 6d $(1,0)$ SCFT admitting a construction
via F-theory. This was achieved in \cite{Heckman:2013pva, Heckman:2015bfa},
where it was concluded that at the fixed point, all bases are given by a choice of orbifold $B=\mathbb{C}^2/\Lambda$, with $\Lambda$ a discrete subgroup of $U(2)$. We now review the procedure whereby elliptic fibrations $\widetilde{Y}$, that lead to elliptic fibrations $Y$ that engineer a 6d $(1,0)$ SCFT, can be constructed.

As is now common in the literature, we denote a curve of
self-interaction $(-n)$ with a non-trivial fiber associated with a gauge
algebra $\mathfrak{g}$ by:
\begin{equation}
		\overset{\mathfrak{g}}{n}\,.
\end{equation}
When the fiber is trivial, $\mathfrak{g}=\varnothing$, we omit it and only
write the curve associated with the tensor multiplet. Furthermore, if two
curves intersect, which they can only do with intersection number $1$, they are depicted side by side. 

The tensor branch of any 6d $(1,0)$ SCFT can then be constructed from
a small number of building blocks associated to non-Higgsable clusters (NHCs)
\cite{Morrison:1996na, Morrison:1996pp, Morrison:2012np}:
\begin{equation}\label{NHC-list}
		\overset{\mathfrak{su}_3}{3}\,,\quad
		\overset{\mathfrak{so}_8}{4}\,,\quad
		\overset{\mathfrak{f}_4}{5}\,,\quad
		\overset{\mathfrak{e}_6}{6}\,,\quad
		\overset{\mathfrak{e}_7}{7}\,,\quad
		\overset{\mathfrak{e}_7}{8}\,,\quad
		\overset{\mathfrak{e}_8}{(12)}\,,\quad
		\overset{\mathfrak{su}_2}{2}\overset{\mathfrak{g}_2}{3}\,,\quad
		2\overset{\mathfrak{su}_2}{2}\overset{\mathfrak{g}_2}{3}\,,\quad
		\overset{\mathfrak{su}_2}{2}\overset{\mathfrak{so}_7}{3}\overset{\mathfrak{su}_2}{2}\,,
\end{equation}
or to ADE Dynkin diagrams constructed out of $(-2)$-curves:
\begin{equation}
		\underbrace{2\dots2}_{N-1}\,,\quad
		\underbrace{2\dots2}_{N-3}\overset{\displaystyle 2}{2}2\,,\quad
		22\overset{\displaystyle 2}{2}22\,,\quad
		222\overset{\displaystyle 2}{2}22\,,\quad
		2222\overset{\displaystyle 2}{2}22\,.\quad
\end{equation}
The fiber over each of these curves may be tuned so that it corresponds to a
larger algebra as long as it leads to a well-defined elliptic fibration. These enhancements may
give rise to additional matter fields on these curves, and there may be a
flavor symmetry $\mathfrak{f}$ rotating them. We denote the presence of
additional flavor symmetries as:
\begin{equation}
		\overset{\mathfrak{g}}{n}[\mathfrak{f}]\,.
\end{equation}

Any SCFT is then obtained by gluing non-Higgsable clusters -- possibly with
enhanced fibers -- via $(-1)$-curves that have a flavor symmetry
$\mathfrak{f}$. For instance, if we consider two curves
$\overset{\mathfrak{g}_L}{m}$ and $\overset{\mathfrak{g}_R}{n}$, we can gauge a
subalgebra of $\mathfrak{f}$ to obtain a new theory:
\begin{equation}\label{gluing-NHC}
		\overset{\mathfrak{g}_L}{m}\,
		\overset{\mathfrak{g}}{1}\,
		\overset{\mathfrak{g}_R}{n}
		\,,\qquad
		\mathfrak{g}_L\oplus \mathfrak{g}_R \subset \mathfrak{f}\,.
\end{equation}
Note that when the subalgebra is not maximal, there might be a residual flavor
symmetry in the new configuration.

This process can then be repeated as many times as necessary to obtain bases
of an elliptic fibration with an arbitrary number of curves subject to the
condition that there are only minimal singularities and that the adjacency
matrix $A^{ij}$ is
negative definite. On the tensor branch, where the curves have finite volume, one
can then find the gauge spectrum straightforwardly. In practice, it is done simply
by reading off the matter content from tables in the large majority of cases.
For a concise review of the tensor branch description of 6d $(1,0)$ SCFTs we have
summarized here and the subtleties that may arise, we defer to
\cite{Heckman:2018jxk}.

A simple example of this pictorial description of the geometry is that of
minimal $(\mathfrak{e}_6, \mathfrak{e}_6)$ conformal matter. In the blown-up
phase, it is constructed out of two $(-1)$-curves with trivial fibers
intersecting a $(-3)$-curve with a type-$IV$ fiber corresponding to an
$\mathfrak{su}_3$ algebra. An inspection of the geometry further reveals the
presence of two non-compact curves with $\mathfrak{e}_6$ fibers, giving rise to
two flavor symmetries. The theory on the tensor branch is therefore denoted by:
\begin{equation}\label{minimal-e6-quiver}
		[\mathfrak{e}_6]1\overset{\mathfrak{su}_3}{3}1[\mathfrak{e}_6]\,.
\end{equation}
The conformal fixed point is then reached by simultaneously shrinking every
curve to zero volume. This example will be used throughout this section to
illustrate some of the features of long quivers.

Depictions of the blown-up geometry -- or equivalently of the tensor branch of a
6d SCFT -- like the one in equation \eqref{minimal-e6-quiver} are called
\emph{generalized quivers} (or often simply quivers, for short), dubbed so due
to their resemblance with those appearing in usual gauge theories, but where
the links symbolizing bifundamental hypermultiplets are now potentially
replaced by more complicated objects, generalizing the notion of matter. A
quiver describing the generic point of tensor branch of a 6d SCFT is unique
in all but a handful of cases; by abuse of language we will often refer to a
specific SCFT and its quiver interchangeably.

The possible curve configurations of the generalized quivers are very
constrained by demanding that the elliptically-fibered Calabi--Yau is well
defined, or equivalently by demanding the absence of gauge anomalies of the field
theory. When the number of curves -- or equivalently tensor multiplets -- is
taken to be large enough, it turns out that they must arrange themselves into a
long linear spine with repeating patterns, up to possible ``decorations'' at
each side. The constituents of the spine are themselves 6d $(1,0)$
SCFTs called minimal $(\mathfrak{g}, \mathfrak{g})$ conformal matter
\cite{DelZotto:2014hpa}, of which equation \eqref{minimal-e6-quiver} is an example. In the
M-theory picture, these correspond to the worldvolume theory on a single
M5-brane probing a $\mathbb{C}^2/\Gamma$ singularity, where $\Gamma$ is a
finite subgroup of $SU(2)$. These theories follow an ADE classification and
have a $\mathfrak{f} = \mathfrak{g}\oplus\mathfrak{g}$ flavor symmetry, where
$\mathfrak{g}$ is of the same ADE type as $\Gamma$. 

When $\Gamma=\mathbb{Z}_K$, the SCFT is nothing else but a single
hypermultiplet transforming in the bifundamental representation of
$\mathfrak{su}_K\oplus\mathfrak{su}_K$. For the other types of simply-laced
algebras, conformal matter can therefore be thought of as a generalization of
ordinary bifundamental matter -- hence their name -- and we will use the
following shorthand to depict them:
\begin{equation}\label{dash-notation}
	[\mathfrak{g}]\text{---}[\mathfrak{g}]\,, 
\end{equation} 
where the brackets indicate the presence of the two flavor symmetries. The ADE
algebra $\mathfrak{g}$ uniquely determines the quiver of minimal conformal
matter, and is summarized in Table \ref{tbl:CM-and-nodes}.

\begin{table}[t]
    \centering
    \begin{threeparttable}   
        \begin{tabular}{ccc}
            \toprule
			$\mathfrak{g}$ & $A_0^\mathfrak{g}:\quad[\mathfrak{g}]\text{---}[\mathfrak{g}]$ & Node $n_{\mathfrak{g}}$\\\midrule
			$\mathfrak{su}_K$ & $[\mathfrak{su}_K]\cdot[\mathfrak{su}_K]$ & $\overset{\mathfrak{su}_K}{2}$\\
			$\mathfrak{so}_{2K}$ & $[\mathfrak{so}_{2K}]\overset{\mathfrak{sp}_{K-4}}{1}[\mathfrak{so}_{2K}]$ & $\overset{\mathfrak{so}_{2K}}{4}$\\
			$\mathfrak{e}_{6}$ & $[\mathfrak{e}_{6}]1\overset{\mathfrak{su}_{3}}{3}1[\mathfrak{e}_{6}]$ & $\overset{\mathfrak{e}_{6}}{6}$\\
			$\mathfrak{e}_{7}$ & $[\mathfrak{e}_{7}]1\overset{\mathfrak{su}_{2}}{2}\overset{\mathfrak{so}_{7}}{3}\overset{\mathfrak{su}_{2}}{2}1[\mathfrak{e}_{7}]$ & $\overset{\mathfrak{e}_{7}}{8}$\\
			$\mathfrak{e}_{8}$ & $[\mathfrak{e}_{8}]1\,2\overset{\mathfrak{su}_{2}}{2}\overset{\mathfrak{g}_2}{3}1\overset{\mathfrak{f}_{4}}{5}1\overset{\mathfrak{g}_2}{3}\overset{\mathfrak{su}_{2}}{2}2\,1[\mathfrak{e}_{8}]$ & $\overset{\mathfrak{e}_{8}}{(12)}$\\
            \bottomrule
        \end{tabular}
    \end{threeparttable}
	\caption{ADE classification of minimal conformal matter and the associated
			nodes of self-intersection $(-n_\mathfrak{g})$. Type $A$ conformal matter corresponds to a
			$(1,0)$ hypermultiplet transforming in the
			bifundamental representation of
			$\mathfrak{su}(K)\oplus\mathfrak{su}(K)$, and is simply denoted by a
			dot $(\cdot)$. 
	}
    \label{tbl:CM-and-nodes}
\end{table}

The interpretation of minimal conformal matter as a generalization of
bifundamental hypermultiplets does not simply come from their flavor
symmetries, but as simple building blocks of more involved theories. Indeed, we
can ``glue'' two minimal conformal matter theories together to obtain a larger
SCFT through a procedure called \emph{fusion} \cite{Heckman:2018pqx}, which
generalizes the usual notion of gauging. Indeed, while in four dimensions
gauging a flavor symmetry by introducing a vector multiplet to mediate the
interaction is sufficient, this is not always the case in six dimensions, and
the new theory may still be plagued by gauge anomalies. However, one of the
features of six-dimensional QFTs is the presence of tensor multiplets, which can
be involved in a Green--Schwarz--West--Sagnotti mechanism \cite{Green:1984sg,
Green:1984bx,Sagnotti:1992qw} curing any such anomalies, and
leading to a well-defined theory. We come back to this point in more detail in
Section \ref{sec:I8}.

From the geometric point of view, the fusion procedure corresponds to identifying the
two non-compact curves associated with the flavor symmetries and make the
resulting curve compact. In the quiver language, these particular curves are
called \emph{nodes}, and their self-intersection numbers are fixed by demanding
consistency of the F-theory geometry. The particular numerology depends on the
ADE algebra, and are given in Table \ref{tbl:CM-and-nodes}.\footnote{ Note that
		for minimal conformal matter of type $\mathfrak{so}(2K)$, the flavor
		symmetry enhances,
		$\mathfrak{so}(2K)\oplus\mathfrak{so}(2K)\to\mathfrak{so}(4K)$, and
		when $\mathfrak{g}=\mathfrak{so}(8)$, we have an undecorated
		$(-1)$-curve associated with the E-string theory endowed with
		$\mathfrak{e}_8$ flavor. However, when two of them are fused together, only a
		$\mathfrak{so}(2K)\oplus\mathfrak{so}(2K)$ flavor symmetry remains. As we
		are only discussing long quivers in this work, we will not encounter
		such enhancements.
}

For instance, in the case $\mathfrak{g}=\mathfrak{e}_6$ 
encountered in equation \eqref{minimal-e6-quiver}, the fusion process applied
to two minimal conformal matter theories leads to the presence of a
$(-6)$-curve with a type-IV${}^\ast$ fiber associated with an $\mathfrak{e}_6$
gauge interaction:
\begin{equation}
		[\mathfrak{e}_6]1\overset{\mathfrak{su}_3}{3}1[\mathfrak{e}_6]\quad\oplus\quad~
	[\mathfrak{e}_6]1\overset{\mathfrak{su}_3}{3}1[\mathfrak{e}_6]\quad = \quad
	[\mathfrak{e}_6]1\overset{\mathfrak{su}_3}{3}1(\overset{\mathfrak{e}_6}{6})1\overset{\mathfrak{su}_3}{3}1[\mathfrak{e}_6]\,.
\end{equation}
Heuristically, we obtain something very similar to quiver theories, namely
links of (conformal) matter and nodes associated with gauging. However, in our
case the links correspond to conformal matter rather than the usual
bifundamental hypermultiplets familiar from 4d $\mathcal{N}=2$ gauge theories, while
the nodes necessitate the presence of tensor multiplets to mediate a
Green--Schwarz--West--Sagnotti mechanism and ensure that the resulting theory
is free of any gauge anomalies. The fusion process can then be concisely
summarized utilizing the notation introduced in equation \eqref{dash-notation}
as:
\begin{equation}
	[\mathfrak{g}]\text{---}[\mathfrak{g}]\quad\oplus\quad
	[\mathfrak{g}]\text{---}[\mathfrak{g}]\quad =\quad
	[\mathfrak{g}]\text{---}\,\mathfrak{g}\,\text{---}[\mathfrak{g}]\,,
\end{equation}
where we used the symbol $\oplus$ to denote fusion. The absence of brackets for
the middle algebra indicates that the flavor symmetry has been gauged, and that
there is an additional compact curve corresponding to a tensor. This notation
for fusion of minimal conformal matter completely determines the quiver, the
links and nodes being those given in Table \ref{tbl:CM-and-nodes}.

This process can of course be repeated \emph{ad nauseam} to obtain patterns of
minimal conformal matter joined by vector and tensor multiplets. When $N$
minimal conformal matter of the same type undergo the fusion process, one
obtains \emph{rank $N$ conformal matter}, $A_{N-1}^{\mathfrak{g}}$. In M-theory, this theory is realized
as a stack of $N$ M5-branes probing a $\mathbb{C}^2/\Gamma$ singularity, where
$\Gamma$ is the McKay-dual discrete group of $\mathfrak{g}$. 

We will refer to the SCFTs that are part of the infinite series described in
the introduction as \emph{long quivers}. Such long quivers can be constructed
by fusing additonal building blocks on each ends of higher-rank conformal
matter. Furthermore, all long quivers can be understood as deformations of
three families of \emph{parent} or ultraviolet theories
\cite{Heckman:2016ssk,Heckman:2018pqx}:
\begin{enumerate}
		\item (Fractional) higher-rank conformal matter, 
                $\mathcal{A}_{N-1;f_L,f_R}^{\mathfrak{g}}$.
		\item Theories whose bases are generalized type-D
				orbifolds, $\mathcal{D}_{N;f}^{\mathfrak{g}}$, existing only
				for a few specific choice of algebras.
		\item Orbi-instantons, $\mathcal{O}_{N;f}^{\mathfrak{g}}$, a class of
				SCFTs with an $\mathfrak{e}_8$ flavor symmetry on one end.
\end{enumerate}

Given a 6d SCFT with a flavor symmetry, one can give a vacuum expectation value
to the associated moment map -- the scalar operator inside the same
supermultiplet as the flavor current -- triggering a renormalization group flow.
In the infrared, one obtains a new conformal fixed point. In the F-theory
picture, these Higgs branch flows correspond to complex-structure deformations
of the Calabi--Yau threefold. For long quivers, the deformations are labelled
by nilpotent orbits of the flavor symmetries, or by embeddings of an ADE discrete
group into $E_8$ in the case of orbi-instanton. In the remainder of this
section, we will review how to construct each of the classes of parent
theories, and then turn to their deformations.

Before doing so however, we further introduce the concept of the \emph{partial
tensor branch} (PTB), which provides an alternative bookkeeping device for the
parent theories. Given the generalized quiver depicting the tensor branch of a
6d SCFT, one can successively blow down all $(-1)$-curves, reaching the so-called endpoint of the base. Ignoring the fiber data,
there are only a handful of possible partial tensor branches, which follow an
ADE classification: 
\begin{equation}
	\begin{gathered}
		n_1\,n_2\,\dots\, n_{N-2}\,n_{N-1}\,,\qquad\qquad
		2\,\overset{\displaystyle 2}{n_3}\,n_4\,\dots\, n_{N-1}\,n_{N}\,,\\
		2\,2\,\overset{\displaystyle 2}{2}\,2\,2\,,\qquad
		2\,2\,\overset{\displaystyle 2}{2}\,2\,2\,2\,,\qquad
		2\,2\,\overset{\displaystyle 2}{2}\,2\,2\,2\,2\,.
	\end{gathered}
\end{equation}
These partial tensor branches were used in the original classification
\cite{Heckman:2013pva} to find the possible bases, see equation
\eqref{eqn:orbifold}, and one can read of the type of generalized orbifold from
the intersection pattern. Generically, partial tensor branches of a
quiver have non-minimal singularities supported over the intersection points of the curves in the base. The original quivers on the generic point
of the tensor branch is then recovered through a series of blow-ups, reversing
the process.\footnote{Technically, orbi-instantons are
		associated with the trivial orbifold and have in principle an empty
		partial tensor branch as we blow down successively all $(-1)$-curves.
		However, one generally defines their partial tensor branch to be
		non-trivial and reproduce the M-theory picture, see Section
\ref{sec:orbi-instantons}.}

This special point of the tensor branch has an interpretation in the M-theory
language: it is the point where all fractional M5-branes recombine into the
maximal number of full M5-branes. This description is particularly useful, as
it makes the fusion processes, as well as Higgs branch deformations, clearer.
Moreover, when discussing the anomalies of the associated 6d SCFTs, once the
anomaly polynomial of minimal conformal matter is known, the partial tensor
branch intersection pattern is -- up to a few additional input data -- enough to
compute that of any long quiver.

\subsection{(Fractional) Conformal Matter}\label{sec:fractional-CM}

The simplest class of parent theories with long quivers is the set of rank $N$
$(\mathfrak{g}, \mathfrak{g})$ conformal matter theories:
$\mathcal{A}_{(N,N-1)}^\mathfrak{g} = A_{N-1}^\mathfrak{g}$. In
M-theory, they are realized as the worldvolume theory of a stack of $N$
M5-brane probing a $\mathbb{C}^2/\Gamma$ orbifold, where $\Gamma\subset SU(2)$
is the McKay-dual discrete group of $\mathfrak{g}$ \cite{DelZotto:2014hpa}. In
the F-theory picture, one engineers them by compactification on a
elliptically-fibered Calabi--Yau three-fold with a base
$B=\mathbb{C}^2/\mathbb{Z}_N$ and fibers of type
$\mathfrak{g}$.\footnote{Note that, while similar in form, the two orbifolds
		appearing in either descriptions are different. In M-theory, the
quotient $\Gamma$ is related to the flavor algebra $\mathfrak{g}$, while in
F-theory, it is associated with the intersection pattern $A_{N-1}$ of the
curves.} The partial tensor branch of these theories is given by a collection
of $(-2)$-curves with $\mathfrak{g}$ fibers intersecting like an $A_{N-1}$
Dynkin diagram:
\begin{equation}\label{PTB-CM}
    \underbrace{\,\overset{\mathfrak{g}}{2} \cdots \overset{\mathfrak{g}}{2} \,}_{N-1} \,.
\end{equation}
Except in cases where $\mathfrak{g}=\mathfrak{su}(K)$, these
geometries have non-minimal singularities over the intersection points of the $(-2)$-curves; thus, we need to perform
a series of blow-ups to move to the generic point of the tensor branch.  As an
example, let us consider the case where $\mathfrak{g} = \mathfrak{e}_6$.  After
blowing-up, one finds a repeating
pattern of minimal conformal matter:
\begin{equation}
	A^{\mathfrak{e}_6}_{N-1}:\qquad
	\underbrace{\,\overset{\mathfrak{e}_6}{2} \cdots \overset{\mathfrak{e}_6}{2} \,}_{N-1} 
	\qquad \longrightarrow\qquad
	[\mathfrak{e}_6]1\overset{\mathfrak{su}_3}{3}1 \overset{\mathfrak{e}_6}{6}1\overset{\mathfrak{su}_3}{3}1\overset{\mathfrak{e}_6}{6}\cdots\overset{\mathfrak{e}_6}{6}1\overset{\mathfrak{su}_3}{3}1\overset{\mathfrak{e}_6}{6}1\overset{\mathfrak{su}_3}{3}1\,[\mathfrak{e}_6]\,.
\end{equation}
The sequence $1\overset{\mathfrak{su}_3}{3}1$ occurs $N$ times, and there are
$N-1$ nodes $\overset{\mathfrak{e}_6}{6}$. For the other cases, one can go
through a similar procedure and find a long spine of $N$ minimal conformal
matter $A_0^{\mathfrak{g}}$ that underwent the fusion process. We can
therefore use the notation of links $[\mathfrak{g}]\text{---}[\mathfrak{g}]$ to
denote the resulting curve configurations as:
\begin{equation}\label{CM-quiver-pict}
	A^{\mathfrak{g}}_{N-1}:\qquad
	\underbrace{\,\overset{\mathfrak{g}}{2} \cdots \overset{\mathfrak{g}}{2} \,}_{N-1} 
	\qquad \longrightarrow\qquad
	[\mathfrak{g}]\underbrace{\vphantom{\overset{\mathfrak{g}}{2}}\,\text{---}\,\mathfrak{g}\,\text{---}\dots\text{---}\,\mathfrak{g}\text{---}\,}_{N}\,[\mathfrak{g}]\,,
\end{equation}
where the links and nodes are summarized in Table \ref{tbl:CM-and-nodes}. In
all cases, there is a flavor symmetry associated with either outermost curves
fixed by the choice of $\mathfrak{g}$, and the total flavor symmetry of
higher-rank conformal matter $A_{N-1}^{\mathfrak{g}}$ is:
\begin{equation}
    \mathfrak{f} = \mathfrak{g} \oplus \mathfrak{g} \,,
\end{equation}
which we will often refer to as the ``left'' and ``right'' factors,
respectively.

\subsubsection{Fractional Conformal Matter}

We have seen that there are additional SCFTs with long quivers that are
associated with generalized $A$-type orbifolds $\mathcal{A}_{(p,q)}$. Moreover,
for a large-enough number of curves and given an ADE algebra, there are only a
few possibilities. The associated long quivers are again made out of conformal
matter as in equation \eqref{CM-quiver-pict}, but either ends of the spine are
now truncated versions of conformal matter, where one or more curves have been
removed. As mentioned in the introduction, when discussing long quivers the
choices of $(p,q)$ and $\mathfrak{g}$ is quite constrained, and it is more
convenient to use the fractions $(f_L, f_R)$ introduced around equation
\eqref{eqn:possfractions}. We denote the base of the elliptic fibration as
$\mathcal{A}_{N-1;f_L,f_R}$. The triplet $(f_L, f_R, \mathfrak{g})$ then
constrains the form of the quiver for the theory
$\mathcal{A}_{N-1;f_L,f_R}^{\mathfrak{g}}$ uniquely.

To illustrate this, let us come back to the case where
$\mathfrak{g}=\mathfrak{e}_6$. All possible A-type long quivers for this choice
of algebra take the form:
\begin{equation}
		\mathcal{A}_{N-1;f_L,f_R}^{\mathfrak{g}}:\qquad	\mathcal{L}_{f_L}\underbrace{1\overset{\mathfrak{su}_3}{3}1\overset{\mathfrak{e}_6}{6} \cdots \overset{\mathfrak{e}_6}{6} 1\overset{\mathfrak{su}_3}{3}1}_{N-2}\mathcal{L}_{f_R} \,,
\end{equation}
where there are $N-2$ minimal conformal matter in the center, and
$\mathcal{L}_L$ and $\mathcal{L}_R$ denote an extra link that can be attached
at each ends of the spine, and which can be chosen among the following
possibilities:
\begin{equation}\label{e6-fractions-curves}
				\mathcal{L}_{1} = \overset{\mathfrak{e}_6}{6}\,1\overset{\mathfrak{su}_3}{3}1[\mathfrak{e}_6]\,,\qquad
				\mathcal{L}_{\frac{2}{3}} = \overset{\mathfrak{e}_6}{6}\,1\overset{\mathfrak{su}_3}{3}\,,\qquad
				\mathcal{L}_{\frac{1}{2}} = \overset{\mathfrak{e}_6}{6}\,1[\mathfrak{su}_3]\,,\qquad
				\mathcal{L}_{\frac{1}{3}} = \overset{\mathfrak{e}_6}{6}\,.
\end{equation}
Note that for $\mathcal{L}_{f_L}$, the curve configuration is understood as
being read in such a way that the node $\overset{\mathfrak{e}_6}{6}$ is
connected to the rest of the spine. We can see that $f_{L,R}=1$ gives a
complete conformal matter, and smaller fraction corresponds to removing some of the curves.

Similar patterns occur for all other ADE algebras. The possible fractions $f$
depends on the algebra $\mathfrak{g}$, and are of the form:
\begin{equation}
		f= \frac{2	k}{n_\mathfrak{g}}\,, \qquad k= 1, 2, \dots, \frac{n_\mathfrak{g}}{2}\,,
\end{equation}
where $n_\mathfrak{g} = 2, 4, 6, 8, 12$ for $\mathfrak{g} = \mathfrak{su}(K)$,
$\mathfrak{so}(2K)$, $\mathfrak{e}_6$, $\mathfrak{e}_7$, $\mathfrak{e}_8$,
respectively. Note that $n_\mathfrak{g}$ is (minus) the self-intersection of
the corresponding nodes, see Table \ref{tbl:CM-and-nodes}. In the case of
$\mathfrak{g}=\mathfrak{su}(K)$, there are no new possibilities as the only
allowed fraction is $f=1$, while $\mathfrak{g} = \mathfrak{e}_8$ has the
largest number of choices. As the fraction becomes smaller, we are retaining fewer and
fewer curves to obtain only part of a full conformal matter theory, giving its
name to this class of theories. This further means that the choice of $f$ is
also changing the flavor symmetry of the theory to a -- possibly trivial --
subalgebra $\mathfrak{g}_f\subseteq\mathfrak{g}$. All combinations of fractions
of a given algebra, their curve configurations, flavor symmetry
$\mathfrak{g}_f$, and the associated data that will be useful for the
compuation of their anomaly polynomials are collated in Table
\ref{tbl:eattachments}.

\begin{table}[p]
    \centering
    \begin{threeparttable}
        \begin{tabular}[t]{ccrcrcr}
            \toprule
			$\mathfrak{g}$ & $f$ & $[\mathfrak{g}_f]\,\text{---}\,\mathfrak{g}$ & $\mathfrak{g}_f$ & \multicolumn{1}{c}{PTB} & $d_f$ & $e$\\\midrule
			\multirow{2}{*}{$\mathfrak{so}_{2K}$}
			& $\frac{1}{2}$ & $\overset{\mathfrak{so}_{2K}}{4}$ & $\mathfrak{sp}_{K-4}$ & $\overset{\mathfrak{so}_{2K}}{3}\cdots$ & $2$ & $0$  \\
			& $1$ & $\overset{\mathfrak{sp}_{K-4}}{1}\overset{\mathfrak{so}_{2K}}{4}$ & $\mathfrak{so}_{2K}$ & $\overset{\mathfrak{so}_{2K}}{2}\cdots$ & $1$ & $0$  \\
			\midrule
			\multirow{4}{*}{$\mathfrak{e}_6$} & 
			$\frac{1}{3}$ & $\overset{\mathfrak{e}_6}{6}$ & $\varnothing$ & $\overset{\mathfrak{e}_6}{4}\cdots$ & $3$ & $-4$  \\
			& $\frac{1}{2}$ & $1\overset{\mathfrak{e}_6}{6}$ & $\mathfrak{su}_3$ & $\overset{\mathfrak{e}_6}{3}\cdots$ & $2$  & $0$  \\
			& $\frac{2}{3}$ & $\overset{\mathfrak{su}_3}{3}1\overset{\mathfrak{e}_6}{6}$ & $\varnothing$ & $\overset{\mathfrak{su}_3}{2}\overset{\mathfrak{e}_6}{3}\cdots$ & $3$  & $4$  \\
            & $1$ & $1\overset{\mathfrak{su}_3}{3}1\overset{\mathfrak{e}_6}{6}$ & $\mathfrak{e}_6$ & $\overset{\mathfrak{e}_6}{2}\cdots$ & $1$ & $0$\\
            \midrule
            \multirow{6}{*}{$\mathfrak{e}_7$} 
			& $\frac{1}{4}$ & $\overset{\mathfrak{e}_7}{8}$ & $\varnothing$ & $\overset{\mathfrak{e}_7}{5}\cdots$ & $4$ & $-9$ \\
			& $\frac{1}{3}$ & $1\overset{\mathfrak{e}_7}{8}$ & $\mathfrak{su}_2$ & $\overset{\mathfrak{e}_7}{4}\cdots$ & $3$ & $-4$ \\
			& $\frac{1}{2}$ & $\overset{\mathfrak{su}_2}{2}1\overset{\mathfrak{e}_7}{8}$ & $\mathfrak{so}_7$ & $\overset{\mathfrak{e}_7}{3}\cdots$ & $2$ & $0$ \\
            & $\frac{2}{3}$ & $\overset{\mathfrak{so}_7}{3}\overset{\mathfrak{su}_2}{2}1\overset{\mathfrak{e}_7}{8}$ & $\mathfrak{su}_2$ & $\overset{\mathfrak{so}_7}{2}\overset{\mathfrak{e}_7}{3}\cdots$ & $3$ & $4$ \\
            & $\frac{3}{4}$ & $\overset{\mathfrak{su}_2}{2}\overset{\mathfrak{so}_7}{3}\overset{\mathfrak{su}_2}{2}1\overset{\mathfrak{e}_7}{8}$ & $\varnothing$ & $\overset{\mathfrak{su}_2}{2}\overset{\mathfrak{so}_7}{2}\overset{\mathfrak{e}_7}{3}\cdots$ & $4$ & $9$ \\
			& $1$ & $1\overset{\mathfrak{su}_2}{2}\overset{\mathfrak{so}_7}{3}\overset{\mathfrak{su}_2}{2}1\overset{\mathfrak{e}_7}{8}$ & $\mathfrak{e}_7$ & $\overset{\mathfrak{e}_7}{2}\cdots$ & $1$ & $0$ \\
            \midrule
            \multirow{12}{*}{$\mathfrak{e}_8$} 
            & $\frac{1}{6}$ & $\overset{\mathfrak{e}_8}{(12)}$ & $\varnothing$ & $\overset{\mathfrak{e}_8}{7}\cdots$ & $6$ & $-20$  \\
			& $\frac{1}{5}$ & $1\overset{\mathfrak{e}_8}{(12)}$ & $\varnothing$ & $\overset{\mathfrak{e}_8}{6}\cdots$ & $5$ & $-\frac{72}{5}$ \\
            & $\frac{1}{4}$ & $2\,1\overset{\mathfrak{e}_8}{(12)}$ & $\mathfrak{su}_2$ & $\overset{\mathfrak{e}_8}{5}\cdots$ & $4$ & $-9$ \\
            & $\frac{1}{3}$ & $\overset{\mathfrak{su}_2}{2}2\,1\overset{\mathfrak{e}_8}{(12)}$ & $\mathfrak{g}_2$ & $\overset{\mathfrak{e}_8}{4}\cdots$ & $3$ & $-4$ \\
			& $\frac{2}{5}$ & $\overset{\mathfrak{g}_2}{3}\overset{\mathfrak{su}_2}{2}2\,1\overset{\mathfrak{e}_8}{(12)}$ & $\varnothing$ & $\overset{\mathfrak{g}_2}{2}\overset{\mathfrak{e}_8}{4}\cdots$ & $5$ & $0$ \\
            & $\frac{1}{2}$ & $1\overset{\mathfrak{g}_2}{3}\overset{\mathfrak{su}_2}{2}2\,1\overset{\mathfrak{e}_8}{(12)}$ & $\mathfrak{f}_4$ & $\overset{\mathfrak{e}_8}{3}\cdots$ & $2$ & $0$ \\
            & $\frac{3}{5}$ & $\overset{\mathfrak{f}_4}{5}1\overset{\mathfrak{g}_2}{3}\overset{\mathfrak{su}_2}{2}2\,1\overset{\mathfrak{e}_8}{(12)}$ & $\varnothing$ & $\overset{\mathfrak{f}_4}{3}\overset{\mathfrak{e}_8}{3}\cdots$ & $5$ & $0$ \\
            & $\frac{2}{3}$ & $1\overset{\mathfrak{f}_4}{5}1\overset{\mathfrak{g}_2}{3}\overset{\mathfrak{su}_2}{2}2\,1\overset{\mathfrak{e}_8}{(12)}$ & $\mathfrak{g}_2$ & $\overset{\mathfrak{f}_4}{2}\overset{\mathfrak{e}_8}{3}\cdots$ & $3$ & $4$  \\
            & $\frac{3}{4}$ & $\overset{\mathfrak{g}_2}{3}1\overset{\mathfrak{f}_4}{5}1\overset{\mathfrak{g}_2}{3}\overset{\mathfrak{su}_2}{2}2\,1\overset{\mathfrak{e}_8}{(12)}$ & $\mathfrak{su}_2$ & $\overset{\mathfrak{g}_2}{2}\overset{\mathfrak{f}_4}{2}\overset{\mathfrak{e}_8}{3}\cdots$ & $4$ & $9$ \\
			& $\frac{4}{5}$ & $\overset{\mathfrak{su}_2}{2}\overset{\mathfrak{g}_2}{3}1\overset{\mathfrak{f}_4}{5}1\overset{\mathfrak{g}_2}{3}\overset{\mathfrak{su}_2}{2}2\,1\overset{\mathfrak{e}_8}{(12)}$ & $\varnothing$ & $\overset{\mathfrak{su}_2}{2}\overset{\mathfrak{g}_2}{2}\overset{\mathfrak{f}_4}{2}\overset{\mathfrak{e}_8}{3}\cdots$ & $5$ & $\frac{72}{5}$ \\
            & $\frac{5}{6}$ & $2\overset{\mathfrak{su}_2}{2}\overset{\mathfrak{g}_2}{3}1\overset{\mathfrak{f}_4}{5}1\overset{\mathfrak{g}_2}{3}\overset{\mathfrak{su}_2}{2}2\,1\overset{\mathfrak{e}_8}{(12)}$ & $\varnothing$ & $2\overset{\mathfrak{su}_2}{2}\overset{\mathfrak{g}_2}{2}\overset{\mathfrak{f}_4}{2}\overset{\mathfrak{e}_8}{3}\cdots$ & $6$ & $20$ \\
			& $1$ & $1\,2\overset{\mathfrak{su}_2}{2}\overset{\mathfrak{g}_2}{3}1\overset{\mathfrak{f}_4}{5}1\overset{\mathfrak{g}_2}{3}\overset{\mathfrak{su}_2}{2}2\,1\overset{\mathfrak{e}_8}{(12)}$ & $\mathfrak{e}_8$ & $\overset{\mathfrak{e}_8}{2}\cdots$ & $1$ & $0$\\
            \bottomrule
        \end{tabular}
    \end{threeparttable}
	\caption{Data for fractional conformal matter,
			$\mathcal{A}_{N-1;f_L,f_R}^{\mathfrak{g}}$. For ease of
			visualization, the building blocks for minimal conformal matter includes the node. The partial
			tensor branch (PTB) gives the curve configuration obtained after
			successively blowing down all $(-1)$-curves.
	}
    \label{tbl:eattachments}
\end{table}

In the same way that rank $N$ conformal matter can be built out of minimal
conformal matter via the fusion process, we can define a new type of building
block which is a fractional version of minimal conformal matter,
$\mathcal{A}_{0;f,1}^{\mathfrak{g}}$. Pictorially, we will denote them as:
\begin{equation}
	\mathcal{A}^{\mathfrak{g}}_{0;f,1}:\qquad [\mathfrak{g}_f]\text{---}[\mathfrak{g}]\,.
\end{equation}
This theory can be glued to an arbitrary number of complete conformal matter
(that is, those with $f=1$) through fusion to obtain the theory
$\mathcal{A}^{\mathfrak{g}}_{N-1;f_L,f_R}$. 
\begin{equation}\label{pict-frac-CM}
		\mathcal{A}^{\mathfrak{g}}_{N-1;f_L,f_R}:\qquad
		[\mathfrak{g}_{f_L}]\,\text{---}\,\mathfrak{g}\,\overbrace{\vphantom{\overset{\mathfrak{g}}{2}}\text{---}\,\mathfrak{g}\,\text{---}\dots\text{---}\,\mathfrak{g}\,\text{---}}^{N-2}\,\mathfrak{g}\,\text{---}\,[\mathfrak{g}_{f_R}]\,.
\end{equation}
Each link can be considered as a theory in its own right. As they are only
part of a full-fledged minimal conformal matter, we refer to these theories as
\emph{fractional} conformal matter.  We can therefore unambiguously define the
parent theories of all long quivers associated with an A-type generalized
orbifolds. Indeed, given a quiver in the notation above, the complete curve
configurations can be read out from Tables \ref{tbl:CM-and-nodes} and
\ref{tbl:eattachments}. Of course, when $f_L=1=f_R$, we recover rank $N$
conformal matter as shown in equation \eqref{CM-quiver-pict} and we have
$\mathcal{A}^{\mathfrak{g}}_{N-1,1,1} = A^{\mathfrak{g}}_{N-1}$.

While the assignment of a fractional number to each of the curve configurations
of fractional minimal conformal matter may at first seem somewhat \emph{ad
hoc}, it finds its origin in the M-theory construction. There,
these SCFTs are realized as the worldvolume of M5-branes probing a
(partially-)frozen version of the $\mathbb{C}^2/\Gamma$ singularity
\cite{deBoer:2001wca, Witten:1997bs, Tachikawa:2015wka, Ohmori:2015pua,
Ohmori:2015pia, Mekareeya:2017sqh}, where there exist BPS solutions
characterized by discrete three-form fluxes:
\begin{equation}
		\int_{S^3/\Gamma} C_3 =f \text{ mod }1\,.
\end{equation}
These frozen versions of the singularity are associated with the (possibly empty)
Lie algebra $\mathfrak{g}_f\subset\mathfrak{g}$, that depends on the
value of $f$, and can now be non-simply-laced. The possible values of the
fractions and the algebras arising in this way perfectly match the numerology of the
F-theory picture, as expected. 

The partial tensor branch of fractional conformal matter
$\mathcal{A}^{\mathfrak{g}}_{N-1;f_L,f_R}$ is slightly different that those of
higher-rank conformal matter, see equation \eqref{PTB-CM}. When the fraction is
of the form $f=\frac{1}{d_f}$, we obtain a curve of self-intersection
$-(d_f+1)$ with an algebra $\mathfrak{g}$ rather than $-2$. When the numerator
is different than one, we may obtain different patterns, which in the M-theory
picture can be understood as a recombination of the M5-branes into fractional
branes \cite{Mekareeya:2017sqh}. For instance the theory
$\mathcal{A}^{\mathfrak{g}}_{N-1;\frac{1}{2},\frac{2}{3}}$ has a partial tensor
branch of the form
\begin{equation}
		[\mathfrak{su}_3]1\,\overset{\mathfrak{e}_6}{6}\,1\overset{\mathfrak{su}_3}{3}1\overset{\mathfrak{e}_6}{6}\,1\overset{\mathfrak{su}_3}{3}1\,\cdots\,\overset{\mathfrak{e}_6}{6}\,1\overset{\mathfrak{su}_3}{3}1\overset{\mathfrak{e}_6}{6}\,1\overset{\mathfrak{su}_3}{3}
		\qquad\longrightarrow\qquad
		\overset{\mathfrak{e}_6}{3}\,\overset{\mathfrak{e}_6}{2}\,\overset{\mathfrak{e}_6}{2}\cdots \overset{\mathfrak{e}_6}{2}\,\overset{\mathfrak{e}_6}{2}\,\overset{\mathfrak{e}_6}{3}\,\overset{\mathfrak{su}_3}{2} \,.
\end{equation}
For completeness, we have given the curve configuration of the partial tensor
branch at one side of the quiver given the choice of fraction and algebra in Table
\ref{tbl:eattachments}.

\subsection{Type-D Bases}

Let us now move to long quivers whose F-theory bases are given by generalized
type-D orbifolds that are part of the infinite series
$\mathcal{D}_{(p,q)}=\mathcal{D}_{N,f}$. As with generalized A-type orbifolds,
the tensor branch of these SCFTs can again be constructed through fusion of
minimal conformal matter theories together, but now they exhibit the trivalent
vertex typical to type-D Dynkin diagram on one end. In M-theory, they are the
worldvolume theory of a stack of $N$ M5-branes probing a $\mathbb{C}^2/\Gamma$ singularity,
but contrary to ordinary conformal matter, the stack also contains OM5-branes
\cite{Hanany:1999sj,Hanany:2000fq,Chen:2019njf}. 

Focusing for a moment on theories with a $D_N$ base, on the partial tensor branch we have a collection of $(-2)$-curves arranging like the associated Dynkin diagram:
\begin{equation}
	2 \overset{\displaystyle 2}{2}\underbrace{2 \cdots 2 \,}_{N-3} \,.
\end{equation}

Undecorated, this quiver corresponds to the type-D $(2,0)$ theory. As in the case
of conformal matter, the fibers over each curve can be tuned to obtain
theories with minimal supersymmetry, which lead to more general quivers after
blowing-up non-minimal singularities. However, demanding that there are only
minimal singularities after this procedure is very
restrictive. There are indeed only a few cases for which this is possible,
corresponding to a choice of algebras $\mathfrak{g}=\mathfrak{su}(3),
\mathfrak{su}(2L), \mathfrak{so}(8), \mathfrak{e}_{6}$. We will not go into
the details of the blow-up procedure, but rather enumerate all the possible
theories of type $D_{N}^{\mathfrak{g}}$, and argue that the generalized-orbifold versions follow immediately.

The first class of D-shaped quivers is obtained by considering hypermultiplets
transforming in the bifundamental representation of $\mathfrak{su}(K)\oplus
\mathfrak{su}(K)$ and a collection of $(-2)$-curves. Anomaly cancellation then
dictates that on the trivalent intersection $K$ is even. This fixes the algebras
on the spine and up to decorations on the other end, one finds:
\begin{equation}\label{D_su2L}
	D_N^{\mathfrak{su}_{2L}}:\qquad
	\overset{\mathfrak{su}_{L}}{2} \overset{\displaystyle \overset{\mathfrak{su}_{L}}{2}}{\overset{\mathfrak{su}_{2L}}{2}}\underbrace{\,\overset{\mathfrak{su}_{2L}}{2} \cdots \overset{\mathfrak{su}_{2L}}{2} \,}_{N-3}[\mathfrak{su}_{2L}] \,.
\end{equation}

There is however an exception to this condition. When dealing with the low-rank
algebras $\mathfrak{su}(2)$ and $\mathfrak{su}(3)$, the absence of independent quartic
Casimir invariants relaxes the constraints and one finds that the following quiver leads
to a well-defined theory:
\begin{equation}\label{D_su3}
	D_N^{\mathfrak{su}_{3}}:\qquad
    2 \overset{\displaystyle 2}{\overset{\mathfrak{su}_{2}}{2}}\underbrace{\,\overset{\mathfrak{su}_{3}}{2} \cdots \overset{\mathfrak{su}_{3}}{2} \,}_{N-3}[\mathfrak{su}_3] \,.
\end{equation}
Note that in both cases, there are versions of these theories with algebras of
lower rank at the end of the spine. As we will explain shortly, these can be
obtained as deformations of these quivers.

One may wonder why we are not considering a variation of the quivers above, but
where there is a ramp of $\mathfrak{su}(K)$ algebras of ever-increasing ranks:
\begin{equation}\label{D_spurious}
    2 \overset{\displaystyle 2}{\overset{\mathfrak{su}_{2}}{2}}\overset{\mathfrak{su}_{3}}{2}\overset{\mathfrak{su}_{4}}{2}\cdots \qquad?
\end{equation}
Naively, it appears that the $\mathfrak{su}(3)$ gauge algebra has the correct
number of fundamental hypermultiplets (i.e., six) to cancel the gauge anomaly
induced by the vector multiplet. However, due to the trivalent pattern, the
matter arising from the intersection of two curves
$\overset{\mathfrak{su}_{2}}{2}\overset{\mathfrak{su}_{3}}{2}$ transforms in
the representation $(\bm{2}\oplus\bm{1},\bm{3})$ rather than the usual
bifundamental of $\mathfrak{su}(2)\oplus\mathfrak{su}(3)$
\cite{Morrison:2016djb}. An $\mathfrak{su}(4)$ algebra is therefore not allowed,
and quivers with such a ramp do not lead to consistent theories. 

A similar phenomenon occurs for algebra of type $\mathfrak{so}(2K)$. There,
the only allowed parent theory is given by $(\mathfrak{so}(8), \mathfrak{so}(8))$
conformal matter attached to one of the non-Higgsable clusters given in
equation \eqref{NHC-list} to obtain the type-D shape:
\begin{equation}
	D_N^{\mathfrak{so}_{8}}:\qquad
	\overset{\mathfrak{su}_2}{2}\overset{\displaystyle\overset{\mathfrak{su}_2}{2}}{\overset{\mathfrak{so}_7}{3}}1\overset{\mathfrak{so}_8}{4}1\overset{\mathfrak{so}_8}{4}\cdots\overset{\mathfrak{so}_8}{4}1[\mathfrak{so}_8]\,.
\end{equation}
One could once again imagine a ramp of increasing rank, but an analysis of the
associated geometry reveals that having anything but an $\mathfrak{so}(8)$ algebra on the
left-most $(-4)$-curve does not lead to a consistent F-theory model
\cite{Merkx:2017jey}.

Finally, if one consider exceptional algebras, there is a single case for which
a trivalent intersection can be obtained using the gluing procedure of
non-Higgsable clusters discussed around equation \eqref{gluing-NHC}, and
involves the algebra $\mathfrak{g}=\mathfrak{e}_6$:
\begin{equation}
	D^{\mathfrak{e}_6}_N:\qquad
	\overset{\mathfrak{su}_3}{3}1 \overset{\displaystyle \overset{\mathfrak{su}_3}{3}}{\overset{\displaystyle 1}{\overset{\mathfrak{e}_6}{6}}} 1\overset{\mathfrak{su}_3}{3}1\overbrace{\overset{\mathfrak{e}_6}{6} 1\overset{\mathfrak{su}_3}{3}1 \cdots \overset{\mathfrak{e}_6}{6} 1\overset{\mathfrak{su}_3}{3}1}^{N-3}[\mathfrak{e}_6] \,.
\end{equation}
This quiver can be understood as fusing a rank $(N-2)$ conformal matter theory,
$A^{\mathfrak{e}_6}_{N-3}$, with two $\frac{1}{3}$-fractional theories
$\mathcal{A}^{\mathfrak{e}_6}_{0;1;\frac{1}{3}}$. 

In fact, all type-D families can be constructed in a similar way: the trivalent
node is obtained by fusing a higher-rank conformal matter with two fractional
(possibly-deformed) minimal ones. To respect the symmetry of the type-D Dynkin
diagram, the latter two are the same. We are also of course free to choose the
other end to be associated with fractional conformal matter. Using fusion, we
then easily find the fractional version associated with generalized type-D
orbifolds:
\begin{equation}
		\mathcal{D}_{N,f}^{\mathfrak{g}} = D_{p}^{\mathfrak{g}} \oplus \mathcal{A}^{\mathfrak{g}}_{N-p-1;1,f}\,.
\end{equation}
In the pictorial description we have used, any theory with a
type-D endpoint can therefore be summarized as:
\begin{equation}
		\mathcal{D}^{\mathfrak{g}}_{N;f}: \qquad
		\mathfrak{g}_1\text{---}\overset{\displaystyle\mathfrak{g}_1}{\overset{\displaystyle|}{\mathfrak{g}_2}}\,\text{---}\mathfrak{g}\,
		\overbrace{\vphantom{\overset{\mathfrak{g}}{2}}\,\text{---}\,\mathfrak{g}\,\text{---}\dots\text{---}\,}^{N-4}\,\mathfrak{g}\text{---}\,[\mathfrak{g}_f]\,
\end{equation}
The allowed combinations $\mathfrak{g}_1\text{---}\mathfrak{g}_2$ are
summarized in Table \ref{tab:d-type}, while the fractional conformal matter at
the other end can be found in Table \ref{tbl:eattachments}. As expected, a
choice of algebra $\mathfrak{g}$ and a fractional number $f$ (or equivalently the
parameters $(p,q)$ of the orbifold) completely defines the quiver.

\begin{table}
		\centering
		\begin{tabular}{cccc}
                \toprule
				$\mathfrak{g}$ &  $\mathfrak{g}_1\text{---}\mathfrak{g}_2\text{---}\mathfrak{g}$\\
				\midrule
				$\mathfrak{su}_{2K}$ & $\overset{\mathfrak{su}_L}{2}\overset{\mathfrak{su}_{2L}}{2}\overset{\mathfrak{su}_{2L}}{2}$\\
				$\mathfrak{su}_{3}$  & $2\,\overset{\mathfrak{su}_2}{2}\overset{\mathfrak{su}_3}{2}$\\
				$\mathfrak{so}_{8}$   & $\overset{\mathfrak{su}_2}{2}\overset{\mathfrak{so}_7}{3}1\overset{\mathfrak{so}_8}{4}$\\
				$\mathfrak{e}_{6}$   & $\overset{\mathfrak{su}_3}{3}1\overset{\mathfrak{e}_6}{6}1\overset{\mathfrak{su}_3}{3}1\overset{\mathfrak{e}_6}{6}$\\
				\bottomrule
		\end{tabular}
		\caption{Quiver description of the trivalent attachments used 6d $(1,0)$ SCFTs of type $\mathcal{D}_{N;f}^\mathfrak{g}$.}
		\label{tab:d-type}
\end{table}

\subsection{Orbi-instantons}\label{sec:orbi-instantons}

The last class of long quivers corresponds to orbi-instanton theories,
$\mathcal{O}_{N;f}^{\mathfrak{g}}$. In the M-theory picture, they arise as the
worldvolume theories on a stack of $N$ M5-branes probing a
$\mathbb{C}^2/\Gamma$ orbifold, and inside an end-of-the-world M9-brane. When there are
no frozen singularities, they are realized in F-theory by an
elliptically-fibered Calabi--Yau with a trivial base $B=\mathbb{C}^2$. On the
partial tensor branch, their quivers take the following form \cite{DelZotto:2014hpa}:
\begin{equation}
		[\mathfrak{e}_8]\,\overset{\mathfrak{g}}{1}\underbrace{\overset{\mathfrak{g}}{2}\,\overset{\mathfrak{g}}{2}\,\dots\,\overset{\mathfrak{g}}{2}\,\overset{\mathfrak{g}}{2}}_{N-1}\,[\mathfrak{g}]\,.
\end{equation}
On a generic point of the tensor branch, we once again recover a long spine of
fused minimal $(\mathfrak{g}, \mathfrak{g})$ conformal matter, but demanding
the $\mathfrak{e}_8$ symmetry associated with the M9-brane at one end
gives a more involved curve configuration. For instance, when
$\mathfrak{g}=\mathfrak{su}(K)$, we obtain a single undecorated $(-1)$-curve
followed by a long ramp of $\overset{\mathfrak{su}_p}{2}$ curves where the rank
increases until it reaches $p=K$:
\begin{equation}\label{eqn:AtypeOI}
		\mathcal{O}^{\mathfrak{su}_K}_{N}:\qquad 1\,2\overset{\mathfrak{su}_2}{2}\overset{\mathfrak{su}_3}{2}\cdots\overset{\mathfrak{su}_K}{2}\underbrace{\,\overset{\mathfrak{su}_K}{2} \cdots \overset{\mathfrak{su}_K}{2} \,}_{N-1} \,.
\end{equation}
This pattern extends to any of the ADE algebras: a rank $N$ $(\mathfrak{g},
\mathfrak{g})$ conformal matter is attached to the \emph{minimal
orbi-instanton}, another building block we depict by $[\mathfrak{e}_8]
\photon[\mathfrak{g}]$ and whose curve configuration is given in Table
\ref{tab:attachment-orbi-instanton} for any algebra $\mathfrak{g}$. If there are frozen
singularities in the M-theory realization of these theories, in F-theory the
base of the elliptic Calabi--Yau will not be trivial, but rather a generalized
type-A orbifold of low order associated with the presence of the fractional
conformal matter. The fusion point of view is again very useful, as the curve
configuration of any orbi-instanton can be uniquely depicted as: 
\begin{equation}
	\mathcal{O}^{\mathfrak{g}}_{N;f}: \qquad
	[\mathfrak{e}_8] \photon\,\,\mathfrak{g}\underbrace{\vphantom{\frac{1}{2}}\,\,\text{---}\dots\text{---}}_{N-2}\,\mathfrak{g}\text{---}\,\,[\mathfrak{g}_f]\,.
\end{equation}
The links and nodes can then be read off from Tables \ref{tbl:CM-and-nodes},
\ref{tbl:eattachments}, and \ref{tab:attachment-orbi-instanton}. The non-Abelian flavor
symmetry of these theories is generically given by
\begin{equation}
		\mathfrak{f} = \mathfrak{e}_8 \oplus \mathfrak{g}_f\,.
\end{equation}

\begin{table}[H]
    \centering
    \begin{threeparttable}
        \begin{tabular}[t]{cccc}
            \toprule
			$\mathfrak{g}$ & $[\mathfrak{e}_8]\photon [\mathfrak{g}]$ &  $f_\text{OI}$ & $\alpha_0$\\\midrule
			$\mathfrak{su}_K$ & $1\,2\overset{\mathfrak{su}_2}{2}\overset{\mathfrak{su}_3}{2}\cdots\overset{\mathfrak{su}_{K-1}}{2}\overset{\mathfrak{su}_{K}}{2}$ & $K+1$ & $\frac{2K(K^2-1)}{15}(3 K^2 - 32)$\\
			$\mathfrak{so}_{2K}$ & $1 \, 2 \overset{\mathfrak{su}_2}{2} \overset{\mathfrak{g}_2}{3} 1 \overset{\mathfrak{so}_9}{4} \overset{\mathfrak{sp}_1}{1} \overset{\mathfrak{so}_{11}}{4} \cdots \overset{\mathfrak{sp}_{K-4}}{1} \overset{\mathfrak{so}_{2K}}{4}\overset{\mathfrak{sp}_{K-4}}{1}$ & $K+1$ & $\frac{8K(K-1)}{15}(82 - 203 K + 27 K^2 + 12 K^3)$\\
			$\mathfrak{e}_6$ & 	$1 \, 2 \overset{\mathfrak{su}_2}{2} \overset{\mathfrak{g}_2}{3} 1 \overset{\mathfrak{f}_4}{5} 1 \overset{\mathfrak{su}_3}{3} 1 \overset{\mathfrak{e}_6}{6}1 \overset{\mathfrak{su}_3}{3} 1$ & $6$ & $93120$\\
			$\mathfrak{e}_7$ & 	$1 \, 2 \overset{\mathfrak{su}_2}{2} \overset{\mathfrak{g}_2}{3} 1 \overset{\mathfrak{f}_4}{5} 1 \overset{\mathfrak{g}_2}{3} \overset{\mathfrak{su}_2}{2} 1 \overset{\mathfrak{e}_7}{8} 1 \overset{\mathfrak{su}_2}{2} \overset{\mathfrak{so}_7}{3} \overset{\mathfrak{su}_2}{2} 1 $ & $6$ & $575232$\\
			$\mathfrak{e}_8$ & 	$1 \, 2 \overset{\mathfrak{su}_2}{2} \overset{\mathfrak{g}_2}{3} 1 \overset{\mathfrak{f}_4}{5} 1 \overset{\mathfrak{g}_2}{3} \overset{\mathfrak{su}_2}{2} 2 1 \overset{\displaystyle 1}{\overset{\mathfrak{e}_8}{(12)}} 1 \, 2 \overset{\mathfrak{su}_2}{2} \overset{\mathfrak{g}_2}{3} 1 \overset{\mathfrak{f}_4}{5} 1 \overset{\mathfrak{g}_2}{3} \overset{\mathfrak{su}_2}{2} 2 1$ & $6$ & $5204096$\\
           \bottomrule
        \end{tabular}
    \end{threeparttable}
	\caption{Quiver for the minimal orbi-instantons, used as a building block for 6d (1,0) SCFTs of type $\mathcal{O}^{\mathfrak{g}}_{N;f}$. The numbers $f_\text{OI}$ and $\alpha_0$ are coefficients appearing in their anomaly polynomial.}
    \label{tab:attachment-orbi-instanton}
\end{table}

\subsection{Nilpotent Orbits and Deformations}\label{sec:nilpotent-deformations}

So far, we have seen that there are three classes of long quivers associated
with generalized ADE orbifolds: higher-rank (fractional) conformal matter
$\mathcal{A}_{N-1;f_L,f_R}^{\mathfrak{g}}$, orbi-instanton theories
$\mathcal{O}_{N;f}^{\mathfrak{g}}$, and SCFTs with generalized type-D bases,
$\mathcal{D}_{N;f}^{\mathfrak{g}}$, for which only a handful of algebras are
allowed. In the pictorial notation we have utilized throughout this section, the
parent theories can all be summarized as:
\begin{equation}
		\begin{aligned}
			\mathcal{A}^{\mathfrak{g}}_{N-1;f_L,f_R}:& \qquad
			[\mathfrak{g}_{f_L}]\,\text{---}\,\mathfrak{g}\,\text{---}\,\mathfrak{g}\,\text{---}\dots\text{---}\,\mathfrak{g}\,\text{---}\,\mathfrak{g}\,\text{---}\,[\mathfrak{g}_{f_R}]\,,\vphantom{\frac{1}{1}}\\
			\mathcal{D}^{\mathfrak{g}}_{N;f}:& \qquad
			\mathfrak{g}_1\text{---}\overset{\displaystyle\mathfrak{g}_1}{\overset{\displaystyle|}{\mathfrak{g}_2}}\,\text{---}\mathfrak{g}\,
			\,\text{---}\,\mathfrak{g}\,\text{---}\dots\text{---}\,\mathfrak{g}\text{---}\,[\mathfrak{g}_f]\,,\vphantom{\frac{1}{1}}\\
			\mathcal{O}^{\mathfrak{g}}_{N;f}:& \qquad
			\quad[\mathfrak{e}_8] \photon\,\,\mathfrak{g}\vphantom{\frac{1}{2}}\,\,\text{---}\dots\text{---}\,\mathfrak{g}\text{---}\,\,[\mathfrak{g}_f]\,.\vphantom{\frac{1}{1}}
		\end{aligned}
\end{equation}
The curve configuration of these quivers can be inferred from Tables
\ref{tbl:CM-and-nodes}--\ref{tab:attachment-orbi-instanton} collated throughout this
section without any ambiguity, as a choice of base and algebra fixes it
uniquely if $N$ is large.

Given a parent theory, one can perform a complex-structure deformation of the
associated elliptically fibered Calabi--Yau threefold, and reach a curve
configuration corresponding to the tensor branch description of a new 6d
$(1,0)$ SCFT \cite{Heckman:2013pva, Heckman:2015bfa}. In the field
theory, these are equivalent to Higgs branch renormalization group flow.
Starting from the parent theory in the ultraviolet, and giving a
particular non-trivial vacuum expectation value to a gauge-invariant operator,
we are led to a new interacting 6d SCFT in the deep infrared. 

For instance, starting with the quiver of the theory $\mathcal{T}^\text{UV} =
A^{\mathfrak{e}_6}_{N-1}$, we can consider the complex-structure deformation of the geometry at the superconformal fixed point in such a way that in passing to the tensor branch geometry it was no longer necessary to perform the blow-up creating the final $(-1)$-curve on the left. Thus, we obtain
a new quiver for the tensor branch of an
SCFT $\mathcal{T}^\text{IR}$ satisfying all the required properties:
\begin{equation}
	\mathcal{T}^\text{UV}:\qquad [\mathfrak{e}_6]1\overset{\mathfrak{su}_{3}}{3}1\overset{\mathfrak{e}_{6}}{6}1\overset{\mathfrak{su}_{3}}{3}1\cdots
	\qquad\longrightarrow\qquad
	\mathcal{T}^\text{IR}:\qquad [\mathfrak{su}_6]\overset{\mathfrak{su}_{3}}{2}1\overset{\mathfrak{e}_{6}}{6}1\overset{\mathfrak{su}_{3}}{3}1\cdots \,.
\end{equation}
In the IR theory, there are six hypermultiplets arising from the left-most
curve, which are rotated by a new flavor symmetry
$\mathfrak{su}(6)\subset\mathfrak{e}_6$. This is in fact a generic feature of
deformed theories: if the generalized quiver is long enough, the remnant flavor symmetry is always a subalgebra of that
of its parent. Indeed, a property of the parent theories
$\mathcal{T}^\text{UV}$ is that each family has a (possibly trivial) flavor
symmetry $\mathfrak{f}^\text{UV}$ arising at each end of the quiver. Then, if
one deforms it to a new theory $\mathcal{T}^\text{IR}$ its flavor satisfies:
\begin{equation}
	\mathfrak{f}^\text{IR}\subseteq\mathfrak{f}^\text{UV}\,.
\end{equation}
Moreover, if one end of the quiver of a parent theory has no flavor, either
because it corresponds to the trivalent end of a type-D theory, or there is a
fraction with a trivial flavor, there exists no deformation preserving the
algebra $\mathfrak{g}$ on the spine or the rank $N$ of the base, and there are
no new SCFTs descending from that end of the quiver.

Based on these observations it has been proposed that in the field theory, the
operator given a vacuum expectation value should be the moment map, the
adjoint-valued scalar belonging to the same supermultiplet as a flavor
conserved current \cite{DelZotto:2014hpa, Heckman:2015bfa}. The flavor
$\mathfrak{f}^\text{IR}$ is then understood as being generated by the unbroken
generators. The notion of parent theory was in fact introduced precisely
because any other long theories can be understood in such a fashion.

As the deformations are associated to each side of the quiver, we can
distinguish two types: those related to an end of a quiver whose flavor come
from (fractional) conformal matter, or with the $\mathfrak{e}_8$ symmetry of an
of orbi-instanton theory $\mathcal{O}_{N;f}^{\mathfrak{g}}$. For the latter, it
has been argued from M-theory that since the $\mathfrak{e}_8$ flavor symmetry
arises from an end-of-the-world M9-brane, deformations should correspond to
choices of boundary condition on the $S^3/\Gamma$ boundary of the
$\mathbb{C}^2/\Gamma$ orbifold, amounting to embeddings of $\Gamma$
into $E_8$ \cite{DelZotto:2014hpa, Heckman:2015bfa}. The quiver description of
the resulting theory can then be mapped from the choice of embedding
\cite{Mekareeya:2017jgc, Frey:2018vpw}.

On the other hand, when a deformation is performed on the side of the quiver
associated with fractional conformal matter with flavor $\mathfrak{g}_f$, the resulting
theories can be labelled by nilpotent orbits of $\mathfrak{g}_f$. As we review
in more detail in Section $\ref{sec:nilpotent-RG}$, nilpotent orbits $O$ of a
simple Lie algebra $\mathfrak{g}_f$ are equivalent to embeddings $\rho_O:
\mathfrak{su}(2)\to\mathfrak{g}_f$. It was then shown that the centralizer
$\mathfrak{f}$ of the embedding precisely matches the flavor symmetry of the
deformed theories \cite{Heckman:2016ssk}.

We can therefore denote all possible SCFTs obtained by deforming one of the
three parent families by either a nilpotent orbit of the corresponding flavor
symmetry, or in the case of orbi-instantons a choice of embedding of the
corresponding discrete group into $E_8$:
\begin{equation}\label{long-quiver-names-with-deformations}
		\mathcal{A}^{\mathfrak{g}}_{N-1;f_L,f_R}(O_L, O_R)\,,\qquad
		\mathcal{D}^{\mathfrak{g}}_{N;f}(O)\,,\qquad
		\mathcal{O}^{\mathfrak{g}}_{N;f}(\sigma, O)\,,\qquad
\end{equation}
where
\begin{equation}
		\sigma \in \text{Hom}(\Gamma,E_8)\,,\qquad O \Leftrightarrow \rho_O\in \text{Hom}(\mathfrak{su}_2, \mathfrak{g}_f)\,.
\end{equation}
The discrete group $\Gamma$ is the McKay-dual of the simple algebra
$\mathfrak{g}$ defining the parent theory. When the choice of embedding  is
trivial, which we denote $\sigma=\varnothing$, $O=\varnothing$, corresponding
to the undeformed theory, we write simply e.g.
$\mathcal{O}^{\mathfrak{g}}_{N;f}(\varnothing,
\varnothing)=\mathcal{O}^{\mathfrak{g}}_{N;f}$ and similarly for the other two
families. 

In this work, we will only focus on theories associated with a nilpotent orbit
of the corresponding possible flavor, and will refer to them as obtained
through \emph{nilpotent deformations}, or \emph{nilpotent RG flows} when
discussing the corresponding field theory description. Moreover, a possible
deformation might propagate through the spine of the parent theory.  We will
therefore only consider cases where the parameter $N$ of the orbifold base is
taken to be large enough such that a deformation on one end does not affect the
other. We will refer to the quivers defined through equations
\eqref{long-quiver-names-with-deformations} satisfying this property as
\emph{long quivers}. In this work, we will focus on the anomaly polynomial of
long quivers and their nilpotent deformations. We leave a more detailed
analysis of short quivers and orbi-instantons associated with non-trivial
choices $\sigma \in \text{Hom}(\Gamma,E_8)$ for future work.

What is more, nilpotent orbits are equipped with a partial ordering given by
the Zariski closure operation. One says that for two nilpotent orbits
$O,O^\prime\subset\mathfrak{g}$ of a given simple algebra, then $O\leq
O^\prime$ if $\overline{O}\subseteq \overline{O}^\prime$, see, e.g.,
\cite{Chacaltana:2012zy, collingwood2017nilpotent} and references therein for a
detailed exposition.  This partial ordering then enables one to arrange the
corresponding 6d $(1,0)$ SCFTs into a Hasse diagram, establishing a hierarchy
of the possible complex-structure deformations / nilpotent RG flows
\cite{Heckman:2016ssk, Hassler:2019eso, Heckman:2018pqx}. These Hasse diagrams
refine the above classification as it tells us that, starting with, e.g., a
theory $A_{N-1}^{\mathfrak{g}}(\varnothing, O_R)$, we can reach the
SCFT $A_{N-1}^{\mathfrak{g}}(\varnothing, O_R')$ by further
deformations only if $O_R\leq O_R'$. In Appendix \ref{app:hasse}, we give the
Hasse diagrams of all the simple algebras appearing in conformal matter
theories of exceptional types and their fractions.

Coming back to $(\mathfrak{e}_6, \mathfrak{e}_6)$ conformal matter, in Figure
\ref{fig:Hasse-E6-quivers} we show all possible deformations, and how they fit
in the Hasse diagram of $\mathfrak{e}_6$. One can see that as one goes deeper
in the diagram, there are fewer and fewer curves at the generic point of the tensor branch, and more than one minimal
conformal matter is ultimately removed from the quiver. This occurs in the vast
majority of cases, as nilpotent deformations will usually propagate throughout
a portion of the spine, and justifies our focus on long quivers.

\begin{figure}[p] 
		\centering
		\resizebox{!}{.7\paperheight}{
			\begin{tikzpicture}[>=latex',line join=bevel,]
\node (0) at (193.94bp,992.96bp) [draw,rectangle] {\begin{tabular}{c}$0$\qquad$1\overset{\mathfrak{su}_3}{3}1\overset{\mathfrak{e}_6}{6}1\overset{\mathfrak{su}_3}{3}1\overset{\mathfrak{e}_6}{6}1\overset{\mathfrak{su}_3}{3}1\overset{\mathfrak{e}_6}{6}1\overset{\mathfrak{su}_3}{3}1\overset{\mathfrak{e}_6}{6}\cdots$\end{tabular}};
  \node (1) at (193.94bp,933.49bp) [draw,rectangle] {\begin{tabular}{c}$A_1$\qquad$\overset{\mathfrak{su}_3}{2}1\overset{\mathfrak{e}_6}{6}1\overset{\mathfrak{su}_3}{3}1\overset{\mathfrak{e}_6}{6}1\overset{\mathfrak{su}_3}{3}1\overset{\mathfrak{e}_6}{6}1\overset{\mathfrak{su}_3}{3}1\overset{\mathfrak{e}_6}{6}\cdots$\end{tabular}};
  \node (2) at (193.94bp,874.03bp) [draw,rectangle] {\begin{tabular}{c}$2A_1$\qquad$\overset{\mathfrak{su}_2}{2}1\overset{\mathfrak{e}_6}{6}1\overset{\mathfrak{su}_3}{3}1\overset{\mathfrak{e}_6}{6}1\overset{\mathfrak{su}_3}{3}1\overset{\mathfrak{e}_6}{6}1\overset{\mathfrak{su}_3}{3}1\overset{\mathfrak{e}_6}{6}\cdots$\end{tabular}};
  \node (3) at (193.94bp,814.56bp) [draw,rectangle] {\begin{tabular}{c}$3A_1$\qquad$21\overset{\mathfrak{e}_6}{6}1\overset{\mathfrak{su}_3}{3}1\overset{\mathfrak{e}_6}{6}1\overset{\mathfrak{su}_3}{3}1\overset{\mathfrak{e}_6}{6}1\overset{\mathfrak{su}_3}{3}1\overset{\mathfrak{e}_6}{6}\cdots$\end{tabular}};
  \node (4) at (193.94bp,750.4bp) [draw,rectangle] {\begin{tabular}{c}$A_2$\qquad$1\overset{\displaystyle 1}{\overset{\mathfrak{e}_6}{6}}1\overset{\mathfrak{su}_3}{3}1\overset{\mathfrak{e}_6}{6}1\overset{\mathfrak{su}_3}{3}1\overset{\mathfrak{e}_6}{6}1\overset{\mathfrak{su}_3}{3}1\overset{\mathfrak{e}_6}{6}\cdots$\end{tabular}};
  \node (5) at (193.94bp,686.23bp) [draw,rectangle] {\begin{tabular}{c}$A_2 + A_1$\qquad$1\overset{\mathfrak{e}_6}{5}1\overset{\mathfrak{su}_3}{3}1\overset{\mathfrak{e}_6}{6}1\overset{\mathfrak{su}_3}{3}1\overset{\mathfrak{e}_6}{6}1\overset{\mathfrak{su}_3}{3}1\overset{\mathfrak{e}_6}{6}\cdots$\end{tabular}};
  \node (6) at (84.943bp,625.85bp) [draw,rectangle] {\begin{tabular}{c}$A_2 + 2A_1$\qquad$\overset{\mathfrak{e}_6}{4}1\overset{\mathfrak{su}_3}{3}1\overset{\mathfrak{e}_6}{6}1\overset{\mathfrak{su}_3}{3}1\overset{\mathfrak{e}_6}{6}1\overset{\mathfrak{su}_3}{3}1\overset{\mathfrak{e}_6}{6}\cdots$\end{tabular}};
  \node (7) at (302.94bp,625.85bp) [draw,rectangle] {\begin{tabular}{c}$2A_2$\qquad$1\overset{\mathfrak{f}_4}{5}1\overset{\mathfrak{su}_3}{3}1\overset{\mathfrak{e}_6}{6}1\overset{\mathfrak{su}_3}{3}1\overset{\mathfrak{e}_6}{6}1\overset{\mathfrak{su}_3}{3}1\overset{\mathfrak{e}_6}{6}\cdots$\end{tabular}};
  \node (8) at (84.943bp,564.55bp) [draw,rectangle] {\begin{tabular}{c}$A_3$\qquad$\overset{\mathfrak{so}_{10}}{4}1\overset{\mathfrak{su}_3}{3}1\overset{\mathfrak{e}_6}{6}1\overset{\mathfrak{su}_3}{3}1\overset{\mathfrak{e}_6}{6}1\overset{\mathfrak{su}_3}{3}1\overset{\mathfrak{e}_6}{6}\cdots$\end{tabular}};
  \node (9) at (301.94bp,564.55bp) [draw,rectangle] {\begin{tabular}{c}$2A_2 + A_1$\qquad$\overset{\mathfrak{f}_{4}}{4}1\overset{\mathfrak{su}_3}{3}1\overset{\mathfrak{e}_6}{6}1\overset{\mathfrak{su}_3}{3}1\overset{\mathfrak{e}_6}{6}1\overset{\mathfrak{su}_3}{3}1\overset{\mathfrak{e}_6}{6}\cdots$\end{tabular}};
  \node (10) at (192.94bp,504.17bp) [draw,rectangle] {\begin{tabular}{c}$A_3 + A_1$\qquad$\overset{\mathfrak{so}_{9}}{4}1\overset{\mathfrak{su}_3}{3}1\overset{\mathfrak{e}_6}{6}1\overset{\mathfrak{su}_3}{3}1\overset{\mathfrak{e}_6}{6}1\overset{\mathfrak{su}_3}{3}1\overset{\mathfrak{e}_6}{6}\cdots$\end{tabular}};
  \node (11) at (192.94bp,444.71bp) [draw,rectangle] {\begin{tabular}{c}$D_4(a_1)$\qquad$\overset{\mathfrak{so}_{8}}{4}1\overset{\mathfrak{su}_3}{3}1\overset{\mathfrak{e}_6}{6}1\overset{\mathfrak{su}_3}{3}1\overset{\mathfrak{e}_6}{6}1\overset{\mathfrak{su}_3}{3}1\overset{\mathfrak{e}_6}{6}\cdots$\end{tabular}};
  \node (12) at (91.94bp,321.08bp) [draw,rectangle] {\begin{tabular}{c}$D_4$\qquad$\overset{\mathfrak{su}_{3}}{3}1\overset{\displaystyle 1}{\overset{\mathfrak{e}_6}{6}}1\overset{\mathfrak{su}_3}{3}1\overset{\mathfrak{e}_6}{6}1\overset{\mathfrak{su}_3}{3}1\overset{\mathfrak{e}_6}{6}\cdots$\end{tabular}};
  \node (13) at (265.94bp,385.25bp) [draw,rectangle] {\begin{tabular}{c}$A_4$\qquad$\overset{\mathfrak{so}_{7}}{3}\overset{\mathfrak{su}_2}{2}1\overset{\mathfrak{e}_6}{6}1\overset{\mathfrak{su}_3}{3}1\overset{\mathfrak{e}_6}{6}1\overset{\mathfrak{su}_3}{3}1\overset{\mathfrak{e}_6}{6}\cdots$\end{tabular}};
  \node (14) at (101.94bp,256.0bp) [draw,rectangle] {\begin{tabular}{c}$D_5(a_1)$\qquad$\overset{\mathfrak{su}_{3}}{3}1\overset{\mathfrak{e}_6}{5}1\overset{\mathfrak{su}_3}{3}1\overset{\mathfrak{e}_6}{6}1\overset{\mathfrak{su}_3}{3}1\overset{\mathfrak{e}_6}{6}\cdots$\end{tabular}};
  \node (15) at (294.94bp,321.08bp) [draw,rectangle] {\begin{tabular}{c}$A_4 + A_1$\qquad$\overset{\mathfrak{g}_{2}}{3}\overset{\mathfrak{su}_2}{2}1\overset{\mathfrak{e}_6}{6}1\overset{\mathfrak{su}_3}{3}1\overset{\mathfrak{e}_6}{6}1\overset{\mathfrak{su}_3}{3}1\overset{\mathfrak{e}_6}{6}\cdots$\end{tabular}};
  \node (16) at (197.94bp,194.7bp) [draw,rectangle] {\begin{tabular}{c}$E_6(a_3)$\qquad$\overset{\mathfrak{su}_{3}}{3}1\overset{\mathfrak{f}_4}{5}1\overset{\mathfrak{su}_3}{3}1\overset{\mathfrak{e}_6}{6}1\overset{\mathfrak{su}_3}{3}1\overset{\mathfrak{e}_6}{6}\cdots$\end{tabular}};
  \node (17) at (284.94bp,256.0bp) [draw,rectangle] {\begin{tabular}{c}$A_5$\qquad$\overset{\mathfrak{g}_{2}}{3}1\overset{\mathfrak{f}_4}{5}1\overset{\mathfrak{su}_3}{3}1\overset{\mathfrak{e}_6}{6}1\overset{\mathfrak{su}_3}{3}1\overset{\mathfrak{e}_6}{6}\cdots$\end{tabular}};
  \node (18) at (197.94bp,134.32bp) [draw,rectangle] {\begin{tabular}{c}$D_5$\qquad$\overset{\mathfrak{su}_{2}}{2}\overset{\mathfrak{so}_7}{3}\overset{\mathfrak{su}_2}{2}1\overset{\mathfrak{e}_6}{6}1\overset{\mathfrak{su}_3}{3}1\overset{\mathfrak{e}_6}{6}\cdots$\end{tabular}};
  \node (19) at (197.94bp,73.942bp) [draw,rectangle] {\begin{tabular}{c}$E_6(a_1)$\qquad$\overset{\mathfrak{su}_{2}}{2}\overset{\mathfrak{g}_2}{3}1\overset{\mathfrak{f}_4}{5}1\overset{\mathfrak{su}_3}{3}1\overset{\mathfrak{e}_6}{6}\cdots$\end{tabular}};
  \node (20) at (197.94bp,12.647bp) [draw,rectangle] {\begin{tabular}{c}$E_6$\qquad$2\overset{\mathfrak{su}_{2}}{2}\overset{\mathfrak{g}_2}{3}1\overset{\mathfrak{f}_4}{5}\cdots$\end{tabular}};
  \draw [->] (0) ..controls (193.94bp,974.04bp) and (193.94bp,965.03bp)  .. (1);
  \draw [->] (1) ..controls (193.94bp,914.58bp) and (193.94bp,905.56bp)  .. (2);
  \draw [->] (2) ..controls (193.94bp,855.11bp) and (193.94bp,846.1bp)  .. (3);
  \draw [->] (3) ..controls (193.94bp,795.8bp) and (193.94bp,787.04bp)  .. (4);
  \draw [->] (4) ..controls (193.94bp,726.44bp) and (193.94bp,717.66bp)  .. (5);
  \draw [->] (5) -- (6);
  \draw [->] (5) -- (7);
  \draw [->] (6) -- (8);
  \draw [->] (6) -- (9);
  \draw [->] (7) -- (9);
  \draw [->] (8) -- (10);
  \draw [->] (9) ..controls (257.03bp,542.97bp) and (237.57bp,531.49bp)  .. (10);
  \draw [->] (10) ..controls (192.94bp,485.26bp) and (192.94bp,476.25bp)  .. (11);
  \draw [blue,->] (11) ..controls (170.05bp,413.11bp) and (129.99bp,372.94bp)  .. (12);
  \draw [blue,->] (11) ..controls (217.22bp,424.6bp) and (231.23bp,413.57bp)  .. (13);
  \draw [->] (12) ..controls (95.66bp,296.61bp) and (97.11bp,287.46bp)  .. (14);
  \draw [->] (13) -- (15);
  \draw [->] (14) ..controls (140.22bp,235.5bp) and (157.03bp,223.91bp)  .. (16);
  \draw [->] (15) ..controls (235.85bp,298.53bp) and (184.04bp,283.29bp)  .. (14);
  \draw [->] (15) -- (17);
  \draw [blue,->] (16) ..controls (197.94bp,174.69bp) and (197.94bp,165.79bp)  .. (18);
  \draw [->] (17) ..controls (255.19bp,234.72bp) and (238.81bp,223.55bp)  .. (16);
  \draw [->] (18) ..controls (197.94bp,115.59bp) and (197.94bp,106.69bp)  .. (19);
  \draw [blue,->] (19) ..controls (197.94bp,54.031bp) and (197.94bp,45.107bp)  .. (20);
\end{tikzpicture}
		}
		\caption{Hasse diagram for the nilpotent orbits of $\mathfrak{e}_6$ and
				the associated quivers.  Each SCFT arising from a nilpotent
				deformation of the theory $A^{\mathfrak{e}_6}_{N-1}$ can be
				associated with a nilpotent orbit, and each arrow can be
				understood as a possible RG flow. Blue arrows
				indicate that a new minimal conformal matter is affected with
				respect to the previous quiver.
		}
		\label{fig:Hasse-E6-quivers}
\end{figure}

The notation for the SCFT defined in equation
\eqref{long-quiver-names-with-deformations} is sufficient to (essentially)
fully determine the curve configuration of any long quiver.\footnote{In the special
		the case of $\mathfrak{g}=\mathfrak{so}_{2K}$, very-even partitions
		lead to the same quiver, but are in fact different theories.  This can
		be shown by computing the Schur index of their $T^2$ compactification \cite{Distler:2020tub,
		Distler:2022yse}.} The
dictionary between the choice of orbit and the quiver describing the tensor
branch of the SCFT depends on the type of algebra -- and the fraction, when
applicable. For classical flavor algebras, nilpotent orbits are labelled by integer partitions, from which the quiver can be found
straightforwardly. In the exceptional cases, we will use Bala--Carter labels \cite{bala1976classesI, bala1976classesII}; there are only a finite number of possibilities, which we have
tabulated in Appendix \ref{app:nilp2TB}, where we also give the procedure to obtain the quivers from integer partitions.

We close this section by noting that nilpotent deformations of parent theories
and those associated with embeddings of discrete groups into $E_8$ do not
exhaust all possibilities. While SCFTs with the same choice of base
$\mathbb{C}^2/\Lambda$ are on the same Higgs branch, there exists more types of
deformations of the geometry or, equivalently, Higgs branch RG flows. Indeed,
as mentioned above, there also exist Higgs flows between deformed theories,
which are described by transverse Slowdowy slices in the associated
Higgs branch moduli space. For completeness, we have labelled each of these
transitions using the notation introduced by Kraft and Procesi
\cite{kraft1981minimal, fu2017generic} in the Hasse diagrams given in Appendix
\ref{app:hasse}. Furthermore, there also exists so-called semi-simple
deformations \cite{Heckman:2018pqx}, which, while keeping the same choice of
base in the geometry, changes the algebra, e.g.,
$A_{N-1}^{\mathfrak{g}}\to A_{N-1}^{\mathfrak{g}^\prime}$
with $\mathfrak{g}^\prime\subset\mathfrak{g}$. We will focus here only on
nilpotent deformations of parent theories, as we will show that the anomaly
polynomial of these theories is encoded uniquely in the notation given in
equation \eqref{long-quiver-names-with-deformations}.

\section{Anomaly Polynomial of Long 6d SCFTs}\label{sec:I8}

Having reviewed the possible 6d $(1,0)$ SCFTs associated with long quivers that
can be realized via F-theory engineering, we can now study part of their
conformal data. We will focus on the anomaly polynomial, which encodes several
features of its protected sector, such as the central charges. In six
dimensions, the anomaly polynomial of a $(1,0)$ quantum field theory takes the general form:
\begin{equation}\label{eqn:I8general}
  \begin{aligned}
		  I_8 &=  \frac{\alpha}{24} c_2(R)^2+ \frac{\beta}{24}  c_2(R) p_1(T) + \frac{\gamma}{24}  p_1(T)^2 + \frac{\delta}{24} p_2(T) \cr &\quad + \sum_a \text{Tr}F_a^2 \left(\kappa_a p_1(T) + \nu_a c_2(R) + \sum_b \rho_{ab} \text{Tr}  F_b^2\right) + \sum_a \mu_a \text{Tr}F_a^4  \,,
  \end{aligned}
\end{equation}
where $c_2(R)$ is the second Chern class of the background R-symmetry bundle,
while $p_{1,2}(T)$ are the Pontryagin classes of the spacetime tangent bundle.
The index $a$ runs over all simple non-Abelian flavor symmetry factors with
background field strength $F_a$ and the traces $\text{Tr}F_a^n$ are chosen to
be one-instanton normalized by convention.

The coefficients appearing in the anomaly polynomial of a 6d theory at the SCFT
point can be determined from the tensor branch theory using a variation of 't
Hooft anomaly matching \cite{Ohmori:2014kda, Intriligator:2014eaa}. Indeed,
moving onto the tensor branch does not break supersymmetry, nor the possible
flavor symmetries. The anomaly polynomial, being a topological quantity, is
preserved under this type of deformations and we can therefore reach a point
where a gauge theory description in terms of weakly-coupled supermultiplets is
available. There, the procedure to compute the anomaly polynomial is purely
algorithmic, and was described concisely in \cite{Baume:2021qho} where it was
applied to a large number of theories, closing potential loopholes of the
original derivation in the presence of flavor symmetries. Let us now summarize
the salient points of the algorithm needed to find equation
\eqref{eqn:I8general} for any 6d SCFT.

First, one distinguishes between two different terms, a ``one-loop'' part, and
a Green--Schwarz (GS) contribution:
\begin{equation}
		I_8 = I_8^\text{1-loop} + I_8^\text{GS}\,.
\end{equation}
The first contains the individual contributions of fermionic and tensor fields
inside $(1,0)$ supermultiplets, and can be understood in terms of four-point
correlators of particular protected operators. On the tensor branch where a
weakly-coupled description is available, these can be computed as
one-loop-exact square Feynman diagrams involving the energy-momentum tensor or
conserved currents -- hence its name. The result of such a computation shows
that it is equivalent to study the index of the appropriate differential operator
\cite{AlvarezGaume:1983ig}. Indeed, for left-handed chiral fermions
transforming in the complex representation $\bm{R}$ of a symmetry algebra, it
is well-known that the formal eight-form related to the Dirac operator is given
by:\footnote{The result of the computation of the Dirac index takes into
account both the representation and its conjugate: $\bm{R}\oplus\bar{\bm{R}}$.
This explain the factor of $\frac{1}{2}$ in equation \eqref{fermion-I8-d}.}
\begin{equation}\label{fermion-I8-d}
		I_8^\text{fermion} = \left.\frac{1}{2}\widehat{A}(T) \text{ch}_{\bm{R}}(F)\right|_{\text{8-form}}\,,\qquad \text{ch}_{\bm{R}}(F) = \text{tr}_{\bm{R}} e^{i F}\,,
\end{equation}
where the A-roof genus $\widehat{A}(T)$ is associated with gravitational
anomalies while the Chern character $\text{ch}_{\bm{R}}(F)$ encodes gauge
or flavor anomalies. The definitions of these characteristic classes and
related quantities are collected in Appendix \ref{app:representations}.

Note that the traces appearing in the Chern character are performed over a
given representation, $\text{tr}_{\bm{R}} F^n$, whereas the anomaly
polynomial in equation \eqref{eqn:I8general} is defined in terms of
one-instanton normalized traces $\text{Tr}F^n$. The conversion between the two
is related to Casimir invariants of the symmetry algebras, and up to quartic
order we have:
\begin{equation}\label{trace-relations-ABC}
	\text{tr}_{\bm{R}}F^2 = A_{\bm{R}}\,\text{Tr}F^2\,,\qquad
	\text{tr}_{\bm{R}}F^4 = B_{\bm{R}}\,\text{Tr}F^4 + C_{\bm{R}} \,(\text{Tr}F^2)^2\,.
\end{equation}
The coefficients $A_{\bm{R}}\,,B_{\bm{R}}$, and $C_{\bm{R}}$
depend on the representation $\bm{R}$; for representations appearing in the matter
spectrum needed in the study of 6d SCFTs, these values have been tabulated in
\cite{Heckman:2018jxk}.\footnote{A word of caution to the reader: while most of
		the recent literature on the anomaly polynomial of 6d SCFTs follow the
		same conventions as in this work, there has been a myriad of different
		choices, which sometimes make comparisons arduous. To wit, our
		normalizations for the traces are: \begin{equation}
				\text{tr}_{\textbf{adj}}F^2 = h^\vee\text{Tr}F^2\,,\quad\quad
				\text{tr}_{\mathcal{F}}F^4 = 1\cdot \text{Tr}F^4\,,
		\end{equation} where $\mathcal{F}$ refers to the defining
		representation of classical algebras. For exceptional cases,
		$B_{\bm{R}}=0$ due to the absence of an independent order-four
		Casimir invariant, and the normalization $A_{\textbf{adj}}=h^\vee$ fixes
		that of $C_{\bm{R}}$. Additionally, the case of $\mathfrak{so}_8$
		needs special attention as it has two independent quartic Casimir
		invariants. More details can be found in Appendix
		\ref{app:representations} and references therein.

Moreover, since there is no uniform normalization of the trace-relation
coefficients, their names also vary greatly across different fields of both
mathematics and physics. In the review of 6d SCFTs \cite{Heckman:2018jxk},
which tabulates them in Appendix F for representations appearing in generalized
quivers, they are defined as: $h_{\bm{R}}=A_{\bm{R}}$,
$x_{\bm{R}}=B_{\bm{R}}$, and $y_{\bm{R}}=C_{\bm{R}}$.  }
The study of nilpotent deformations and the associated breaking patterns
however involves representations that go beyond those usually encountered in
the quiver description. The trace-relation coefficients of a representation can
nonetheless be found in a straightforward manner knowing the weight system of the Lie algebra.
This procedure is reviewed in Appendix \ref{app:indices}.

In addition to contributions from the standard weakly-coupled hypermultiplets,
the ``one-loop'' term $I_8^\text{1-loop}$ of a theory might also involve
$(2,0)$ tensor multiplets and E-strings, which may arise in the
presence of undecorated $(-1)$- and $(-2)$-curves, respectively. The anomaly
polynomial of the former can be decomposed in terms of its $(1,0)$ supermultiplet
content, while the contribution of an E-string can be computed via anomaly
inflow \cite{Ohmori:2014pca}. The explicit expression of each individual
contribution to the ``one-loop'' term can be found in, e.g.,
\cite{Ohmori:2014pca, Ohmori:2014kda, Baume:2021qho}. The particular cases of
the tensor and vector multiplets, which are the supermultiplets most relevant
in this work, are written explicitly later in this section, see equations \eqref{I8-tensor}
and \eqref{I8-vector}.

After summing the contributions of each multiplet in the matter
spectrum -- which can be read directly from the tensor branch quiver -- the
``one-loop'' term will generically not be free of all gauge anomalies. However,
due to the presence of tensor multiplets in the spectrum, these can be cured
via a six-dimensional Green--Schwarz--West--Sagnotti mechanism
\cite{Green:1984sg, Green:1984bx,Sagnotti:1992qw}, leading to a well-defined
theory. This term is a generalized version of the celebrated Green--Schwarz
mechanism in ten dimensions, and is therefore commonly referred to simply as
the Green--Schwarz or ``GS'' term.

The contribution $I_8^\text{GS}$ is found at a non-generic point of the tensor
branch reached by successively blowing down all undecorated $(-1)$-curves. At
that point the quiver is described by an adjacency matrix $\widetilde{A}^{ij}$,
and the GS term is given by:
\begin{equation}
		I_8^\text{GS} = -\frac{1}{2}\widetilde{A}_{ij} I^i I^j\,,\qquad I^i = -\widetilde{A}^{ij}c_2(F_j) - B^{ia}c_2(F_a) - (2+\widetilde{A}^{ii})p_1(T) + y^i c_2(R)\,,
\end{equation}
where $\widetilde{A}_{ij} = (\widetilde{A}^{-1})_{ij}$, and
$c_2(F)=\frac{1}{4}\text{Tr}F^2$ denotes the second Chern class of the gauge
and flavor bundles. Note that we will always define them in terms of
one-instanton normalized traces, see equation \eqref{trace-relations-ABC}. The
four-form $I^i$ is furthermore related to the Bianchi identity for the
three-form field strength of the tensor multiplet $H_i$ on the $i$-th curve:
\begin{equation}
		dH^i = I^i\,.
\end{equation}
At the generic point of the tensor branch, the coefficients $y^i$ are given by
the dual Coxeter number $h^\vee_{\mathfrak{g}_i}$ of the algebra on the curve, and
their changes must be tracked when blowing down $(-1)$-curves. Finally, the
matrix $B^{ia}$ encodes the intersection of non-compact flavor curves with
those associated with gauge algebras and is often either zero or one, although
special care must be taken when the flavor involves E-strings or (pseudo)-real
representations. We defer to the algorithm described in \cite{Baume:2021qho}
for a detailed explanation of how to obtain the GS contribution. 

Once both the ``one-loop'' and Green--Schwarz--West--Sagnotti terms have been
determined, one obtains a final expression free of any gauge anomaly. While it
may seem very cumbersome to track the various coefficients associated to the
potentially quite large number of curves for either terms given a quiver, under
the classification scheme described in Section \ref{sec:bestiary}, it turns
out that the resulting anomaly polynomial always takes a relatively simple form
depending only on the parent theory and the nilpotent orbit describing the
Higgs mechanism applied to the moment map.

\subsection{Anomaly Polynomial of Conformal Matter}\label{sec:anomaly-conformal-matter}

To emphasize the simplicity of the resulting expressions, let us first consider
the case of $A_{N-1}^\mathfrak{g}$, the rank $N$ $(\mathfrak{g}, \mathfrak{g})$
conformal matter theory, see Section \ref{sec:fractional-CM}. Going through the
algorithm summarized above, one finds expressions that depend solely on
quantities related to $\mathfrak{g}$ \cite{Ohmori:2014kda}:
\begin{equation}\label{I8-CM-expanded}
		\begin{aligned}
			I_8(A_{N-1}^\mathfrak{g}) =& 
			\frac{\alpha_\text{CM}}{24}c_2(R)^2
			+ \frac{\beta_\text{CM}}{24} c_2(R) p_1(T)
			+\frac{\gamma_\text{CM}}{24} p_1(T)^2
			+\frac{\delta_\text{CM}}{24} p_2(T)\\
			& + \frac{B_{\textbf{adj}}}{48}\left(\text{Tr}F^4_L + \text{Tr}F^4_R\right)
			- \left(\frac{1}{32N} - \frac{C_{\textbf{adj}}}{48} \right)\left( (\text{Tr}F_L^2)^2 + (\text{Tr}F_R^2)^2\right)\\
			& + \frac{1}{16N}\text{Tr}F_L^2\text{Tr}F_R^2 + \left(\frac{h^\vee_\mathfrak{g}}{4}\frac{p_1(T)}{24}-\frac{1}{8}(\Gamma N-h^\vee_{\mathfrak{g}})c_2(R)\right)\left(\text{Tr}F_L^2 + \text{Tr}F_R^2\right)\,.
		\end{aligned}
\end{equation}
By abuse of notation, $\Gamma$ will denote the order of the McKay-dual discrete
group of $\mathfrak{g}$ -- that is, the ADE discrete subgroup of $SU(2)$ of the
same type -- when appearing in expressions related to the
anomaly polynomial. Its value as well as some of the main quantities of simple
Lie algebras are collated in Table \ref{tab:algebra-values} for convenience.
The coefficients involving purely R-symmetry or spacetime terms are given by:
\begin{equation}
	\begin{aligned}
		\alpha_\text{CM} &= \Gamma^2 N^3 - 2 \Gamma\chi  N + (\text{dim}(\mathfrak{g}) - 1) \,,\\
		\beta_\text{CM} &= -\frac{1}{2}(\Gamma\chi - 1) N + \frac{1}{2}(\text{dim}(\mathfrak{g}) - 1)\,,\\
		\gamma_\text{CM} &= \frac{N}{8} + \frac{7 \text{dim}(\mathfrak{g}) - 23}{240}\,,\\
		\delta_\text{CM} &= -\frac{N}{2} - \frac{\text{dim}(\mathfrak{g}) -29}{60} \,.
	\end{aligned}
\end{equation}
The form of the tensor branch description of conformal matter depends on the
choice of $\mathfrak{g}$, see Table \ref{tbl:CM-and-nodes}, and the gauge
spectrum entails various representations of the involved algebras, in
particular for exceptional algebras. However, note that the above expressions
only depend on standard group-theoretical quantities associated with
$\mathfrak{g}$ and its adjoint representation, rather than the details of the
quiver. In particular, $\chi$ is a combination of the rank of the algebra and the order of the associated discrete group, and is defined as
\begin{equation}
    \chi = r_\mathfrak{g} + 1 - \frac{1}{\Gamma} \,.
\end{equation}
This is our first hint that the moment map -- falling in the same
multiplet as the flavor current -- plays a special r\^ole in the anomaly
polynomial of conformal matter and its deformations.
\begin{table}
		\centering
		\begin{tabular}{ccccccc}
				\toprule
				$\mathfrak{g}$ & $r_\mathfrak{g}$ & $\text{dim}(\mathfrak{g})$ & $h^\vee_\mathfrak{g}$ & $\Gamma$ & $B_{\textbf{adj}}$ & $C_{\textbf{adj}}$\\\midrule
				$\mathfrak{su}_2$ & $1$ & $3$ & $2$ & $2$ & $0$ & $2$\\
				$\mathfrak{su}_3$ & $2$ & $8$ & $3$ & $3$ & $0$ & $\frac{9}{4}$\\
				$\mathfrak{su}_{n\geq4}$ & $n-1$ & $n^2-1$ & $n$ & $n$ & $2n$ & $\frac{3}{2}$ \\
				$\mathfrak{so}_{8}$ & $4$ & $28$ & $6$ & $8$ & $6$ & $3$\\
				$\mathfrak{so}_{p\neq8}$ & $\lfloor \frac{p}{2}\rfloor$ & $\frac{1}{2}p(p-1)$ & $p-2$ & $2p-8$ & $p-8$ & $3$\\
				$\mathfrak{sp}_{k}$ & $k$ & $k(2k+1)$ & $k+1$ & --- & $2k+8$ & $\frac{3}{4}$\\
				$\mathfrak{g}_{2}$ & $2$ & $14$ & $4$ & --- & $0$ & $\frac{5}{2}$\\
				$\mathfrak{f}_{4}$ & $4$ & $52$ & $9$ & --- & $0$ & $\frac{15}{4}$\\
				$\mathfrak{e}_{6}$ & $6$ & $78$ & $12$ & $24$ & $0$ & $\frac{9}{2}$\\
				$\mathfrak{e}_{7}$ & $7$ & $133$ & $18$ & $48$ & $0$ & $\frac{6}{2}$\\
				$\mathfrak{e}_{8}$ & $8$ & $248$ & $30$ & $120$ & $0$ & $\frac{9}{2}$\\
				\bottomrule
		\end{tabular}
		\caption{Group-theoretic quantities associated with simple Lie algebras
				relevant in this work. Note that the order of the McKay-dual
				discrete group, $\Gamma$, is defined only for ADE algebra
				$\mathfrak{su}_n\,,\mathfrak{so}_{2k}\,,\mathfrak{e}_{6,7,8}$.
				Furthermore, exceptional algebras, including
				$\mathfrak{su}_2\,,\mathfrak{su}_3$ do not have quartic Casimir
				invariant, while $\mathfrak{so}_8$ has two. This changes the
				values of $B_{\textbf{adj}}$ and $C_{\textbf{adj}}$ for these algebras
				as a result, see Appendix \ref{app:representations} for
		additional details.} \label{tab:algebra-values}
\end{table}

The expression given in equation \eqref{I8-CM-expanded} was first obtained from
the tensor branch description in \cite{Ohmori:2014kda}, where the result of the
algorithm was also cross-checked with an anomaly-inflow computation from the
M-theory realization, namely as the worldvolume theory of a stack $N$ M5-branes
probing a $\mathbb{C}^2/\Gamma$ singularity. Through this method, one obtains a
more elegant and compact expression for the anomaly polynomial:
\begin{equation}\label{I8-CM}
		\begin{aligned}
			I_8(A^{\mathfrak{g}}_{N-1}) =&\,  
		\frac{N^3}{24}(c_2(R)\Gamma)^2
		- \frac{N}{2}(c_2(R)\Gamma) \left(J(F_L) + J(F_R)\right) - \frac{1}{2N}\left(J(F_L) - J(F_R)\right)^2\\
		&
		+ N I_\text{sing} - I_8^\text{tensor} - \frac{1}{2}\left(I_8^\text{vec}(F_L)- I_8^\text{vec}(F_L)\right)\,,
		\end{aligned}
\end{equation}
where we defined the following quantities:
\begin{equation}\label{J-CM}
	\begin{aligned}
			I_\text{sing} &= \frac{1}{24}\left(\frac{1}{2}p_1(T)c_2(R) + \frac{1}{8}p_1(T)^2- \frac{1}{2}p_2(T)\right)\,,\\
			J(F) &= \frac{\chi}{48}(4c_2(R)+ p_1(T)) + c_2(F) \,,
	\end{aligned}
\end{equation}
which find their origin in the reduction of the M-theory Chern--Simons terms,
and lead to contributions from degrees of freedom localized at the orbifold
singularity.\footnote{The term proportional to $N^{-1}$ is associated with
center-of-mass contributions and was written in reference
\cite{Ohmori:2014kda} simply as $- \frac{1}{8N}\left(\text{Tr}F^2_L -
\text{Tr}F^2_L\right)^2$. While \emph{a priori} unrelated to the
reduction of terms near the singularity, in equation \eqref{I8-CM} we
write it in terms of $J(F)$ for later convenience.} In the M-theory
realization, one must also remove contributions from the center-of-mass
tensor multiplet and those associated with boundary conditions for
directions normal to the M5-branes. These are proportional to the
anomaly polynomial of tensor and vector multiplets, respectively. As
mentioned above, the contribution of these two types of multiplet are
given in terms of characteristic classes, see
Appendix \ref{app:representations}: 
\begin{align}
		I_8^\text{tensor} =& \left.\left(\frac{1}{2}\text{ch}_{\bm{2}}(R)\widehat{A}(T) - \frac{1}{8} L(T)\right)\right|_\text{8-form} \nonumber\\
				=& \frac{1}{24}c_2(R)^2 + \frac{1}{48}c_2(R)p_1(T) + \frac{1}{5760}\big(23p_1(T)^2 - 116 p_2(T)\big)\,,\label{I8-tensor}\\
				I_8^\text{vec}(F) =& - \frac{1}{2}\left.\widehat{A}(T)\text{ch}_{\bm{2}}(R)\text{ch}_{\textbf{adj}}(F)\right|_\text{8-form} \nonumber\\
						=& -\frac{1}{24}\big(\text{tr}_{\textbf{adj}}F^4 + 6 c_2(R)\text{tr}_{\textbf{adj}}F^2 + \text{dim}(\mathfrak{g}) c_2(R)^2\big)\nonumber\\
		& - \frac{1}{48}p_1(T)\big(\text{tr}_{\textbf{adj}}F^2 + \text{dim}(\mathfrak{g})c_2(R)\big) - \frac{\text{dim}(\mathfrak{g})}{5760}\big(7p_1(T)^2 - 4p_2(T)\big)\,.\label{I8-vector}
\end{align}

Equation \eqref{I8-CM} therefore neatly repackages every flavor contribution
either in the Chern character of the adjoint representation, or quantities
scaling with $N$, and encoded in $J(F)$. The only remaining explicit dependence
on $\mathfrak{g}$ appears through the order of the McKay dual discrete group
$\Gamma$. 

Beyond its elegant form, the usefulness of this expression for the anomaly
polynomial is its behavior under fusion, as it makes clear that the resulting
theory will be free of any gauge anomaly. Indeed, we have reviewed in the
previous section that fusion allows us to ``glue'' quivers together by gauging
common flavor symmetries. Equivalently, in the field theory we are gauging this
flavor, and using the tensor multiplet to mediate a
Green--Schwarz--West--Sagnotti mechanism so as to ensure a gauge-anomaly-free
result. Therefore, the form of the anomaly polynomial in equation \eqref{I8-CM}
teaches us that we can in fact use it as if it was a ``one-loop'' contribution!

To illustrate this, let us consider two minimal, i.e., rank one, $(\mathfrak{g},
\mathfrak{g})$ conformal matter theories,
$A_0^{\mathfrak{g}}$. If we fuse
them together by gauging one of the common flavor factors, we obtain rank two conformal
matter, $A_1^{\mathfrak{g}}$:
\begin{equation}
	[\mathfrak{g}_L]\text{---}[\mathfrak{g}_1]\quad\oplus\quad
	[\mathfrak{g}_1]\text{---}[\mathfrak{g}_R]\quad =\quad
	[\mathfrak{g}_L]\text{---}\,\mathfrak{g}_1\,\text{---}[\mathfrak{g}_R]\,,
\end{equation}
where we differentiated the different factors of the same algebra $\mathfrak{g}$
by their respective subscript. By gauging the common the flavor factor
$\mathfrak{g}_1$, we introduce a vector multiplet in the spectrum, as well as a
tensor multiplet to mediate the Green--Schwarz--West--Sagnotti mechanism. The
effective ``one-loop'' term of the new theory is then given by:
\begin{equation}\label{2CM-1loop}
		I_8^\text{1-loop} = I_8(A_0^{\mathfrak{g}}; F_L,F_1) + I_8(A_0^{\mathfrak{g}}; F_1,F_R) + I_8^\text{vec}(F_1) + I_8^\text{tensor}\,,
\end{equation}
where for clarity, we have shown each of the field strengths explicitly. The
form of the anomaly polynomial for $A_{N-1}^{\mathfrak{g}}$, given in
equation \eqref{I8-CM}, makes it clear that the ``one-loop'' contribution
cannot depend on quartic traces of the gauge algebra, $\text{Tr}F_1^4$, as all
contributions $I_8^\text{vec}(F_1)$ cancel, and only potentially dangerous
terms involving the second Chern class $c_2(F_1)$ remain. These can however be
cancelled via Green--Schwarz--West--Sagnotti mechanism, using a term of the
form
\begin{equation}\label{2CM-GS}
		I_8^\text{GS} = -\frac{1}{2} (\widetilde{A}^{11})^{-1} (I^1)^2\,,\qquad I^1 = -\widetilde{A}^{11} J(F_1) - B^{1L}J(F_L) - B^{1R}J(F_R) + \chi e^1 c_2(R)\,.
\end{equation}
A short inspection of the involved terms reveals that we must choose $\widetilde{A}^{11} =
-2$, $B^{1L}=B^{1R}=e^1=1$. This is precisely what we expect from the
partial tensor branch description. In the F-theory construction, when fusing
two conformal matter links together we must introduce a compact curve of
self-intersection $(-2)$, possibly leading to non-minimal fibers, see the discussion around equation \eqref{PTB-CM}. We therefore find the
base corresponding to the algebra $A_1$. Moreover, $B^{1L}$ and $B^{1R}$ can also be understood as the intersection numbers between this $(-2)$-curve and the non-compact curves
supporting the flavor symmetries.

Adding the two contributions from equations \eqref{2CM-1loop} and
\eqref{2CM-GS} together, we obtain an expression free of gauge anomalies:
\begin{equation}
		I_8(A_1^{\mathfrak{g}}) = I_8(A_0^{\mathfrak{g}};F_L,F_1) + I_8(A_0^{\mathfrak{g}}; F_1,F_R) + I_8^\text{vec}(F_1) + I_8^\text{tensor} + I_8^\text{GS}\,.
\end{equation}
Comparing this result with equation \eqref{I8-CM}, one can check that this
exactly reproduces the anomaly polynomial of rank two conformal matter,
$A_1^{\mathfrak{g}}$. By recursion, this is true also for higher-rank
conformal matter; we simply need to fuse $N$ of them together as if they were
simple hypermultiplets. The ``one-loop'' term is 
\begin{equation}
    \begin{aligned}
    I_8^\text{1-loop} = &\,
    I_8(A_{0}^{\mathfrak{g}};F_L,F_1) 
    + I_8(A_0^{\mathfrak{g}};F_{N-1},F_R) 
    + (N-1)I_8^\text{tensor}\\
    &+ \sum_{i=1}^{N-1}\bigg(I_8(A_0^{\mathfrak{g}};F_i,F_{i+1}) + I^\text{vec}_8(F_i)\bigg) 
    \,,
    \end{aligned}
\end{equation}
and the remaining gauge terms are cancelled by a similar
Green--Schwarz--West--Sagnotti mechanism:
\begin{equation}\label{GS-CM}
		I_8^\text{GS} = -\frac{1}{2} \widetilde{A}_{ij} I^iI^j\,,\qquad I_i = -\widetilde{A}^{ik} J(F_k) - B^{ia}J(F_a) + \chi e^i c_2(R)\,,
\end{equation}
where $e^i=1\,\forall i$, $\widetilde{A}_{ij} = (\widetilde{A}^{ij})^{-1}$, and $B^{ia} =
\delta_{i-a=1}$. It is easy to see the adjacency matrix $\widetilde{A}^{ij}$ must be
exactly (minus) the Cartan matrix of the algebra $A_{N-1}$ associated
with the base. The quantity $J(F)$ is then interpreted as a contribution to the
Bianchi identity related to the associated tensionless string:
\begin{equation}
		d H \sim J(F)\,,
\end{equation}
where $H$ is the field strength of the associated tensor field. This should not
be too surprising, as in the anomaly inflow computation in M-theory, the
quantity $J$ is by construction associated with a solution for the four-flux
form in eleven dimensions: $G_4\sim J$ \cite{Ohmori:2014kda}. In equation
\eqref{I8-CM}, we can therefore think of the second line as a ``one-loop''
contribution, essentially depending simply on characteristic classes, while the
first originates from the GS term.

This simple argument teaches us that while 't Hooft anomaly matching enables us
to find the anomaly polynomial via the full tensor branch description, once we
know the contribution of a single elementary building block at the singular
point, namely minimal conformal matter $A_0^{\mathfrak{g}}$, the rest
follows as if we were in a weakly-coupled regime. It is in spirit much the same
as the case of E-strings: these theories are themselves non-perturbative
objects, but once their contribution to the anomaly polynomial has been
determined \cite{Ohmori:2014pca}, when they are coupled to other weakly-coupled
supermultiplets to form a more complicated theory, they follow the same rules
as if they were regular supermultiplets. 

One might think that the case of conformal matter is quite special, and this
line of thought should not generalize to more complicated theories. However,
since long quivers are made up by fusing minimal conformal matter together up
to decorations at each ends, we will now see that a similar reasoning can be
applied not only to any parent theory of long quivers, but their nilpotent
breakings as well.

Let us first consider the case of fractional conformal matter
$\mathcal{A}_{N-1;f_L,f_R}^{\mathfrak{g}}$. Since the fractions are only
allowed for algebras $\mathfrak{g}$ of type DE and their number limited, it
is straightforward to apply the algorithm and compute the anomaly polynomial
of the long quivers in each case. When $f_L=f_R$, this analysis was already
performed in \cite{Mekareeya:2017sqh} by fusing together multiple copies of
fractional theories.\footnote{We note that we use different conventions than
		in \cite{Mekareeya:2017sqh} to label the theories. There, a long
		quiver is obtained by fusing different types of quiver than the one
		we have defined in Table \ref{tbl:eattachments}. For instance, the
		theory we defined as $\mathcal{A}_{N-1;\frac{1}{3},
		\frac{1}{3}}^{\mathfrak{e}_6}$ is obtained by fusing the $N$ times
the quiver $1\overset{\mathfrak{e}_6}{6}1$ with $N-1$ ``nodes''  $\overset{\mathfrak{su}_3}{3}$. This notation unfortunately
obscures the possibility of having different fractions at each end, and makes
breaking patterns harder to track.} Based on their results, we have obtained
an expression that also applies in cases where $f_L\neq f_R$.  As we have
seen in the previous section, fractional conformal matter can be depicted as 
\begin{equation}
\mathcal{A}^{\mathfrak{g}}_{N-1; f_L,f_R}:\qquad
	[\mathfrak{g}_{f_L}]\,\text{---}\,\mathfrak{g}\,\overbrace{\vphantom{\overset{\mathfrak{g}}{2}}\text{---}\,\mathfrak{g}\,\text{---}\dots\text{---}\,\mathfrak{g}\,\text{---}}^{N-2}\,\mathfrak{g}\,\text{---}\,[\mathfrak{g}_{f_R}]\,,
\end{equation}
and we can see that there are $N-2$ ``full'' minimal conformal matter forming
the long spine, and two $\mathcal{A}_{0;1, f_{L,R}}^{\mathfrak{g}}$ theories
with a fraction at each end. Defining the ``effective'' total number of
conformal matter as
\begin{equation}\label{eqn:Neff}
	N_\text{eff} = (N-2) + f_L + f_R\,,
\end{equation}
the anomaly polynomial of $\mathcal{A}_{N,f_L,f_R}^{\mathfrak{g}}$ takes a form
that, barring the use of $N_\text{eff}$ rather than $N$, closely resembles that
of its non-fractional cousin -- see equation \eqref{I8-CM}:
\begin{equation}\label{I8-fCM}
	\begin{aligned}
			I_8(\mathcal{A}^{\mathfrak{g}}_{N-1;f_L,f_R}) &=  
			\frac{N_\text{eff}^3}{24}(c_2(R)\Gamma)^2
			- \frac{N_\text{eff}}{2}(c_2(R)\Gamma) \left(J_\text{fr}(F_L) + J_\text{fr}(F_R)\right)\\
			&- \frac{1}{2N_\text{eff}}(J_\text{fr}(F_L) - J_\text{fr}(F_R))^2
			+ N_\text{eff} I_\text{sing}
			- I_8^\text{tensor}  \\
			&- \frac{1}{2}\left(I_8^\text{vec}(F_L)+I_8^\text{vec}(F_R)\right)- (e_L+e_R)I_\text{fr}\,,
	\end{aligned}
\end{equation}
where the coefficients $e_{L,R}$ are collated in Table \ref{tbl:eattachments} and $I_\text{fr}$ is defined in equation \eqref{def-Ifr} below. The main difference between this case and the original anomaly polynomial of conformal matter is that the contribution of the vector multiplets are now
taken with respect to the algebras associated with the fractions,
$\mathfrak{g}_{f_L}$ and $\mathfrak{g}_{f_R}$, see Table
\ref{tbl:eattachments}, and the four-form related to the Bianchi identity at
the fixed point needs to be slightly modified:
\begin{equation}\label{Jf-fractional}
	\begin{gathered}
			J_\text{fr}(F) = \frac{\chi_\text{fr}}{48}(4c_2(R) + p_1(T)) + \frac{1}{d_f} c_2(F)\,,\\
			\chi_\text{fr} = \chi + 12 (1-\frac{1}{d_f})\,,\qquad d_f = \text{denom}(f)\,.
	\end{gathered}
\end{equation}
In equation \eqref{I8-fCM}, we stress that $r_{\mathfrak{g}}, \Gamma$ are those
associated with the ``full'' algebra $\mathfrak{g}$, rather than
$\mathfrak{g}_{f_{L,R}}\subseteq\mathfrak{g}$, and $J_\text{fr}(F_{L,R})$ is
defined with respect to the left, respectively right, fraction. In addition,
there is also an extra contribution whose prefactor depends on the fractions:
\begin{equation}\label{def-Ifr}
		I_\text{fr} = \frac{1}{24}\left(c_2(R)^2 + \frac{1}{6}c_2(R)p_1(T) - \frac{1}{48}\big(p_1(T)^2 - 4p_2(T)\big)\right)\,,
\end{equation}
This terms can be understood as coming from the frozen singularity, or
equivalently as descending from the contribution of the full, non-fractional, vector multiplet $I_8^\text{vec}(F_{\mathfrak{g}})$, which is now decomposed into the fractional part of the remnant flavor symmetry and $I_\text{fr}$. The particular
values of the coefficients $e_{L,R}$ are given in Table \ref{tbl:eattachments}.
While we have not found a closed-form expression in terms of the fraction
numbers and characters such as the A-roof genus $\widehat{A}$, we observe that they do not depend on the full algebra $\mathfrak{g}$
but rather on the fractional number itself, and have a reflection symmetry around
$f=\frac{1}{2}$. 

As a sanity check, we can see that when $f_L=1=f_R$, we recover the result for
conformal matter: $J_\text{fr}(F)=J(F)$ and $e=0$ in that case, and the above
expression is identical to that of equation \eqref{I8-CM}. 

The anomaly polynomial of higher-rank fractional conformal matter
$\mathcal{A}^{\mathfrak{g}}_{N-1;f_L,f_R}$ given in equation \eqref{I8-fCM}
once again enables us to understand this theory as coming from quantities
computed directly at the partial tensor branch by fusing rank one building blocks:
\begin{equation}
		\begin{aligned}
			I_8(A^{\mathfrak{g}}_{N-1;f_L,f_R}) =&\, I_8(A^{\mathfrak{g}}_{0;f_L,1}; F_L, F_1) + \sum_{i=1}^{N-2}\bigg( I_8(A^{\mathfrak{g}}_{0}; F_i, F_{i+1}) + I_8^\text{vec}(F_i)\bigg)\\
			& + I_8(A^{\mathfrak{g}}_{0;1, f_R}; F_{N-1}, F_R)+ (N-1)I_8^\text{tensor} + I_8^\text{GS}\,.
		\end{aligned}
\end{equation}
The GS terms takes the same form as in equation \eqref{GS-CM} with the
replacement $J(F)\to J_\text{fr}(O)$ for the contribution at each
end, and the pairing matrix $\widetilde{A}^{ij}$ must be slightly modified with respect to
that of conformal matter. Taking the fractions with unit numerators, $f_{L,R} =
\frac{1}{d_{L,R}}$, the presence of incomplete minimal conformal matter at
either sides will change the charge of the strings associated with the gauge
algebras at both ends of the quiver, and a short computation shows that the
pairing matrix is that of the quiver:
\begin{equation}\label{PTB-fraction}
		\overset{\mathfrak{g}}{(1+d_L)}\underbrace{\overset{\mathfrak{g}}{2}\dots\overset{\mathfrak{g}}{2}}_{N-2}\overset{\mathfrak{g}}{(1+d_R)}\,.
\end{equation}
This is again exactly what one would expect for the geometry, as it is the
partial tensor branch quiver obtained after successively shrinking all
$(-1)$-curves, see Table \ref{tbl:eattachments}.

When the numerators of the fractions are not one, this interpretation gets
slightly obscured and the form of the partial tensor branch in equation
\eqref{PTB-fraction} needs to be modified. Indeed, as we have discussed in
Section \ref{sec:bestiary}, the bases in those cases involve more curves
than for the complete conformal matter theories of the same rank.
These can however be understood as being realized by fusing further rank one
fractional theories to the main spine until the correct fractional number has
been achieved. The self-intersection of the additional branes is then
understood as above: coming from the Green--Schwarz--West--Sagnotti mechanism
the singular point. In the M-theory picture, this phenomenon can also be
understood as the recombination of fractional M5-branes into larger
fractional -- or even full -- M5-branes \cite{Mekareeya:2017sqh}.

Beyond conformal matter, other long quivers with type-D bases
$\mathcal{D}^{\mathfrak{g}}_{N; f }$ and orbi-instantons $\mathcal{O}^{\mathfrak{g}}_{N;f}$ can be understood in a similar fashion. We come back to those cases in
Section \ref{sec:other-I8}.

\subsection{Nilpotent Higgs Branch Flows in Six Dimensions}\label{sec:nilpotent-RG}

The interpretation of the anomaly polynomial of conformal matter as obtained
by fusing minimal conformal matter theories together to obtain arbitrary
large quivers is quite suggestive, and we will now show that there are
similar expressions when a parent theory is deformed to another with
deformations associated to nilpotent orbits of the flavor symmetries.

For a given F-theory base, $\mathcal{A}_{N-1;f_L,f_R}$, $\mathcal{D}_{N;f}$
and $\mathcal{O}_{N;f}$, there can be multiple 6d $(1,0)$ SCFTs, which from the
geometric point of view are reached through complex-structure deformations.
As alluded to above, in the case of long quivers -- that is for large-enough
values of $N$ -- it has been proposed that a partial classification scheme is
given by nilpotent orbits $O$ of simple components of the flavor symmetry \cite{Heckman:2016ssk, Mekareeya:2016yal}. 

Field theoretically, these are interpreted as renormalization group flows.
More precisely, the presence of flavor implies the existence of a conserved
flavor current, which is part of a $\mathcal{D}[2]$ superconformal multiplet in
the nomenclature of \cite{Cordova:2016emh, Buican:2016hpb}. Its superconformal
primary, called the moment map, is a scalar operator $\phi$ of
conformal dimension $\Delta=4$, transforming in the adjoint representation of
the flavor symmetry, as well as a doublet of the $SU(2)$ R-symmetry. Through
the usual arguments, giving a vacuum expectation value to the moment map will
trigger an RG flow, ultimately  reaching another conformal fixed point in the
infrared. The flows of interest are then associated with those where the
vacuum expectation values of $\phi$ is chosen to be a representative of the
nilpotent orbit:
\begin{equation}
		\left<\phi\right> = X \in O\,.
\end{equation}
We will refer to such deformations as nilpotent RG flows. Note that in the case
of orbi-instantons, nilpotent RG flows are not sufficient to classify all long
quivers, and must be supplemented by a choice of embedding of a discrete group
$\Gamma\subset SU(2)$ into $E_8$. The associated flows are more involved that
those coming from nilpotent orbits, and will not be considered here -- see
however \cite{Heckman:2013pva, Heckman:2015bfa, Frey:2018vpw, Fazzi:2022yca, Fazzi:2022hal}. We leave a
systematic analysis of their anomaly polynomial for future work.

The case of (fractional) conformal matter corresponds to an unbroken
$\mathfrak{g}_L\oplus\mathfrak{g}_R$ flavor symmetry, and therefore the trivial
nilpotent orbits $O_{L,R}=\varnothing$ are used:
$\mathcal{A}_{N-1;f_L,f_R}^\mathfrak{g} =
\mathcal{A}^{\mathfrak{g}}_{N-1;f_L,f_R}(\varnothing,\varnothing)$. As we have
discussed above, at the level of the anomaly polynomial, these theories can be
understood without referring to the details of the F-theory geometry. One may
therefore ask whether the proposed classification scheme from nilpotent orbits
can also be understood fully in terms of the gauge-invariant conformal data. 

This is indeed the case, and as we will now show, knowing the anomaly
polynomial of a parent theory, that of an SCFT reached via an RG flow
described by a nilpotent orbit can be obtained directly from simple
group-theoretic data, without going through the tensor branch description and
the associated algorithm invoking 't Hooft anomaly-matching arguments.

Before describing the prescription to obtain the IR anomaly polynomial, let us
first summarize some of the relevant properties of nilpotent orbits of a simple
algebra $\mathfrak{g}$. For an in-depth treatment of the topic, we refer to
\cite{collingwood2017nilpotent}. Given a nilpotent orbit $O$, by the
Jacobson--Morozov theorem one can construct a triplet $(X,Y,H)$ of generators
of $\mathfrak{g}$ satisfying the standard $\mathfrak{su}_2$ commutation
relations, with $X\in O$ and $H$ in the Cartan subalgebra of $\mathfrak{g}$. A
nilpotent orbit therefore defines a homomorphism $\rho_O:
\mathfrak{su}(2)\hookrightarrow\mathfrak{g}$.

It follows that for any generator $E_i$ of $\mathfrak{g}$ associated with a simple root, $H$-eigenvalues of these generators can only take
values $0, 1$, and $2$ (up to conjugacy). They can then be arranged into a
\emph{weighted Dynkin diagram} labelling uniquely the nilpotent orbit
$O$:\footnote{The converse is not true, not all weighted Dynkin diagrams with
$w_i=0,1,2$ are associated with a nilpotent orbit.} 
\begin{equation}\label{def-wdd}
	O:\qquad  [w_1\dots,w_{r_{\mathfrak{g}}}]\,,\qquad
	[H,E_i] =w_i\,E_i\,,\qquad w_i=0,1,2\,.
\end{equation}

While this labelling exists for any type of simple Lie algebra, it is sometimes
useful to use an algebra-specific scheme. For instance, nilpotent orbits of
$\mathfrak{su}_K$ are in one-to-one correspondence with partitions of $K$,
while those of $\mathfrak{so}_{2K}$ are associated with \emph{even} partitions
of $K$. For exceptional algebras, there are no classification in terms of
partitions, and it is common to use so-called Bala--Carter labels
\cite{bala1976classesI, bala1976classesII} -- see
\cite{collingwood2017nilpotent} for a discussion of the different ways to label
nilpotent orbits. While there is a procedure to find the quiver from a
partition for classical algebras, we are not aware of a simple connection
between the weighted Dynkin diagram of a nilpotent orbit or its Bala--Carter
labels and its curve configuration. We recall that the procedure for partitions
and the tables giving the mapping between Bala--Carter labels of exceptional
algebras and the associated quivers are given in Appendix \ref{app:nilp2TB}.

As the triplet $(X,Y,H)$ defines the embedding $\rho_O$, it also induces a branching
rule for the adjoint (or any other) representation:
\begin{equation}\label{nilpotent-BR}
	\begin{aligned}
			\rho_O:\qquad\mathfrak{g}~\, &\longrightarrow~ \mathfrak{su}(2)_X \oplus \mathfrak{f}\,,\\
			\textbf{adj} &\longrightarrow~ \bigoplus_{\ell}~(\bm{d}_\ell,\bm{R}_\ell)\,,
	\end{aligned}
\end{equation}
where $\mathfrak{su}(2)_X$ is written so to emphasize that it is related to the
nilpotent element $X$, rather than the R-symmetry, and $\mathfrak{f}$ is its
centralizer in $\mathfrak{g}$. When $X$ is interpreted as the vacuum expectation
value of the moment map, $\mathfrak{f}$ is the (possibly semi-simple,
or even trivial) remnant flavor symmetry of the infrared theory. For nilpotent
deformations of 6d $(1,0)$ SCFT, they perfectly match the ones that can be read
off the tensor branch quivers described in the previous section
\cite{Heckman:2016ssk, Hassler:2019eso, Heckman:2018pqx}. 

In the mathematics literature, the branching rule defined in equation
\eqref{nilpotent-BR} is known as the Jacobson--Morozov decomposition. While the
weighted Dynkin diagrams defined in equation \eqref{def-wdd} are in principle
enough to obtain the branching rule of the adjoint representation of
$\mathfrak{g}$, we have tabulated them in Appendix \ref{app:nilps} for all
exceptional algebras. For the classical series, the Jacobson--Morozov
decomposition of the defining representation can be read off directly from
partitions, and that of the adjoint can then be found using tensor products.
This procedure is explained in detail in the same appendix.

We will now show that the Jacobson--Morozov decomposition is not only enough to
find find the flavor of a given 6d $(1,0)$ theory, but the complete anomaly
polynomial as well. We will focus first on conformal matter theories with only
a single nilpotent orbit turned on, that is the class of SCFT given by
$\mathcal{A}_{N-1;1,1}^{\mathfrak{g}}(O, \varnothing)$. If $N$ is large, the
breaking on the left will not affect the right-hand side of the quiver. 

The anomaly polynomial of the parent theory -- that is when $O$ is trivial --
in the form given in equation \eqref{I8-CM}, in addition to its compact form
and making it easier to implement the fusion process, has the property that the
flavor dependence appears only through $J(F)$ and $I^\text{vec}_{8}$. These two
quantities have the advantage that by construction, they behave nicely under
branching rules, and will be the basis for our prescription to obtain the
anomaly polynomial of the IR theory. It is therefore easy to implement the
branching rule given in equation \eqref{nilpotent-BR} and rewrite the anomaly
polynomial of the parent theory in terms of the flavor data of the IR
description. Of course, along the RG flows, there are modes that will decouple,
and the IR anomaly polynomial must take this into account. As we will
see shortly, once we have found the prescription for conformal matter, we may
argue that in the same way that other long quivers are obtained through fusion
of lower-rank conformal matter, the prescription also applies to any theory with
large-enough $N$.

To illustrate how the Jacobson--Morozov decomposition encodes data in the IR
theory, let us first focus on the anomaly coefficient $\delta$. As the
Green--Schwarz--West--Sagnotti mechanism described above does not involve terms
proportional to $p_2(T)$, this ensures that from the tensor branch point of
view, $\delta$ can only arise from ``one-loop'' contributions. In
\cite{Mekareeya:2016yal}, it was observed that in the case of conformal matter,
the difference in this coefficient between the UV parent theory and the IR SCFT
obtained by turning on a nilpotent orbit $O$ was given by
\begin{equation}\label{deltadelta}
		\delta^\text{UV} - \delta^\text{IR} = -\frac{1}{120}\text{dim}(O)\,.
\end{equation}
Through the tensor branch description, it can be understood as decoupling the
various vector and hypermultiplets getting massive after performing a
complex-structure deformation. In fact, this coefficient is related to the
dimension of the Higgs branch of the SCFTs \cite{Ohmori:2015pua, Ohmori:2015pia,
Mekareeya:2016yal}:
\begin{equation}
	\text{dim}(\text{HB}^\text{UV}) - \text{dim}(\text{HB}^\text{IR})=-60(\delta^\text{UV} - \delta^\text{IR})\,,
\end{equation}
and it is therefore natural to expect that it decreases as we go to the IR
fixed point. Note that the complex dimension of a nilpotent orbit is always even, and thus
that the change in the quaternionic dimension of the Higgs branch is an integer.

One can however interpret equation \eqref{deltadelta} from a slightly
different, but ultimately equivalent, point of view that does not involve
fields charged under a gauge symmetry and is agnostic about the details of the
parent theory: the appearance of the dimension of the orbit is a consequence of
Goldstone's theorem. In the UV theory, the moment map transforms in
the adjoint representation of the flavor symmetry and there are therefore
$\text{dim}(\mathfrak{g})$ degrees of freedom. After turning on a nilpotent
vacuum expectation value $\left<\phi\right>=X$, there is a massless field
associated with each of the unbroken generators. That is, they are part of the
commutant of the image of $\rho_O$:
\begin{equation}
	\mathcal{C}(O)=\left\{ Z\in\textbf{adj}\,\middle|\, [X,Z]=0 \right\}\,.
\end{equation}
Using that $\text{dim}(\mathfrak{g})=\text{dim}(O)+\text{dim}(\mathcal{C}(O))$,
there are therefore $\text{dim}(O)$ degrees of freedom that become massive, and
whose contributions must be removed from the anomaly polynomial.

More generally, given a representation $(\bm{d}_\ell,\bm{R}_\ell)$ appearing in
the Jacobson--Morozov decomposition of $O$, there are $\text{dim}(\bm{R}_\ell)$
elements commuting with $X$: by definition, given an irreducible
representations of $\mathfrak{su}_2$, only the highest-weight state commutes
with $X$. In the infrared the coefficient $\delta$ is therefore given by:
\begin{equation}\label{deltadelta2}
		\delta^\text{UV} - \delta^\text{IR} = 
		-\frac{1}{120}\left(\text{dim}(\mathfrak{g}) - \sum_\ell \text{dim}(\bm{R}_\ell)\right)
        =-\frac{1}{120}\text{dim}(O) \,.
\end{equation}
Note that while the original argument \cite{Ohmori:2015pua, Ohmori:2015pia,
Mekareeya:2016yal} used the tensor branch description to argue the validity of
equation \eqref{deltadelta}, here we reinterpret this result by invoking solely
the Jacobson--Morozov decomposition of the nilpotent orbit, and in terms of
gauge-invariant data of the SCFTs. The number of unbroken generators in the
adjoint representation are thus explained as a consequence of Goldstone's
theorem applied to the moment-map superconformal multiplet.

Since we are giving a nilpotent vacuum expectation to the moment map,
it is therefore not surprising that all complex-structure deformations of
conformal matter arrange themselves into Hasse diagrams of the relevant
nilpotent orbits. We however need to argue that the remaining coefficients of
the anomaly polynomial can be explained in a similar fashion. A key point to
obtain such a result is that the modification of the coefficient $\delta$ can
be traced back to the contribution proportional to $I_8^\text{vec}(F_L)$ in
equation \eqref{I8-CM}.

``One-loop'' contributions of vector multiplets depend on the Chern character
of the flavor bundle, see equation \eqref{I8-vector}. As summarized in Appendix
\ref{app:representations} the Chern character has, by construction,
particularly nice properties under branching rules. In the parent UV theory, we
can therefore decompose quantities involving the background field strength of
the flavor symmetry $F_\mathfrak{g}$ into those appearing in the
Jacobson--Morozov decomposition given in equation \eqref{nilpotent-BR}:
\begin{equation}\label{ch-decomp-JM}
		\text{ch}_{\textbf{adj}}(F_{\mathfrak{g}})= \sum_{\ell}\text{ch}_{\bm{d}_\ell}(F_X)\text{ch}_{\bm{R}_\ell}(F_\mathfrak{f})\,,
\end{equation}
where we have used the notation $F_X$ to emphasize that this $\mathfrak{su}_2$
background field strength is associated with the element $X\in O$, rather than
a possible remnant flavor symmetry. 

Note that at this point, the above decomposition simply amounts to a
relabelling of the various roots of the adjoint representation of
$\mathfrak{g}$ in the yet-unbroken parent theory. Now, when one does turn on a
vacuum expectation value for the momentum map, the symmetry associated with
$F_X$ is broken, and only the highest-weight states remain massless.  Moreover,
as $\phi$ also transforms as a doublet of the R-symmetry, supersymmetry is also
broken away from the fixed point. However, in the IR a new R-symmetry emerges
as the unbroken diagonal subgroup of
$\mathfrak{su}(2)_{R^{UV}}\oplus\mathfrak{su}(2)_X$ \cite{Shimizu:2017kzs,
Tachikawa:2015bga}:
\begin{equation}\label{broken-R-sym}
	c_2(R^\text{IR}) = c_2(R^\text{UV}) + c_2(F_X)\,,
\end{equation}
in the obvious notation. The second Chern classes are given in terms of
one-instanton normalized traces, that is $c_2(F) = \frac{1}{4}\text{Tr}F^2$.

While the result for $\delta$, see equation \eqref{deltadelta2}, suggests to
sum over all multiplets in the commutant of $X$, from the discussion above we
must take the new R-symmetry into account. The Chern characters appearing in
the definition of $I_8^\text{vec}(F_{\mathfrak{g}})$ after taking into account
the Jacobson--Morozov decomposition, need to be rewritten in terms of the
one-instanton normalized second Chern class, $c_2(R^\text{IR})$, and only take
into account unbroken modes. We find that the correct combination is given by:
\begin{equation}\label{def:Ivec(rho)}
		\operatorname{ch}_2(R^\text{UV})\text{ch}_{\bm{d}}(F_X)\xrightarrow{\quad \text{IR}\quad}
		\widetilde{\operatorname{ch}}_{\bm{d}}(R^\text{IR}) = 2 - \big(d\,c_2(R^\text{IR})\big) + \frac{1}{12}\big(d\,c_2(R^\text{IR})\big)^2 + \cdots \,.
\end{equation}

Comparing the right-hand side of equation \eqref{def:Ivec(rho)} with the usual
definition of the Chern character, see, e.g., equation
\eqref{def-chern-character}, we see that $\widetilde{\text{ch}}_{\bm{d}_\ell}(R)$
can be understood as a ``twisted'' version that selects only the highest-weight
state of the $\mathfrak{su}_2$ representation with Cartan eigenvalue $d$. This
quantity is enough to completely account for the ``one-loop'' part of the anomaly
polynomial after the nilpotent Higgs branch renormalization flow. Indeed, as
the contribution from the vector multiplet is the only source of the broken
flavor symmetry in second line of equation \eqref{I8-CM}, which we have
interpreted as the ``one-loop'' part of the anomaly polynomial, we must simply
replace
\begin{equation}\label{Irho-def}
		I^\text{vec}_8(F_{\mathfrak{g}},R^\text{UV}) \xrightarrow{\quad \text{IR}\quad}
		I^\text{vec}_8(O) = -\frac{1}{2}\widehat{A}(T)\sum_{\ell}\widetilde{\text{ch}}_{\bm{d}_\ell}(R^\text{IR})\text{ch}_{\bm{R}_\ell}(F_\mathfrak{f})\,,
\end{equation}
where by abuse of language $I^\text{vec}_8(O)$ denotes the contribution in the
infrared after the decoupling of the massive modes, and is written in terms of
the IR R-symmetry. Note that the terms involving the remnant flavor symmetry
must once again be converted to one-instanton normalized traces. Generically,
they will now involve representations beyond the usual singlet, defining, and
adjoint representations. The procedure to find the trace-relation coefficients
for arbitrary representation can be obtained from the associated weight system,
and is reviewed in Appendix \ref{app:representations}.

Having given a prescription on how to find the ``one-loop'' contributions of
the IR theory, let us now turn our attention to the
Green--Schwarz--West--Sagnotti mechanism. There, the flavor only appears via
its second Chern class $c_2(F_\mathfrak{g})$. To rewrite it in terms of the IR
data, we can simply use the branching-rules properties of the Chern character
shown in equation \eqref{ch-decomp-JM}:
\begin{equation}\label{class_to_character}
		c_2(F_{\mathfrak{g}}) = \frac{1}{4}\text{Tr}F_{\mathfrak{g}}^2 = \frac{1}{4h^\vee_{\mathfrak{g}}}\text{tr}_{\textbf{adj}}F_{\mathfrak{g}}^2 =\frac{1}{2h^\vee} \text{ch}_{\textbf{adj}}(F_{\mathfrak{g}})\bigg|_\text{4-form}\,.
\end{equation}
It is then straightforward to see that the decomposition of the second Chern
class depends only on the second-order embedding indices:
\begin{equation}\label{c2F-decomp}
		\begin{aligned}
				c_2(F_\mathfrak{g}) = I_{\mathfrak{su}_2\hookrightarrow\mathfrak{g}}\, c_2(F_X) + \sum_a I_{\mathfrak{f}_a\hookrightarrow\mathfrak{g}}\, c_2(F_{\mathfrak{f}_a})\,,
		\end{aligned}
\end{equation}
where we have allowed for the remnant flavor symmetry to have semi-simple
components: $\mathfrak{f}=\oplus_a\mathfrak{f}_a$. The embedding index of a
subalgebra $\mathfrak{g}'$ into $\mathfrak{g}$ can be computed from the
branching rule of a representation as 
\begin{equation}\label{embedding-index-def}
		I_{\mathfrak{g}'\hookrightarrow\mathfrak{g}} = \frac{1}{A_{\bm{R}}} \sum_\ell m_\ell A_{\bm{R}_\ell}\,,\qquad \bm{R}\to \bigoplus_\ell m_\ell\bm{R}_\ell\,,
\end{equation}
where $m_\ell$ is the multiplicity of $\bm{R}_\ell$ and $A_{\bm{R}_\ell}$ its
Dynkin index, see equation \eqref{trace-relations-ABC} as well as Appendix
\ref{app:representations} for details. It is furthermore possible to show that the
embedding index does not depend on the representation, only the choice of
embedding.\footnote{For a recent review in the physics literature of Dynkin indices, Dynkin embedding indices, and their relevant properties, see \cite{Esole:2020tby}.} Moreover, if $\mathfrak{g}'$ is semi-simple, it is computed
independently for each simple factor, and we sum over each component as in
equation \eqref{c2F-decomp}.

For the particular case of nilpotent orbits, the embedding index can be
computed directly from the orbit data. As reviewed earlier in this section,
every nilpotent orbit can be uniquely labelled by a weighted Dynkin diagram
$w$, seen as a vector in the weight lattice of $\mathfrak{g}$. It satisfies the
property
\begin{equation}\label{su2-embedding-index-wdd}
		I_X = I_{\mathfrak{su}_2\hookrightarrow\mathfrak{g}} = \frac{1}{2}\left<w,w\right>\,,
\end{equation}
where $\left<\cdot,\cdot\right>$ is the pairing on the root lattice. For
simply-laced algebras, it is given via the Cartan matrix; using the conventions
defined in Table \ref{tab:cartan}, we have $\left<\alpha,\beta\right> = \alpha^T\cdot
C^{-1}\cdot\beta$. In practical applications, such as when studying the
central charges of the SCFT, it is often enough to compute terms of the anomaly
polynomial involving only $c_2(R), p_i(T)$, and equation
\eqref{su2-embedding-index-wdd} can therefore give a way of directly finding
them without going through the Jacobson--Morozov decomposition. For the
reader's conveninence, the embedding indices $I_X$ and
$I_{\mathfrak{f}_a\hookrightarrow\mathfrak{g}}$ are tabulated in Appendix
\ref{app:nilps}.

The GS contribution to the anomaly polynomial, namely the first line in
equation $\eqref{I8-CM}$, therefore simply needs to be rewritten in term of the
IR R-symmetry, which corresponds to the replacement:
\begin{equation}\label{c2F-in-IR}
	c_2(F_{\mathfrak{g}}) \xrightarrow{\quad \text{IR}\quad} I_X c_2(R^\text{IR}) + I_{\mathfrak{g}\to\mathfrak{f}} c_2(F) \,.
\end{equation}
As the second Chern class of the flavor symmetry appears through $J(F)$ in the
anomaly polynomial of conformal matter given in equation \eqref{I8-CM}, we can
define an IR version:
\begin{equation}\label{broken-J}
		J(O) = \frac{\chi}{48}(4c_2(R)+p_1(T)) + I_X c_2(R) + \sum_a I_{\mathfrak{f}_a\hookrightarrow\mathfrak{g}}c_2(F_{\mathfrak{f}_a})\,,
\end{equation}
where we have again allowed for the possible presence of semi-simple factors, $\mathfrak{f}=\oplus_a\mathfrak{f}_a$ in the remnant symmetry.

We have now found an IR prescription for every contribution involving the
flavor symmetry after a nilpotent Higgs branch flow. Moreover, since in the GS
term we only need to replace $c_2(R^\text{UV})\to c_2(R^\text{IR})$, see the
discussion around equation \eqref{broken-R-sym}, we conclude that the anomaly
polynomial of a theory obtained by a nilpotent deformation of conformal matter
is given by
\begin{equation}\label{prescription-one-breaking}
		I_8(A_{N-1}^{\mathfrak{g}}(O_L,\varnothing)) = \left.I_8(A_{N-1}^{\mathfrak{g}})\right|_{J(F_L)\to J(O)\,,\, I^\text{vec}_8(F_L)\to I^\text{vec}_8(O_L)}\,,
\end{equation}
where on the right-hand side, everything is understood to be in terms of the IR
R-symmetry.

While we have so far argued for this prescription only for long conformal
matter theories where only one of the flavor is broken, nothing prevents us to
consider a nilpotent breaking on each end. By assumption, the quiver is
long enough so that a deformation on one end cannot affect the other. We can
therefore repeat our argument applied to both sides independently. As can be
seen from equation \eqref{I8-CM}, the anomaly polynomial is symmetric under
exchange of the left and right flavor, and the generalization of equation
\eqref{prescription-one-breaking} is straightforward. We need only consider the
fusion of two different conformal matter theories with only one deformation:
\begin{equation}
		A_{N-1}^{\mathfrak{g}}(O_L,O_R) = A_{N-p-1}^{\mathfrak{g}}(O_L,\varnothing) \oplus A_{p}^{\mathfrak{g}}(\varnothing, O_R)\,.
\end{equation}
We then obtain the same prescription as in equation
\eqref{prescription-one-breaking}, but doing the replacement for nilpotent
orbits on both sides. Moreover, the approach described above is unaffected by
the presence of fractional conformal matter. The anomaly polynomial has only
minimal changes: as long as the fraction does not imply the absence of a flavor
symmetry, there is still a moment map for which a nilpotent vacuum
expectation can be turned on, and our prescription remains valid. The only
difference is the presence of the term $I_\text{fr}$, and the four-form $J(O)$
is changed to:
\begin{equation}\label{broken-Jfr}
		J_\text{fr}(O) = \frac{\chi_\text{fr}}{48}(4c_2(R)+p_1(T)) + \frac{I_X}{d_f} c_2(R) + \frac{1}{d_f}\sum_a I_{\mathfrak{f}_a\hookrightarrow\mathfrak{g}_{f}}c_2(F_{\mathfrak{f}_a})\,,
\end{equation}
where we recall that $d_f$ is the denominator of the associated fraction $f$, and
$O$ is a nilpotent orbit of the flavor symmetry associated with the fraction,
$\mathfrak{g}_f$.

In summary, given a parent theory $\mathcal{A}_{N-1;f_L,f_R}^{\mathfrak{g}}$
for which we turn on (possibly trivial) vacuum expectation values for the
moment maps associated with nilpotent orbits $O_L\,,O_R$, the anomaly polynomial
of the resulting theory $\mathcal{A}_{N-1;f_L,f_R}^{\mathfrak{g}}(O_L,O_R)$ is
given by:
\begin{equation}\label{I8-CM-broken}
	\begin{aligned}
			I_8(\mathcal{A}^{\mathfrak{g}}_{N-1;f_L,f_R}(O_L,O_R)) &=  
		\frac{N_\text{eff}^3}{24}(c_2(R)\Gamma)^2
		- \frac{N_\text{eff}}{2}(c_2(R)\Gamma) \left(J_\text{fr}(O_L) + J_\text{fr}(O_R)\right)\\
		& - \frac{1}{2N_\text{eff}}\left(J(O_L) - J(O_R)\right)^2 + N_\text{eff} I_\text{sing} - I_8^\text{tensor}\\
		&- \frac{1}{2}\left(I_8^\text{vec}(O_L) + I_8^\text{vec}(O_R)\right) - (e_L+e_R)I_\text{fr}\,.
	\end{aligned}
\end{equation}
The R-symmetry is understood to be that of the IR theory, and the decomposition
of the quantities $I_8^\text{vec}(O_L)$ and $J_\text{fr}(O_R)$ are given in equations
\eqref{Irho-def} and \eqref{broken-Jfr}, respectively. The other contributions
have been defined around equation \eqref{I8-fCM}. In all cases, only the
Jacobson--Morozov decomposition of the adjoint representation is required. For
all relevant cases, these branching rules are summarized in Appendix
\ref{app:nilps}.

Note that $I_8^\text{vec}(O_L)$ is associated with the ``one-loop''
contribution of the surviving moment map modes after the nilpotent flow. While
the definition of $\widetilde{\text{ch}}_{\bm{d}_\ell}(R)$ in equation
\eqref{def:Ivec(rho)} taking into account the IR R-symmetry might seem
\emph{ad hoc}, in Section \ref{sec:NG} we will show that it is justified and
equivalent to removing the contributions of the Nambu--Goldstone modes
decoupling from the parent theory. We have indeed checked using the tensor
branch description that equation \eqref{I8-CM-broken} is correct. In the
case of exceptional algebras, the proof is done by exhaustion as the number of
possible cases is finite, and we give a derivation of the validity of this
formula for type-A algebras in the next section. While we have not performed a
similar proof for type-D algebras -- this is left as an exercise for the
diligent reader -- we have checked it exhaustively for $r_{\mathfrak{g}}\leq
20$ and $f_{L,R}= \frac{1}{2}\,,1$, with some additional sporadic crosschecks
at $r_{\mathfrak{g}}=\mathcal{O}(100)$.


We stress once again that while the tensor branch description can be used to
confirm the validity of the formula given in equation \eqref{I8-CM-broken}, the
details of the underlying geometric description does not matter: for long
quivers, the complete data needed is encoded into the notation
$\mathcal{A}_{N-1;f_L,f_R}^{\mathfrak{g}}(O_L,O_R)$, up to the Jacobon--Morozov
decompositions that can be easily computed or read off from tables. The gauge
spectrum and the curve-intersection patterns specific to a given algebra
$\mathfrak{g}$ is completely irrelevant, and the anomaly polynomial is only
given in terms of gauge-invariant data.

\subsection{Other Long Quivers}\label{sec:other-I8}

So far, we have focused solely on (possibly-fractional) conformal matter
theories and their deformations. We have however seen that there are three
families of long quivers:
\begin{equation}
		\mathcal{A}_{N-1;f_L,f_R}^{\mathfrak{g}}(O_L, O_R)\,,\qquad
		\mathcal{D}_{N;f}^{\mathfrak{g}}(O)\,,\qquad
		\mathcal{O}_{N;f}^{\mathfrak{g}}(\rho, O)\,.
\end{equation}
Except for orbi-instanton theories, which also have deformations parameterized
by embedding of ADE discrete groups into $E_8$, all other possible deformations
of the parents theories are nilpotent. As such, the same kind of arguments that
allowed us to find the anomaly polynomial of
$\mathcal{A}_{N-1;f_L,f_R}^{\mathfrak{g}}(O_L, O_R)$ from that of
$\mathcal{A}_{N-1;1,f_R}^{\mathfrak{g}}(\varnothing, O_R)$, can be repeated for
the other types of long quivers. Indeed, as we have described in Section
\ref{sec:bestiary}, we can use the fusion point of view to obtain them from
their undeformed counterparts fused with deformed conformal matter:
\begin{equation}
		\begin{aligned}
				\mathcal{D}_{N,f}^{\mathfrak{g}}(O)\quad &= \quad D_{N-p}^{\mathfrak{g}}(\varnothing)\quad\quad\oplus\quad\mathcal{A}_{p-1;1,f}(\varnothing, O)\,,\\
				\mathcal{O}_{N,f}^{\mathfrak{g}}(\varnothing, O)\quad &= \quad \mathcal{O}_{N-p}^{\mathfrak{g}}(\varnothing, \varnothing)\quad\oplus\quad\mathcal{A}_{p-1;1,f}(\varnothing, O)\,,
		\end{aligned}
\end{equation}
where we used $\oplus$ to indicate fusion of the unbroken common $\mathfrak{g}$
factors. This means that if we can find the anomaly polynomial of
orbi-instantons or type-D SCFTs at low rank, which is achieved easily through
the tensor branch algorithm, that of the broken theory can be found for an
arbitrary quiver length $N$ by recurrence, and the prescription for its
anomaly polynomial is the same sort of replacements showed in equation
\eqref{prescription-one-breaking}.

For type D, the result turns out to be quite simple. We however must
distinguish between $\mathfrak{g}=\mathfrak{su}(2K)$ and the three sporadic
series $\mathfrak{g}=\mathfrak{su}(3)\,,\mathfrak{so}(8)\,,\mathfrak{e}_6$. For
the former, there cannot be fractions, i.e. $f=1$, and we find:
\begin{equation}
    \begin{aligned}
		I_8(D^{\mathfrak{su}_{2K}}_{N}(O)) = &\, \frac{N_\text{eff}^3}{6}(c_2(R)\Gamma)^2 - N_\text{eff}(c_2(R)\Gamma)J(O)\\
		& + N_\text{eff} I_\text{sing} - \frac{1}{2}I_8^\text{vec}(O) + I_8^\text{tensor} + \frac{3}{2}I_8^\text{free}\,,
    \end{aligned}
\end{equation}
with $N_\text{eff}=N-1$, and $I_8^\text{free}$ is the contribution of a free
fermion transforming in the doublet of the R-symmetry, see equation
\eqref{I8-free-hyper}. For the latter three cases, the expression is slightly
different, and of course may depend on the fractions:
\begin{equation}
    \begin{aligned}
		I_8(\mathcal{D}_{\mathfrak{g},N;f}(O)) = &\, \frac{N_\text{eff}^3}{6}(c_2(R)\Gamma)^2 - N_\text{eff}(c_2(R)\Gamma)J_\text{fr}(O)\\
		& + N_\text{eff} I_\text{sing} - \frac{1}{2}I_8^\text{vec}(O) - I_8^\text{tensor} - (e+15) I_\text{fr} + 
		I_{8}^{\mathcal{D},\mathfrak{g}}\,.
    \end{aligned}
\end{equation}
In this case, the parameter $N_\text{eff}$ must be slightly modified:
\begin{equation}
		N_\text{eff} = (N-3) + f + \frac{1}{2}
\end{equation}
This trivalent pattern also gives rise to an extra term depending only on the
$c_2(R)$ and $p_1(T)$ but whose coefficients depend on the choice of algebra
$\mathfrak{g}$; they are given in Table \ref{tab:I8-extra-D}. The quantities
depending explicitly on the nilpotent orbit $O$ are the ones appearing in the
anomaly polynomial of conformal matter, see equations \eqref{def:Ivec(rho)} and
\eqref{broken-J}.

\begin{table}
		\centering
		\begin{tabular}{cc}
				\toprule
				$\mathfrak{g}$    & $I_{8}^{\mathcal{D},\mathfrak{g}}$\\\midrule
				$\mathfrak{su}_3$ & $c_2(R)(\frac{23}{16} c_2(R) + \frac{25}{96}p_1(T))$\\
				$\mathfrak{so}_8$ & $c_2(R)(\frac{5}{2}c_2(R) +\frac{7}{24} p_1(T))$\\
				$\mathfrak{e}_6$  & $c_2(R)(\frac{5}{2}c_2(R) +\frac{7}{24} p_1(T))$\\
				\bottomrule
		\end{tabular}
		\caption{Term appearing in the anomaly polynomial of type-D quiver, originating from the trivalent intersection.}
		\label{tab:I8-extra-D}
\end{table}

Orbi-instantons theories are dealt with in a similar fashion. We note however
that turning on a deformation for the $\mathfrak{e}_8$ flavor associated with
an end-of-the-world brane in the M-theory description substantially complicates
the analysis. 
When $\mathfrak{g}=\mathfrak{su}(K)$,
corresponding to $\Gamma=\mathbb{Z}_K$, the anomaly polynomial of theories with
non-trivial $\rho$ -- but trivial nilpotent orbits of $\mathfrak{su}_K$ -- were
studied in \cite{Mekareeya:2017jgc}. Here, we will conversely allow for any
choice of $\mathfrak{g}$ and its nilpotent orbits, but keep $\rho$ trivial.

While slightly more involved than previous cases, we find a closed-form
expression for the anomaly polynomial that depends once again only
group-theoretical quantities related to $\mathfrak{g}$: 
\begin{equation}
		\begin{aligned}
		I_8(\mathcal{O}^{\mathfrak{g}}_{N;f}(\varnothing,O)) = & \frac{N_\text{eff}^3}{6}(c_2(R)\Gamma)^2 - \frac{N_\text{eff}^2}{2} (c_2(R)\Gamma)\widetilde{J}(F_{\mathfrak{e}_8}) + N_\text{eff} I_\text{sing}\\
		& + N_\text{eff} \left(\frac{1}{2}\widetilde{J}(F_{\mathfrak{e}_8})^2 - (\Gamma c_2(R)) J_\text{fr}(O)\right)\\
		& + \widetilde{J}(F_{\mathfrak{e}_8})\left(J_\text{fr}(O)-\frac{\chi_\text{fr}}{48}(4c_2(R)+p_1(T))\right)- \frac{\alpha_0}{24}c_2(R)^2 + \frac{r_{\mathfrak{g}}}{2} I_\text{free}\\
	& 
			 - \frac{\text{dim}(\mathfrak{g})h^\vee_{\mathfrak{g}}}{6}c_2(R)\big(\frac{1}{6}p_1(T) + c_2(F_{\mathfrak{e}_8})\big)-\frac{1}{2}I_8^\text{vec}(O) - e I_\text{fr}\,.
		\end{aligned}
\end{equation}
The possible values of $e$ given a fraction $f$ are shown in Table
\ref{tbl:eattachments}, and quantities depending on the nilpotent orbit are
given in equations \eqref{Irho-def} and \eqref{broken-Jfr}.  The contribution
of a free fermion in a doublet of the R-symmetry $I_\text{free}$ is given in
equation \eqref{I8-free-hyper}, and we have defined:
\begin{align}
		N_\text{eff} &= (N-2) + f_\text{OI} + f\,,\\
		\widetilde{J}(F_{\mathfrak{e}_8}) &= \frac{1}{2}(\Gamma \chi-1)c_2(R) + \frac{1}{4}p_1(T) + c_2(F_{\mathfrak{e}_8})\,.
\end{align}
The quantity $\widetilde{J}(F_{\mathfrak{e}_8})$ can be understood in the same
way as $J(F_{\mathfrak{g}})$, namely as a contribution to the Green--Schwarz term
appearing in the Bianchi identity for the tensor multiplet dual to the
tensionless strings at the fixed point, see the discussion around equation
\eqref{GS-CM}. Contrary to other long quivers, the ``effective'' number of
conformal matter theories $N_\text{eff}$ now depends on the choice of algebra
$\mathfrak{g}$, and in the anomaly polynomial is encoded in the parameter
$f_\text{OI}$. Additionally, there is a contribution to the $c_2(R)^2$ term,
$\alpha_0$, for which we did not find a simple expression in terms of the data
of $\mathfrak{g}$. We show the values for both these quantities in Table
\ref{tab:attachment-orbi-instanton}.

\subsection{Examples}\label{sec:sec3examples}

To illustrate the simplicity of the formula given in equation
\eqref{I8-CM-broken}, we now turn to two examples. For ease of exposition we
will only consider conformal matter with $f_{L,R}=1$, and a single deformation. 

\paragraph{The minimal orbit of $\mathfrak{e}_6$:} we first consider the theory
obtained by breaking the left flavor of higher-rank $(\mathfrak{e}_6,
\mathfrak{e}_6)$ conformal with the $A_1$ nilpotent orbit of $\mathfrak{e}_6$
in the Bala--Carter notation. It is therefore the
$A_{N-1}^{\mathfrak{e}_6}(A_1, 0)$ theory, described at a generic
point of the tensor branch by the following quiver:
\begin{equation}
		A_{N-1}^{\mathfrak{e}_6}(A_1, 0):\qquad\qquad
		[\mathfrak{su}_6]\overset{\mathfrak{su}_3}{2}1\overset{\mathfrak{e}_6}{6}1\overset{\mathfrak{su}_3}{3}1\,\cdots\,\overset{\mathfrak{e}_6}{6}1\overset{\mathfrak{su}_3}{3}1\overset{\mathfrak{e}_6}{6}1\overset{\mathfrak{su}_3}{3}1[\mathfrak{e}_6]\,.
\end{equation}

This nilpotent orbit is minimal, in the sense that it is smallest with respect
to the partial ordering of nilpotent orbits, i.e., the first non-trivial level
in the corresponding Hasse diagram. In the geometric description, this is
reflected by the fact that the simplest deformation is obtained by blowing down
the left-most $(-1)$-curve of the conformal matter quiver. Furthermore, its
weighted Dynkin diagram is given by $w=[0,0,0,0,0,1]$, and Jacobson--Morozov
decomposition for the adjoint representation yields:
\begin{equation}
	\begin{aligned}
			\rho_{A_1}:\qquad\mathfrak{e}_6~\, &\longrightarrow~ \mathfrak{su}(2)_X \oplus \mathfrak{su}(6)\\
			\bm{78} &\longrightarrow~ (\bm{1},\bm{35})\oplus(\bm{2},\bm{20})\oplus(\bm{3},\bm{1})\,.
	\end{aligned}
\end{equation}
This branching rule is simple, and as can be checked from the tables collated
in Appendix \ref{app:nilps}, or using equation \eqref{embedding-index-def}, the
embedding indices are both equal to one. The second Chern class decomposes as
\begin{equation}\label{c2(A1)}
		c_2(F_{\mathfrak{e}_6})= c_2(F_X) + c_2(F_{\mathfrak{su}_6}) \,.
\end{equation}
Using that after the nilpotent breaking we must replace
$c_2(R^\text{UV})=c_2(R)=c_2(F_X)$, we have
\begin{equation}
		J(A_1) = \frac{455}{288}c_2(R) + \frac{167}{1152}p_1(T) + c_2(F_{\mathfrak{su}_6})\,.
\end{equation}

To use the decomposition of $I_8^{\text{vec}}(A_1)$ defined in equation
\eqref{Irho-def}, we need to convert the traces over the representations
$\bm{20}$ and $\bm{35}$ of $\mathfrak{su}_6$ into one-instanton normalized
traces. Using the techniques explained above and in appendix
\ref{app:representations}, one finds:
\begin{align}
		\text{ch}_{\bm{20}}(F_{\mathfrak{su}_6}) &= 20 -\frac{1}{2} (3\,\text{Tr}F^2_{\mathfrak{su}_6}) + \frac{1}{4!}(-6\,\text{Tr}F^4_{\mathfrak{su}_6} + \frac{3}{2}(\text{Tr}F^2_{su6})^2) + \cdots \,, \\
		\text{ch}_{\bm{35}}(F_{\mathfrak{su}_6}) &= 35 -\frac{1}{2} (6\,\text{Tr}F^2_{\mathfrak{su}_6}) + \frac{1}{4!}(12\,\text{Tr}F^4_{\mathfrak{su}_6} + \frac{3}{2}(\text{Tr}F^2_{\mathfrak{su}_6})^2)+\cdots \,.
\end{align}
Taking into account the ``twisted'' character for the R-symmetry in the
deformed theory, we have:
\begin{align}
		I_8^{\text{vec}}(A_1) =& -\frac{1}{2}\widehat{A}(T)\bigg(
			\widetilde{\text{ch}}_{\bm{1}}(R)\text{ch}_{\bm{35}}(F_{\mathfrak{su}_6}) + 
			\widetilde{\text{ch}}_{\bm{2}}(R)\text{ch}_{\bm{20}}(F_{\mathfrak{su}_6}) + 
			\widetilde{\text{ch}}_{\bm{3}}(R)\bigg)\\
			=& -\frac{109}{6}c_2(R)^2 - \frac{31}{12}c_2(R)p_1(T)  - \frac{56}{5760}(7p_1(T)^2 -4p_2(T)) \\
			&- \frac{9}{2}c_2(R)\text{Tr}(F_{\mathfrak{su}_6}^2)- \frac{3}{16}p_1(T)\text{Tr}F_{\mathfrak{su}_6}^2  - \frac{1}{8}(\text{Tr}F_{\mathfrak{su}_6})^2 - \frac{1}{4}\text{Tr}F_{\mathfrak{su}_6}^4\,.
\end{align}
Note that the coefficient of $p_2(T)$ is proportional to the dimension of the
commutant of the nilpotent orbit $\text{dim}(\mathfrak{e}_6) - \text{dim}(A_1)
= 56$, as expected. Putting everything together as in equation
\eqref{I8-CM-broken}, the complete anomaly polynomial of the theory after
breaking is given by
\begin{align}
		I_8(A_{N-1}^{\mathfrak{e}_6}(A_1, 0))  = &
				\frac{288N^4 - 311 N^2 + 128 N -6}{12N} c_2(R)^2 
				- \frac{83 N- 50}{24}p_1(T)c_2(R)\nonumber\\
				& + \frac{15N + 223}{2880} p_1(T)^2 - \frac{15N+19}{720}  p_2(T)
				+  p_1(T) \bigg(\frac{1}{8}\text{Tr}F^2_{\mathfrak{e}_6} + \frac{3}{32} \text{Tr}F^2_{\mathfrak{su}_6}\bigg)\nonumber\\
				& -c_2(R) \bigg( \frac{12 N^2 - 6N -1}{4N} \text{Tr}F^2_{\mathfrak{e}_6} + \frac{12 N^2 - 9N + 1}{4N} \text{Tr}F^2_{\mathfrak{su}_6}\bigg) \nonumber\\
				& + \frac{3 N-1}{32N} (\text{Tr}F^2_{\mathfrak{e}_6})^2 + \frac{1}{16N} \text{Tr}F^2_{\mathfrak{e}_6} \text{Tr}F^2_{\mathfrak{su}_6} + \frac{2 N-1}{32N} (\text{Tr}F^2_{\mathfrak{su}_6})^2 
				+ \frac{1}{8}\text{Tr}F^4_{\mathfrak{su}_6}\,.
\end{align}

This result can then be compared against the computation on the tensor branch
description, and one can check that the two results agree.

\paragraph{Partitions of $\mathfrak{su}_8$:} when the flavor algebra is of type
$\mathfrak{su}_K$, conformal matter takes a particularly simple form, and the
nilpotent deformations are described by partitions of $K$.  We now specialize
to $\mathfrak{g}=\mathfrak{su}(8)$, where conformal matter has been deformed by
the nilpotent orbit associated with the partition $[2^4]$ on the left, while
the right flavor is left untouched -- or equivalently described by the trivial
partition $[1^8]$:
\begin{equation}
	A_{N-1}^{\mathfrak{su}_8}([2^4], [1^8]):\qquad\qquad
    [\mathfrak{su}_4]\underbrace{\,\overset{\mathfrak{su}_4}{2} \overset{\mathfrak{su}_8}{2} \overset{\mathfrak{su}_8}{2} \cdots \overset{\mathfrak{su}_8}{2} \overset{\mathfrak{su}_8}{2}\,}_{N-1}[\mathfrak{su}_8] \,.
\end{equation}
The relations between the partitions and the tensor branch quivers are discussed
in Appendix \ref{app:nilp2TB}. As discussed there, the remnant flavor is found
from the multiplicities of each partition, and we therefore have total non-Abelian flavor
$\mathfrak{f} = \mathfrak{su}_4\oplus\mathfrak{su}_8$ in the IR theory.

We once again need to use the Jacobson--Morozov decomposition to find the
anomaly polynomial. For the fundamental representation, it can be read of
directly from the partition, from which that of the adjoint is easily computed,
as $\bm{K}\otimes\bar{\bm{\bm{K}}}=\textbf{adj}\oplus\bm{1}$ for type-A algebras:
\begin{equation}\label{[2^4]-JM-decomposition}
	\begin{aligned}
		\rho_{[2^4]}:\qquad\mathfrak{su}(8) &\rightarrow \mathfrak{su}(2)_X \oplus \mathfrak{su}(4) \,,\\
		\bm{8}  &\rightarrow (\bm{2, 4})\,,\\
		\bm{63} &\rightarrow (\bm{3, 1}) \oplus (\bm{1, 15}) \oplus (\bm{3, 15}) \,.		
	\end{aligned}
\end{equation}
In the IR theory, using equation \eqref{c2F-in-IR}, we therefore have:
\begin{equation}\label{decomp[2^4]}
	\begin{aligned}
			c_2(F_L) &= 4\,c_2(R) + 2\,c_2(F_{\mathfrak{su}_4})\,,\\
			I_8^\text{vec}([2^4]) &= -\frac{1}{2}\widehat{A}(T)\bigg( \widetilde{ch}_{\bm{3}}(R) + \big(\widetilde{ch}_{\bm{1}}(R) + \widetilde{ch}_{\bm{3}}(R)\big)\text{ch}_{\bm{15}}(F_{\mathfrak{su}_4})\bigg)\,,
	\end{aligned}
\end{equation}
where we have written everything in terms of the IR R-symmetry. For
completeness, we recall that the weighted Dynkin diagram of this nilpotent
orbit is given by $w=[0,0,0,2,0,0,0]$ \cite{collingwood2017nilpotent}, and one
can check that the index of the $\mathfrak{su}_2$ embedding is indeed given by
$I_X=4$ via equation \eqref{su2-embedding-index-wdd}.

Note that this case is simpler than the minimal orbit of $\mathfrak{e}_6$, as
the only non-trivial representation is the adjoint, for which the trace
relations are given by:
\begin{equation}
		\text{tr}_{\textbf{adj}}F_{\mathfrak{su}_K}^2 = K\, \text{Tr}F_{\mathfrak{su}_K}\,,\qquad
		\text{tr}_{\textbf{adj}}F_{\mathfrak{su}_K}^4 = 2K\, \text{Tr}F_{\mathfrak{su}_K}^4 + \frac{3}{2}(\text{Tr}F_{\mathfrak{su}_K}^2)^2\,.
\end{equation}
The anomaly polynomial of the theory
$A_{N-1}^{\mathfrak{su}_8}([2^4], [1^8])$ is finally obtained by
plugging back equation \eqref{decomp[2^4]} into the formula given in equation
\eqref{I8-CM-broken}. When the dust settles, we obtain:
\begin{align}
		I_8(A_{N-1}^{\mathfrak{su}_8}([2^4], [1^8]))  = &~
				\frac{32N^4 - 255 N^2 + 343 N -96}{12N} c_2(R)^2 
				- \frac{31 N- 55}{24}p_1(T)c_2(R)\nonumber\\
				& + \frac{5N + 51}{960} p_1(T)^2 - \frac{5N+3}{240}  p_2(T)
				+  p_1(T) \bigg(\frac{1}{12}\text{Tr}F^2_{\mathfrak{su}_8} + \frac{1}{12} \text{Tr}F^2_{\mathfrak{su}_4}\bigg)\nonumber\\
				& -c_2(R) \bigg( \frac{N^2 - N -1}{N} \text{Tr}F^2_{\mathfrak{su}_8} + \frac{2 N^2 - 5N + 2}{N} \text{Tr}F^2_{\mathfrak{su}_4}\bigg) \nonumber\\
				& + \frac{N-1}{32N} (\text{Tr}F^2_{\mathfrak{su}_8})^2 + \frac{1}{8N} \text{Tr}F^2_{\mathfrak{su}_8} \text{Tr}F^2_{\mathfrak{su}_4} + \frac{N-2}{16N} (\text{Tr}F^2_{\mathfrak{su}_4})^2 \nonumber\\
				& + \frac{1}{3}\text{Tr}F^4_{\mathfrak{su}_8}+ \frac{1}{3}\text{Tr}F^4_{\mathfrak{su}_4}\,.
\end{align}
In Section \ref{sec:NG}, we will give the form of the anomaly polynomial for
any theory of type $A_{N-1}^{\mathfrak{su}_K}(O_L,O_R)$, and one can
check that the above result is correct.

\section{Anomalies from Nambu--Goldstone Modes}\label{sec:NG}

In the previous section, we have given closed-form expressions for
the anomaly polynomials of every long 6d $(1,0)$ SCFT. We found them using the
anomaly polynomial of conformal matter and obtained that of its nilpotent
deformations by studying which moment map modes were surviving the flow to the infrared theory. The other cases
were then reached through fusion, and our results confirmed by using the
generic tensor branch geometry. In this section, we will determine the putative
anomaly polynomials from a bottom-up perspective without invoking the tensor
branch geometry and show that they match with the closed-form
expressions.


We consider here the parent theory of a long quiver, which we will refer to as the UV theory, $\mathcal{T}^\text{UV}$. To wit, these are the three classes of infinite series:
\begin{equation}\label{parent-theory-list}
		\mathcal{A}_{N-1;f_L,f_R}^{\mathfrak{g}}\,,\qquad
		\mathcal{D}_{N;f}^{\mathfrak{g}}\,,\qquad
		\mathcal{O}_{N;f}^{\mathfrak{g}}\,.
\end{equation}
We assume that $N$ is large enough such that giving nilpotent vacuum expectation values to the moment maps lead to an interacting SCFT, and, when the parent theory is of conformal matter type, that the Higgsing of the left and right moment maps do not influence each other.\footnote{Recall that we are not considering Higgsing by giving a nilpotent vacuum expectation value to the moment map of the $\mathfrak{e}_8$ flavor symmetry on the left in the parent theory $\mathcal{O}_{N;f}^{\mathfrak{g}}$.} Giving a nilpotent vacuum expectation to these operators, $\left<\phi\right> = X\in O$, where $O$ is a nilpotent orbit, we trigger a Higgs branch renormalization group flow leading to an infrared fixed point corresponding to an interacting SCFT which we label as $\mathcal{T}^\text{IR}$. 

It is natural to ask: can we determine the anomaly polynomial of the SCFT $\mathcal{T}^\text{IR}$ using only the information of the anomaly polynomial of $\mathcal{T}^\text{UV}$ and the information contained in the nilpotent orbit $O$ by which we Higgs. Of course, if we allow ourselves to use the tensor branch description of $\mathcal{T}^\text{UV}$ the path to determining the anomaly polynomial of $\mathcal{T}^\text{IR}$ is straightforward, if circuitous: we know how the tensor branch geometry is modified by the nilpotent orbit $O$, and then it is direct to determine the infrared anomaly polynomial from the known tensor branch description of $\mathcal{T}^\text{IR}$. Instead, we only assume knowledge of the anomaly polynomial of $\mathcal{T}^\text{UV}$, and no additional, microscopic details of the theory. 

Under a nilpotent Higgsing, the superconformal symmetry, and thus the $\mathfrak{su}(2)_R$ R-symmetry, is broken along the flow. At the interacting fixed point a new R-symmetry emerges: $\mathfrak{su}(2)_{R^\text{IR}}$. A nilpotent orbit of a simple flavor symmetry factor $\mathfrak{f}$ corresponds to a homomorphism $\mathfrak{su}(2)_X \rightarrow \mathfrak{g}$, and when a nilpotent Higgsing is performed, the new infrared R-symmetry is simply the diagonal of the original $\mathfrak{su}(2)_R$ and the $\mathfrak{su}(2)_X$. See \cite{Tachikawa:2015bga} for a review of nilpotent Higgsing, particularly in the context of 4d $\mathcal{N}=2$ SCFTs. In addition to the breaking and emergence of the R-symmetry, we must take care of modes which decouple along the flow into the infrared. One class of modes which decouple belong to the moment map (and its superpartners) to which we give the VEV; we refer to these as the Nambu--Goldstone modes inside of the moment map. In particular, the moment map supermultiplet contains chiral fermions, and the decoupling of these fermions in the infrared affects the anomalies of $\mathcal{T}^\text{IR}$.

In six dimensions, a positive-chirality fermion transforming in the representation $\bm{\ell}$ of $\mathfrak{su}(2)_{R^\text{IR}}$ and some representation $\bm{R}$ of a flavor symmetry $\mathfrak{f}$ contributes to the anomaly polynomial as\footnote{The $1/2$-prefactor appears in equation \eqref{eqn:I8fermion} as $\bm{R}$ may be a real irreducible representation of $\mathfrak{f}$.}
\begin{equation}\label{eqn:I8fermion}
    I_8^{\text{fermion}}(\bm{\ell}, \bm{R}) = \frac{1}{2}\widehat{A}(T)\operatorname{ch}_{\bm{\ell}}(R^\text{IR})\operatorname{ch}_{\bm{R}}(F) \bigg|_{\text{8-form}} \,.
\end{equation}
Thus, once we know the representations of $\mathfrak{su}(2)_{R^\text{IR}}$ and $\mathfrak{f}$ under which the Nambu--Goldstone fermions inside of the moment map transform, we can determine the contribution to the anomaly polynomial from these modes, which must be removed in the IR. Conveniently, the representations under the global symmetries of the Nambu--Goldstone modes belonging to the moment map are known \cite{Tachikawa:2015bga}.\footnote{The analysis in \cite{Tachikawa:2015bga} focused on 4d $\mathcal{N}=2$ SCFTs, however the generalization to the structure of the Nambu--Goldstone modes inside of the moment map in 6d $(1,0)$ is clear.} Given the homomorphism $\mathfrak{su}(2)_X \rightarrow \mathfrak{g}$ associated to the nilpotent orbit, there is an induced branching rule
\begin{equation}\label{eqn:inducedBR}
    \begin{aligned}
        \mathfrak{g} \,\, &\longrightarrow \,\, \mathfrak{su}(2)_X \oplus \mathfrak{f} \,,\\
        \textbf{adj} \,\, &\longrightarrow \,\, \bigoplus_\ell (\bm{d}_\ell, \bm{R}_\ell) \,,
    \end{aligned}
\end{equation}
where $\mathfrak{f}$ is the centralizer of the image of the homomorphism. The Nambu--Goldstone fermions transform in the $\mathfrak{su}(2)_{R^\text{IR}} \oplus \mathfrak{f}$ representations:
\begin{equation}
    \bigoplus_\ell \, (\bm{d_\ell-1}, \bm{R}_\ell) \,.
\end{equation}
Thus, we can determine the contribution to the infrared anomaly polynomial that comes from the decoupled Nambu--Goldstone modes inside of the moment map.

In principle, there may be anomaly-contributing modes that decouple along the flow into the infrared that do not belong to the Nambu--Goldstone fermions inside of the moment map supermultiplet. In fact, such modes often exist when considering the nilpotent Higgsing of the flavor symmetry factor $\mathfrak{g}$ in an arbitrary SCFT, as has been noted in both 4d $\mathcal{N}=2$ and 6d $(1,0)$ contexts in \cite{Distler:2022nsn,Distler:2022kjb}. We propose that when the parent theory is one of those in equation \eqref{parent-theory-list}, then the only modes that decouple in the infrared and contribute non-trivially to the anomaly are those that belong to the moment map; we verify this by demonstrating that the anomaly polynomial worked out under such an assumption is identical to the anomaly polynomial worked out using the tensor branch geometry. 

Putting everything together concisely, the anomaly polynomial of the interacting infrared SCFT can be determined via the following algorithm.
\begin{alg}\label{alg:I8IR} 
	Given the anomaly polynomial, $I_8(\mathcal{T}^\text{UV})$, for a long
	quiver, $\mathcal{T}^\text{UV}$, chosen among the theories given in
	equation \eqref{parent-theory-list}, and a nilpotent orbit, $O$, of one of
	the simple flavor factors, $\mathfrak{g}$,\footnote{Again, not including
	the $\mathfrak{e}_8$ flavor symmetry of $\mathcal{O}_{N;f}^\mathfrak{g}$.}
	of $\mathcal{T}^\text{UV}$,  the anomaly polynomial,
	$I_8(\mathcal{T}^\text{IR})$, of the SCFT $\mathcal{T}^\text{IR}$ obtained
	via giving a VEV valued in $O$ to the moment map of $\mathfrak{g}$ is found
	as follows.
    \begin{enumerate}
		\item In $I_8(\mathcal{T}^\text{UV})$, rewrite the
			one-instanton normalized traces of the flavor symmetries in
			terms of the curvatures of the algebras $\mathfrak{su}(2)_X$
			and $\mathfrak{f}$ appearing in equation \eqref{eqn:inducedBR}:
			$\operatorname{Tr}F^n_X$ and $\operatorname{Tr}F^n_{\mathfrak{f}}$.
			\item Replace the second Chern class of the UV R-symmetry
					$c_2(R^\text{UV})$ and $c_2(F_X)$ by that of the IR R-symmetry:
					\begin{equation}
						c_2(R^\text{UV})\,, c_2(F_X)\rightarrow c_2(R^\text{IR})\,.
					\end{equation}
			Steps 1 and 2 define the quantity
			$I_8(\mathcal{T^\text{UV}};R^\text{IR})$.
			\item The IR anomaly polynomial is then obtained by removing the Nambu--Goldstone modes:
					\begin{equation}
							I_8(\mathcal{T}^\text{IR}) = I_8(\mathcal{T^\text{UV}};R^\text{IR}) - \sum_{\ell}I_8^\text{\emph{fermion}}(\bm{d}_\ell\bm{-1},\bm{R}_\ell)\,,
					\end{equation}
                    where the representations of fermions are given from the decomposition in equation \eqref{eqn:inducedBR}.
					The contribution of each fermionic Nambu--Goldstone mode is as defined in equation
					\eqref{eqn:I8fermion}.
\end{enumerate}
	If $\mathcal{T^\text{UV}}=\mathcal{A}^{\mathfrak{g}}_{N-1;f_L,f_R}$, there
	can be a nilpotent vacuum expectation value for both moment maps. Then we
	perform the three steps sequentially for each nilpotent deformations.
\end{alg}

Having given the algorithm, one can show that its result is equivalent to the
closed-form expression presented in Section \ref{sec:I8}. To see this, recall
that we have interpreted these anomaly polynomials as coming from both
``one-loop'' and GS contributions at the fixed point. The quantity
$I_8^\text{vec}(O)$, defined in equation \eqref{def:Ivec(rho)}, was introduced
as a way to sum only over the contribution of massless modes of the adjoint
representation after the nilpotent breaking. It is straightforward to show that
this contribution and that of the Nambu--Goldstone modes as defined above
reorganize into the anomaly polynomial of a full vector multiplet written in
terms of the IR data. This is expected, as $I_8^\text{vec}(O)$ counts
$\mathfrak{su}(2)_X$ highest-weight states appearing in the Jacobson--Morozov,
weighted by the IR R-symmetry through the twisted character
$\widetilde{\operatorname{ch}}_{\bm{d}}$, while the Nambu--Goldstone modes are
by definition those that do not commute with the representative $X$ of the
nilpotent orbit. This takes care of the ``one-loop'' part of the closed-form
expression, while the replacement $J_\text{fr}(F)\to J_\text{fr}(O)$ is
equivalent to steps 1 and 2 of the Algorithm \ref{alg:I8IR} applied to the
Green--Schwarz contribution, when the R-symmetry is understood to be that of
the infrared theory. This therefore justifies \emph{a posteriori} the
introduction of $\widetilde{\operatorname{ch}}_{\bm{d}}$ as a quantity encoding
the R-symmetry of the modes remaining massless in the IR theory.

Note that while equivalent, the results of previous section and Algorithm
\ref{alg:I8IR} have different advantages. Starting from a nilpotent deformation
of conformal matter, we can reach a long quiver of e.g. an orbi-instanton
through fusion, and both the ``one-loop'' and Green--Schwarz terms can be
easily identified with the closed-form expressions. One the other hand, using
Algorithm \ref{alg:I8IR} we can find the anomaly polynomial of any 6d $(1,0)$
SCFT reached by nilpotent RG flows without knowing the precise details of its
quiver. 

For completeness, and as they are usually the most relevant quantities in
practical computations, we give here the shifts that the gravitational and
R-symmetry anomaly coefficients undergo when (fractional) conformal matter
$\mathcal{A}_{N-1;f_L,f_R}^{\mathfrak{g}}$ is Higgsed via nilpotent orbits
$O_L$ and $O_R$:
\begin{equation}\label{shift-grav-coeffs}
		\begin{aligned}
		\alpha^\text{UV} - \alpha^\text{IR}  =&\, 12\Gamma N_\text{eff}\left(\frac{I_{X_L}}{d_{f_L}}+\frac{I_{X_R}}{d_{f_R}}\right) + \frac{12}{N_\text{eff}}\left(\frac{I_{X_L}}{d_{f_L}}-\frac{I_{X_R}}{d_{f_R}}\right)^2 + 4(\beta^\text{UV}-\beta^\text{IR})\\
		& - \big(4\varphi_3(w_L) + \varphi_0(w_L) + 4\varphi_3(w_R) + \varphi_0(w_R)\big)\,,\\
		\beta^\text{UV} - \beta^\text{IR}  = &\, +\frac{6}{N_\text{eff}}\left(\frac{1}{d_{f_L}}-\frac{1}{d_{f_R}}\right)\left(\frac{I_{X_L}}{d_{f_L}}-\frac{I_{X_R}}{d_{f_R}}\right)\\
		& - \big(\varphi_1(w_L) - \frac{1}{2}\varphi_0(w_L) + \varphi_1(w_R) - \frac{1}{2}\varphi_0(w_R)\big)\,,\\
		\gamma^\text{UV} - \gamma^\text{IR} =&\, +\frac{7}{240}\left(\text{dim}(O_L) + \text{dim}(O_R)\right)\,,\\
		\delta^\text{UV} - \delta^\text{IR} =&\, -\frac{1}{120}\left(\text{dim}(O_L) + \text{dim}(O_R)\right)\,.
		\end{aligned}
\end{equation}
Here, $N_\text{eff}$ is as given in equation \eqref{eqn:Neff}, $d_{f_L}$ and
$d_{f_R}$ are the denominators of the fractions $f_L$ and $f_R$, respectively,
and $I_{X_L}$, $I_{X_R}$ are the embedding indices for the $\mathfrak{su}(2)_X$
under the branching of the respective nilpotent orbits. The quantity
$\varphi_n(w)$ can be obtained directly from the weighted Dynkin diagram
associated with $O$, and the set of positive roots $\Lambda^+$ of $\mathfrak{g}_f$
\begin{equation}\label{def-varphi}
		\varphi_n(w) = \sum_{\alpha\in\Lambda^+}\left<\alpha,w\right>^n\,,
\end{equation}
with the weighted Dynkin diagram understood as an element of the weight
lattice, and the scalar product is the pairing on that lattice. We remind the reader than in our convention, for simply-laced algebras, it is given by $\left<\alpha,\beta\right>=\alpha^T\cdot C^{-1}\cdot\beta$, where $C$ is the associated Cartan matrix, see Table \ref{tab:cartan}. Note that $\varphi_0(w)$ is understood as counting the number of positive roots $\alpha$ for which $\langle \alpha, w \rangle \neq 0$. The quantities defined in equation \eqref{shift-grav-coeffs} are
therefore easy to compute from the tables in Appendix \ref{app:nilps} for any
theory without needing to know the complete Jacobson--Morozov decomposition, but only the weighted Dynkin diagrams of the nilpotent orbits.

In Section \ref{sec:generalproof}, we provide a proof that Algorithm
\ref{alg:I8IR} as applied to nilpotent RG flows of rank $N$ $(\mathfrak{su}_K,
\mathfrak{su}_K)$ conformal matter leads to the same anomaly polynomial that is
obtained from an analysis of the tensor branch effective field theory,
following \cite{Ohmori:2014kda,Intriligator:2014eaa,Baume:2021qho}. In essence,
we can consider this as a proof that when a nilpotent vacuum expectation value
is given to the moment map of the flavor symmetry of these theories, the only
modes to decouple along the flow are the Nambu--Goldstone modes arising from
the moment map itself. General proofs that Algorithm \ref{alg:I8IR} reproduces
the known anomaly polynomials (as determined from the tensor branch geometry)
for the theories 
\begin{equation}
    \mathcal{A}^{\mathfrak{g}}_{N-1;f_L,f_R}(O_L, O_R) \,, \qquad \mathcal{D}^\mathfrak{g}_{N;f}(O) \,, \qquad \mathcal{O}^\mathfrak{g}_{N;f}(\varnothing, O) \,,
\end{equation}
where $\mathfrak{g}$ is any allowed simply-laced classical Lie algebra can be
shown in a similar manner; it is tedious to be explicit. When $\mathfrak{g}$ is
an allowed exceptional Lie algebra, then the number of nilpotent orbits is
finite, and the proof of the matching between Algorithm \ref{alg:I8IR} and the
tensor branch analysis can be shown by exhaustion; in Section
\ref{sec:exceptNGeg}, we provide an explicit example:
$A^{\mathfrak{e}_6}_{N-1}(0, A_1)$. In particular, this requires the knowledge
of the branching rules of the adjoint representation for the nilpotent orbits
of the exceptional simple Lie algebras, which we collate for convenience in
Appendix \ref{app:nilps}.

\subsection{General Proof for \texorpdfstring{$A^{\mathfrak{su}_K}_{ N-1}(O_L, O_R)$}{A\^suK\_N-1(OL, OR)}}\label{sec:generalproof}

In this section, we provide a general proof that Algorithm \ref{alg:I8IR} determines the correct anomaly polynomial for arbitrary nilpotent deformations of rank $N$ $(\mathfrak{su}(K), \mathfrak{su}(K))$ conformal matter. We apply the algorithm to $A_{N-1}^{\mathfrak{su}_K}(O_L, O_R)$ generically, and we show that this produces the same anomaly polynomial as can be read off from the known tensor branch geometry. For ease of notation, we assume that $K \geq 4$.

\subsubsection{From the Algorithm}

The first step in Algorithm \ref{alg:I8IR} is to determine the decompositions of the traces under the decomposition of the $\mathfrak{su}(K)$ algebra induced by the choice of nilpotent orbits. We begin by writing the one-instanton normalized traces in terms of traces over the fundamental representation: 
\begin{equation}
    \operatorname{Tr}F^2 = 2 \operatorname{tr}_\textbf{fund} F^2 \,, \qquad \operatorname{Tr}F^4 = \operatorname{tr}_\textbf{fund} F^4 \,.
\end{equation}
Let $O$ be a nilpotent orbit of $\mathfrak{su}(K)$ represented by the integer partition 
\begin{equation}
    P_O = [1^{m_1}, 2^{m_2}, \cdots, K^{m_K}] \qquad \text{ such that } \qquad \sum_{i=1}^K i m_i = K \,.
\end{equation}
The decomposition of Lie algebras associated to this nilpotent orbit is
\begin{equation}\label{eqn:ZZZ}
    \mathfrak{su}(K) \,\, \rightarrow \,\, \mathfrak{su}(2)_X \oplus \bigoplus_{i=1}^K \mathfrak{su}(m_i) \,,
\end{equation}
where we have ignored Abelian factors.\footnote{We also assume that the Higgsing is not the trivial Higgsing, i.e., $O \neq [1^K]$.} Under this decomposition, the branching rule of the fundamental representation is
\begin{equation}
    \bm{K} \,\, \rightarrow \,\, \bigoplus_{i=1}^K (\bm{i}, \bm{m_i}) \,,
\end{equation}
where $\bm{i}$ is the $i$-dimensional irreducible representation of $\mathfrak{su}(2)_X$ and $\bm{m_i}$ is the fundamental representation of $\mathfrak{su}(m_i)$.\footnote{More precisely, $\bm{m_i}$ is the representation of $\bigoplus_{j=1}^K \mathfrak{su}(m_j)$ obtained by taking the tensor product of the fundamental representation of $\mathfrak{su}(m_i)$ with the trivial representation of all $\mathfrak{su}(m_{j\neq i})$ factors.} Recalling the following simple identities regarding the traces:
\begin{equation}
  \begin{aligned}
    \operatorname{tr}_{A \oplus B}F^k &= \operatorname{tr}_{A}F^k + \operatorname{tr}_{B}F^k \,, \\
    \operatorname{tr}_{A \otimes B}F^2 &= \operatorname{dim}(B)\operatorname{tr}_{A}F^2 + \operatorname{dim}(A)\operatorname{tr}_{B}F^2 \,, \\
    \operatorname{tr}_{A \otimes B}F^4 &= \operatorname{dim}(B)\operatorname{tr}_{A}F^4 + \operatorname{dim}(A)\operatorname{tr}_{B}F^4 + 6\operatorname{tr}_{A}F^2\operatorname{tr}_{B}F^2  \,,
    \end{aligned}
\end{equation}
it is then straightforward to determine the decompositions of the traces. In particular, let $F_X$ denote the curvature of the $\mathfrak{su}(2)_X$ bundle and $F_i$ the curvature of the $\mathfrak{su}(m_i)$ bundle, then 
\begin{equation}
    \operatorname{Tr} F^2 \,\, \rightarrow \,\, \sum_{i=1}^K \left( i \operatorname{Tr} F_i^2 + \frac{m_i i(i^2 - 1)}{6} \operatorname{Tr}F_X^2 \right)  \,,
\end{equation}
where we have used that
\begin{equation}
    \operatorname{tr}_{\bm{d}} F^2 = \frac{d(d^2-1)}{12} \operatorname{Tr} F^2 \,,
\end{equation}
for an arbitrary irreducible representation $\bm{d}$ of $\mathfrak{su}(2)$. We can now utilize a similar procedure to work out the decomposition of $\operatorname{Tr} F^4$. We find
\begin{equation}\label{eqn:doratheexplorer}
    \operatorname{Tr} F^4 \,\, \rightarrow \,\, \sum_{i=1}^K \left( i \operatorname{tr}_{\bm{m_i}} F_i^4 + m_i \operatorname{tr}_{\bm{i}} F_X^4 + 6 \operatorname{tr}_{\bm{m_i}} F_i^2 \operatorname{tr}_{\bm{i}} F_X^2 \right) \,.
\end{equation}
To convert all the traces to one-instanton normalized traces, we need to know several identities. First, for an arbitrary $d$-dimensional irreducible representation of $\mathfrak{su}(2)$, we have
\begin{equation}
    \operatorname{tr}_{\bm{d}} F^4 = \frac{d(3d^4 - 10d^2 + 7)}{240} \left(\operatorname{Tr}F^2\right)^2 \,.
\end{equation}
For the $\operatorname{tr}_{\bm{m_i}} F_i^4$, the trace converts differently depending on whether $m_i = 2, 3$ or $m_i \geq 4$. We have
\begin{equation}
    \operatorname{tr}_{\bm{m_i}} F_i^4 = \begin{cases}
        \operatorname{Tr} F_i^4 &\qquad \text{ if $m_i \geq 4$} \,, \\
        \frac{1}{8} \left(\operatorname{Tr} F_i^2\right)^2 &\qquad \text{ if $m_i = 2, 3$} \,.
    \end{cases}
\end{equation}
Thus, we have completed the first step of decomposing the traces. 

Next, we need to determine the anomaly polynomial of the Nambu--Goldstone modes arising from the moment map. Under the algebras after the decomposition in equation \eqref{eqn:ZZZ}, where we recall that $\mathfrak{su}(2)_X$ is replaced with $\mathfrak{su}(2)_{R^\text{IR}}$, the Nambu--Goldstone modes transform in the following reducible representation:
\begin{equation}\label{eqn:NGs}
    \bm{R} = \bigoplus_{\substack{i, j = 1 \\ j \neq i}}^K \bigoplus_{k=1}^{\operatorname{min}(i,j)} (\bm{i + j - 2k}, \bm{m_i}, \overline{\bm{m_j}}) \oplus \bigoplus_{i = 1}^K \bigoplus_{k=1}^{i} (\bm{2i - 2k}, \textbf{adj}_i \oplus \bm{1}) \,,
\end{equation}
where $\textbf{adj}_i$ is the adjoint representation of $\mathfrak{su}(m_i)$. The anomaly contribution from the Nambu--Goldstone modes is simply
\begin{equation}\label{eqn:NGI8}
    I_8^\text{NG} = \frac{1}{2} \operatorname{ch}_{\bm{R}}(\{R^\text{IR}, F_i\}) \widehat{A}(T) \,,
\end{equation}
where we use $\{R^\text{IR}, F_i\}$ to collectively denote the curvatures of the $\mathfrak{su}(2)_{R^\text{IR}}$ bundles and all the $\mathfrak{su}(m_i)$ bundles. Recalling that the Chern character can be written in terms of the traces as
\begin{equation}
   \operatorname{ch}_\rho(F) = \operatorname{tr}_\rho F^0 - \frac{1}{2} \operatorname{tr}_\rho F^2 + \frac{1}{24} \operatorname{tr}_\rho F^4 + \cdots \,,
\end{equation}
where we have ignored the terms of odd form-degree, it is straightforward to expand the anomaly polynomial in equation \eqref{eqn:NGI8} in terms of the irreducible components of the representation in equation \eqref{eqn:NGs}. As we will see, certain expressions appear regularly in the expansions of the traces; for convenience we define the following matrices
\begin{equation}
    \begin{gathered}
        X_{i,j} = ij - \operatorname{min}(i,j) \,, \quad Y_{i,j} = (ij - \operatorname{min}(i,j))(i^2 + j^2 - 2(i + j - \operatorname{min}(i,j)) - 1) \,, \\ 
        Z_{i,j} =  \frac{1}{15} \sum_{k=1}^{\operatorname{min}(i,j)} (i + j - 2k + 1)(i + j - 2k)(i + j - 2k - 1)(3(i+j-2k)^2 - 7)  \,.
    \end{gathered}
\end{equation}

Now we consider the traces appearing in the expansion of the Nambu--Goldstone anomaly polynomial. We begin with
\begin{equation}
    \operatorname{tr}_{\bm{R}} F^0 = \sum_{i, j = 1}^K m_i m_j X_{i,j} = \text{dim}(\bm{R})\,,
\end{equation}
which is simply the dimension of the representation $\bm{R}$. We now turn to the second term, which decomposes as
\begin{equation}
    \operatorname{tr}_{\bm{R}} F^2 = \sum_{\substack{i, j = 1}}^K \frac{1}{2}X_{i,j} (m_i\operatorname{Tr}F_j^2 + m_j\operatorname{Tr}F_i^2) + \sum_{i, j = 1}^K \frac{1}{3}m_im_jY_{i,j} c_2(R^\text{IR}) \,.
\end{equation}
Here we have used that
\begin{equation}
    c_2(R^\text{IR}) = \frac{1}{4} \operatorname{Tr} F_{R^\text{IR}}^2 \,.
\end{equation}
The quartic trace is the most tedious term to decompose, however, after a little algebra, one finds the following: 
\begin{equation}
  \begin{aligned}
    \operatorname{tr}_{\bm{R}} F^4 &= \sum_{i, j = 1}^K m_i m_j Z_{i,j} c_2(R^\text{IR})^2 + X_{i,j} (m_i\operatorname{tr}_{\bm{m_j}}F_j^4 + m_j\operatorname{tr}_{\bm{m_i}}F_i^4) \\ &\qquad\qquad + Y_{i,j}(m_i\operatorname{Tr}F_j^2 + m_j\operatorname{Tr}F_i^2)c_2(R^\text{IR}) + \frac{3}{2} X_{i,j} \operatorname{Tr}F_i^2 \operatorname{Tr}F_j^2 \,.
  \end{aligned}
\end{equation}

Now that we have determined the appropriate decompositions of the traces and the anomaly contribution for the Nambu--Goldstone modes, we are ready to apply Algorithm \ref{alg:I8IR} to determine the anomaly polynomial of $A_{N-1}^{\mathfrak{su}(K)}(O_L, O_R)$. Let the integer partitions of $K$ associated to the nilpotent orbits $O_L$ and $O_R$ be
\begin{equation}\label{eqn:swiper}
    P_{O_L} = [1^{m_1}, 2^{m_2}, \cdots, K^{m_K}] \,, \qquad P_{O_R} = [1^{m_1'}, 2^{m_2'}, \cdots, K^{m_K'}] \,.
\end{equation}
To aid in the comparison with the tensor branch 't Hooft anomaly matching approach to determining the anomaly polynomial, we will be extremely explicit here, and go coefficient-by-coefficient.

First, the contributions to the coefficients $\gamma^\text{IR}$ and $\delta^\text{IR}$ come only from the Nambu--Goldstone modes:
\begin{equation}\label{gamma-delta-NB}
  \begin{aligned}
		  \gamma^\text{IR} &= \gamma^\text{UV} - \frac{7}{480} \sum_{i,j=1}^K (m_i m_j + m_i'm_j') X_{i,j}= \gamma^\text{UV} - \frac{7}{480} \left(\text{dim}(O_L)+\text{dim}(O_R)\right) \,, \\ 
    \delta^\text{IR} &= \delta^\text{UV} + \frac{1}{120} \sum_{i,j=1}^K (m_i m_j + m_i'm_j')X_{i,j} =\delta^\text{UV} + \frac{1}{120} \left(\text{dim}(O_L)+\text{dim}(O_R)\right) \,.
  \end{aligned}
\end{equation}
The relation between the partition data and the dimension of the orbit can be
found in e.g. \cite{collingwood2017nilpotent, Mekareeya:2016yal}, and this
result is therefore consistent with equation \eqref{shift-grav-coeffs}.

Next, we turn to $\beta^\text{IR}$; there are three contributions to the
infrared anomaly from the ultraviolet anomaly polynomial, coming from the terms
\begin{equation}
    c_2(R)p_1(T) \,, \qquad p_1(T) \operatorname{Tr}F_L^2 \,, \qquad \text{ and } \qquad p_1(T) \operatorname{Tr}F_R^2 \,,
\end{equation}
together with the Nambu--Goldstone contribution. To succinctly write how these UV terms enter in the infrared anomaly coefficients, we define the embedding index for the $\mathfrak{su}(2)_X$ factor associated to the nilpotent orbit $O_L$, following equation \eqref{su2-embedding-index-wdd}, as
\begin{equation}
    I_{X_L} = \sum_{i=1}^K m_i \frac{i(i^2-1)}{6} \,,
\end{equation}
and similarly we define $I_{X_R}$. Putting all four contributions together, we find
\begin{equation}\label{beta-suk-NBG}
  \begin{aligned}
    \beta^\text{IR} &= \beta^\text{UV} + 96(I_{X_L} \kappa_L^\text{UV} + I_{X_R} \kappa_R^\text{UV})  - \frac{1}{12} \sum_{i, j = 1}^K (m_im_j + m_i'm_j')Y_{i,j} \,. 
  \end{aligned}
\end{equation}
Finally, we turn to the $c_2(R^\text{IR})^2$ anomaly coefficient $\alpha^\text{IR}$. In addition to the Nambu--Goldstone modes, there exist eight terms in the UV anomaly polynomial that contribute to this IR coefficient; these are:
\begin{equation}
  \begin{gathered}
    c_2(R)^2 \,, \quad c_2(R) \operatorname{Tr} F_L^2 \,, \quad c_2(R) \operatorname{Tr} F_R^2 \,, \quad \operatorname{Tr} F_L^4 \,, \quad \operatorname{Tr} F_R^4 \,, \\
    \operatorname{Tr} F_L^2 \operatorname{Tr} F_L^2  \,, \quad  \operatorname{Tr} F_L^2 \operatorname{Tr} F_R^2  \,, \quad \operatorname{Tr} F_R^2 \operatorname{Tr} F_R^2 \,.
  \end{gathered}
\end{equation}
For ease of notation, we define the following quantity:
\begin{equation}
    I_{O_L^2} = \sum_{i=1}^K m_i \frac{i(3i^4 - 10i^2 + 7)}{240} \,,
\end{equation}
and again analogously for $I_{O_R^2}$. Then 
\begin{equation}\label{alpha-suk-NBG}
  \begin{aligned}
    \alpha^\text{IR} &= \alpha^\text{UV} + 96 (I_{X_L} \nu_L^\text{UV} + I_{X_R} \nu_R^\text{UV}) + 384 (I_{O_L^2} \mu_L^\text{UV} + I_{O_R^2} \mu_R^\text{UV}) \\ &\qquad + 384(I_{X_L}^2 \rho_{LL}^\text{UV} + I_{X_L}I_{X_R} \rho_{LR}^\text{UV} + I_{X_R}^2 \rho_{RR}^\text{UV}) - \frac{1}{2} \sum_{i,j = 1}^K (m_i m_j + m_i' m_j') Z_{i,j} \,.
  \end{aligned}
\end{equation}

The anomalies involving the infrared non-Abelian flavor symmetries are our next port of call. We begin with the mixed flavor-gravitational anomalies. We find
\begin{equation}\label{eqn:kappaIR}
    \kappa_{L,i}^\text{IR} = i \kappa_{L}^\text{UV} - \frac{1}{96} \sum_{j=1}^K m_j X_{i,j} \,, \qquad
    \kappa_{R,i}^\text{IR} = i \kappa_{R}^\text{UV} - \frac{1}{96} \sum_{j=1}^K m_j' X_{i,j} \,.
\end{equation}
Next, we consider the quartic anomalies of the $\mathfrak{su}(m_i)$ flavor symmetries -- obviously these $\operatorname{Tr}F^4_i$ terms can only exist if the flavor algebra admits an independent quartic Casimir, i.e., $m_i > 3$. Assuming that this condition is satisfied, we find
\begin{equation}\label{eqn:muIR}
    \mu_{L,i}^\text{IR} = i \mu_{L}^\text{UV}  - \frac{1}{24} \sum_{j=1}^K m_j X_{i,j} \,, \qquad
    \mu_{R,i}^\text{IR} = i \mu_{R}^\text{UV}  - \frac{1}{24} \sum_{j=1}^K m_j' X_{i,j} \,.
\end{equation}
Let us now consider the mixed R-flavor anomalies. In addition to the Nambu--Goldstone modes which must be subtracted, these anomalies have contributions from three distinct UV anomalies. We have
\begin{equation}\label{nuL-NBG}
    \nu_{L,i}^\text{IR} = i \nu_{L}^\text{UV} + 8 i I_{X_L} \rho_{LL}^\text{UV}  + i(i^2 - 1)\mu_L^\text{UV} - \frac{1}{24} \sum_{j=1}^K m_j Y_{i,j} \,, 
\end{equation}
and analogously for $\nu_{R,i}^\text{IR}$. Finally, we consider the mixed flavor-flavor anomalies. We begin with the anomalies from the infrared flavor algebras that come from either the left or the right UV flavor algebra. We have
\begin{equation}\label{eqn:rhoLL1}
    \rho_{LL,ij}^\text{IR} = ij \rho_{LL}^\text{UV} - \frac{1}{32} X_{i,j} \,.
\end{equation}
If $i$ is such that $m_i = 2, 3$, then there are two extra contributions to the $j = i$ anomaly coefficient:
\begin{equation}\label{eqn:rhoLL2}
    \rho_{LL,ii}^\text{IR} = i^2 \rho_{LL}^\text{UV} + \frac{i}{8} \mu_L^\text{UV} - \frac{1}{32} X_{i,i} - \frac{1}{192} \sum_{j=1}^K X_{i,j} m_j  \,.
\end{equation}
The obvious modifications hold for the anomaly coefficients $\rho_{RR,ij}$. We also consider the flavor-flavor anomalies that mix the flavor symmetry factors localized on the left and the right; such anomalies lack a Nambu--Goldstone contribution. We find
\begin{equation}\label{eqn:rhoLRIR}
    \rho^\text{IR}_{LR,ij} = ij \rho^\text{UV}_{LR} \,.
\end{equation}

We have now used Algorithm \ref{alg:I8IR} to determine the generic form of the anomaly polynomial $A_{N-1}^{\mathfrak{su}_K}(O_L, O_R)$ written in terms of the partitions in equation \eqref{eqn:swiper} defining the nilpotent orbits $O_L$ and $O_R$. While we have focused on the cases where $K > 3$, the special cases of $K = 2, 3$ are governed by identical formulae, except that one needs to formally set $\mu_L^\text{UV} = \mu_R^\text{UV} = 0$, to account for the absence of an independent quartic Casimir for $\mathfrak{su}(2)$ and $\mathfrak{su}(3)$.


\subsubsection{From the Tensor Branch}

We now determine the anomaly polynomial of $A_{N-1}^{\mathfrak{su}_K}(O_L, O_R)$ from the geometric description of the effective field theory at the generic point of the tensor branch. The tensor branch configuration takes the form
\begin{equation}
    A_{N-1}^{\mathfrak{su}_K}(O_L,O_R):\qquad
    \underset{[m_1]}{\overset{\mathfrak{su}_{k_1}}{2}}
    \underset{[m_2]}{\overset{\mathfrak{su}_{k_2}}{2}}
    \cdots
    \underset{[m_K]}{\overset{\mathfrak{su}_{k_K}}{2}}
    \underbrace{\overset{\mathfrak{su}_K}{2}\cdots \overset{\mathfrak{su}_K}{2}}_{N-2K-1}
    \underset{[m_K']}{\overset{\mathfrak{su}_{k_K'}}{2}}
    \cdots
    \underset{[m_2']}{\overset{\mathfrak{su}_{k_2'}}{2}}
    \underset{[m_1']}{\overset{\mathfrak{su}_{k_1'}}{2}} \,,
\end{equation}
as described in Appendix \ref{app:nilp2TB}. We will assume that neither $O_L$ nor $O_R$ are the nilpotent orbits associated to the $[K]$ partition; these special cases can be handled individually. The anomaly polynomial can be written as
\begin{equation}
    I_8 = I_8^\text{1-loop} + I_8^\text{GS} \,,
\end{equation}
where the one-loop contribution is the sum of the contributions of the vector, tensor and hypermultiplets. We refer in particular to \cite{Baume:2021qho}, where the algorithm to determine the anomaly polynomial from the tensor branch geometry is given explicitly, and for the anomaly contributions from each multiplet. The Green--Schwarz contribution is 
\begin{equation}
    I_8^\text{GS} = - \frac{1}{2} \widetilde{A}_{i,j} I^i I^j \,, \qquad I^i = -\frac{1}{4}B^{ia} \operatorname{Tr}F^2_a + h_i^\vee c_2(R) \,.
\end{equation}
Here $i$, $j$ index the $(-2)$-curves, left-to-right. We have not written the gauge field strengths in $I^i$ as they all cancel in the final result, $a$ runs over the simple non-Abelian flavor factors of the SCFT, and $h^\vee_i$ is the dual Coxeter number of the algebra supported over the $i$th curve. Finally, the matrix $\widetilde{A}$ is the inverse of the (negative-definite) Cartan matrix of $A_{N-1}$; it has entries
\begin{equation}
    \widetilde{A}_{i,j} = \frac{ij}{N} - \operatorname{min}(i,j) \,.
\end{equation}

To compare to the anomalies as determined by Algorithm \ref{alg:I8IR}, we need to know the UV anomaly coefficients for conformal matter; these appear in equation \eqref{I8-CM-expanded}. We study the 't Hooft anomalies associated to the flavor symmetries first, as they are they simplest. Consider $\kappa_{L,i}$; the only contribution is from the bifundamental hypermultiplet charged under $\mathfrak{su}(m_i)$. We find that
\begin{equation}
    \kappa_{L, i} = \frac{1}{96} k_i = \frac{1}{96} \left( K - \sum_{j=1}^{K-i} j m_{i+j} \right) = i \left( \frac{K}{96} \right) - \frac{1}{96} \sum_{j=1}^K m_j X_{i,j} \,.
\end{equation}
Here we have used the elementary fact about integer partitions of $K$ that
\begin{equation}\label{elementary-fact}
    K - \sum_{j=1}^{K-i} j m_{i+j} = i K - \sum_{j=1}^K m_j (ij - \operatorname{min}(i,j)) \,.
\end{equation}
We have thus verified equation \eqref{eqn:kappaIR}. Contributions of $\mu_{L,i}$ similarly arise only from the bifundamental hypermultiplet between $\mathfrak{su}(m_i)$ and $\mathfrak{su}(k_i)$, and thus almost identical manipulations reveal that the tensor branch calculation reproduces equation \eqref{eqn:muIR}. Consider now the $\rho_{LR,ij}$ coefficient; the only contribution is the Green--Schwarz term and we find
\begin{equation}
  \begin{aligned}
    \rho_{LR,ij} = - \frac{1}{16} \widetilde{A}_{i, N-j} &= \frac{1}{16} \left( \operatorname{min}(i,N-j) - \frac{i(N-j)}{N} \right) \\
    &= \frac{1}{16} ij \left( 1 - \frac{N-1}{N} \right) = ij \left( -\frac{\widetilde{A}_{1,N-1}}{16}\right) \,.
  \end{aligned}
\end{equation}
This verifies equation \eqref{eqn:rhoLRIR}. Henceforth, we silently use the fact that
\begin{equation}
    \widetilde{A}_{i, N-j} = ij \widetilde{A}_{1, N-1} \,.
\end{equation}
Next, we consider $\rho_{LL,ij}$; we first assume that $i \neq j$ or $m_i > 3$. The only contribution is from the Green--Schwarz term and we find
\begin{equation}
    \rho_{LL,ij} = - \frac{1}{16} \widetilde{A}_{i,j} = \frac{1}{32} \left( \operatorname{min}(i,j) - \frac{ij}{N} \right) = ij \left( \frac{N-1}{32N} \right) - \frac{1}{32} \left( ij - \operatorname{min}(i,j) \right) \,.
\end{equation}
When $i = j$ and $m_i = 2, 3$ then there is an additional contribution from the bifundamental hypermultiplet charged under $\mathfrak{su}(m_i)$. This contributes in the same way as the bifundamental hypermultiplet for the $\kappa_{L,i}$ and $\mu_{L,i}$, and thus we find
\begin{equation}
    \rho_{LL,ii} = i^2 \left( \frac{N-1}{32N} \right) - \frac{1}{32} i(i-1) + \frac{i}{8} \left( \frac{K}{24} \right) - \frac{1}{192} \sum_{j=1}^K X_{i,j} m_j \,.
\end{equation}
Thus, we have shown that equations \eqref{eqn:rhoLL1} and \eqref{eqn:rhoLL2}
from the algorithm agree with the geometric calculation.  Next, we turn to
$\nu_{L,i}$. Only Green--Schwarz terms can contribute to this anomaly
coefficient. We find
\begin{equation}
    \nu_{L,i} = \frac{1}{4} \sum_{j=1}^{N-1}k_i \widetilde{A}_{j,i} = \frac{i}{4N} + (N - 1) I_{X_L} + \frac{K}{24} i(i^2 + 2) - \frac{iKN}{8}-\frac{1}{24}\sum_{j=1}^Km_jY_{i,j}\,,
\end{equation}
where we have rewritten the sum over the whole quiver in terms of the partition data, therefore recovering equation \eqref{nuL-NBG}.

We now move on to the terms that do not involve the flavor symmetry.
Closed-form expression in terms of the gauge group ranks $k_i$ and the pairing
matrix are straightforward to determine, see, e.g., \cite{Mekareeya:2016yal}. We
must therefore convert them into the partition data to compare them with the
expressions found in the previous subsection.

The coefficients $\gamma$ and $\delta$ have contributions from all three types
of multiplets, but not from the Green--Schwarz term. For the former, one finds
that
\begin{equation}
\delta = - \frac{N-1}{2} -\frac{1}{120}\sum_{i=1}^{N-1}k_im_i = -\frac{N}{2} - \frac{(K^2 - 1) - 29}{60}+\frac{1}{120}\sum_{i,j=1}^K (m_i m_j + m_i'm_j') X_{i,j}\,,
\end{equation}
where we have used the relation between the value of $k_i$ for the gauge
algebra and the partition data, see Appendix \eqref{A-k-to-m}, as well as
equation \eqref{elementary-fact} to simplify the result and put it in the same
form the value of $\delta$ given in equation \eqref{gamma-delta-NB}. Up to
small difference in the numerology of the constant factors, one can verify
that the same is also true for $\gamma$.

For $\beta$, the only contributions are from the tensor and vector multiplets
in the one-loop anomaly polynomial. We find
\begin{equation}
    \beta = -\frac{1}{2} \left( \sum_{i=1}^K (k_i^2 + k_i'^2) - 2K^3 - K^2 + 2 + N (K^2-2)  \right) \,,
\end{equation}
where we written the Green--Schwarz contribution in terms of the $k_i$ on the left and the right. These can then be converted to the partition data using the relation
\begin{equation}
		\sum_{i=1}^K k_i^2 =  K^3- 2 I_X K + \frac{1}{6}\sum_{i=1}^Km_im_jY_{i, j}\,,
\end{equation}
and we therefore recover what we have found for in the previous subsection, see equation \eqref{beta-suk-NBG}.
Finally, the $c_2(R)^2$ term arise from contributions of both tensor and vector
multiplets, as well as the Green--Schwarz term:
\begin{equation}
		\alpha = -12 \sum_{i,j=1}^{N-1} k_ik_j\widetilde{A}_{i,j} +2\beta\,.
\end{equation}
Decomposing the first term into the partition orders $m_i$ and $m'_i$ is tedious, but straightforward. When the dust settles, one finds:
\begin{equation}
\begin{aligned}
    \sum_{i,j=1}^{N-1}k_ik_j\widetilde{A}_{i,j} = 
    &-\frac{N(N^2-1)}{12} + \frac{1}{N}(I_{X_L}+I_{X_R}) - (I_{X_L}^2+I_{X_R}^2) \\
    &+ \frac{6N-5}{6}K(I_{X_L}+I_{X_R})-\frac{4}{3}(I_{O^2_L}+I_{O^2_R}) \\
    &+ \frac{1}{24}\sum_{i,j=1}^K(m_im_j + m'_im'_j)(Z_{i,j}-\frac{1}{3}Y_{i,j})\,,
\end{aligned}
\end{equation}
from which we once obtain the same expression as in equation \eqref{alpha-suk-NBG}, as expected. 

Thus, we have proven that the anomaly polynomial of Higgsed rank $N$
$(\mathfrak{su}(K), \mathfrak{su}(K))$ as determined from the geometric
description of the tensor branch is identical to the anomaly polynomial
obtained by following Algorithm \ref{alg:I8IR}. 

\subsection{An Exceptional Example: \texorpdfstring{$A^{\mathfrak{e}_6}_{N-1}(A_1,0)$}{A\_(e6, N-1)(A1,0)}}\label{sec:exceptNGeg}

For the parent theories in equation \eqref{parent-theory-list} where $\mathfrak{g}$ is an exceptional Lie group, there are only a finite number of nilpotent orbits, and thus one can verify that Algorithm \ref{alg:I8IR} produces the same result as the geometry, exhaustively. While we have carried out this exhaustive process, we present here only one example: $A_{N-1}^{\mathfrak{e}_6}(0, A_1)$. This example also appeared in Section \ref{sec:sec3examples}.

First, we consider the anomalies of the ultraviolet theory: $A_{N-1}^{\mathfrak{e}_6}$. These can be determined directly from the geometry, and we find
\begin{equation}\label{eqn:e6egUV}
    \begin{gathered}
        \alpha^\text{UV} = 576N^3 - 334N + 77 \,, \quad \beta^\text{UV} = \frac{77 - 166N}{2} \,, \\[0.3em]
        \gamma^\text{UV} = \frac{30N+523}{240} \,, \quad \delta^\text{UV} = \frac{-30N - 49}{60} \,, \quad \kappa_{L, R}^\text{UV} = \frac{1}{8} \,, \\[0.3em]
        \nu_{L, R}^\text{UV} = \frac{3 - 6N}{2} \,, \quad \mu_{L, R}^\text{UV} = 0 \,, \quad \rho_{LL, RR}^\text{UV} = \frac{3N - 1}{32N} \,, \quad \rho_{LR, RL}^\text{UV} = \frac{1}{16N} \,,
    \end{gathered}
\end{equation}
where the subscripts $L$ and $R$ refer to the ``left'' and ``right'' $\mathfrak{e}_6$ flavor symmetry factors, respectively. Similarly, we can determine the anomalies of the infrared theory, $A_{N-1}^{\mathfrak{e}_6}(0, A_1)$, from the geometry. In this case, the effective field theory on the tensor branch is
\begin{equation}
    \underbrace{\,1\overset{\mathfrak{su}_3}{3}1\overset{\mathfrak{e}_6}{6} \cdots 1\overset{\mathfrak{su}_3}{3}1\overset{\mathfrak{e}_6}{6} \,}_{N-1} 1 \overset{\mathfrak{su}_3}{2} \,,
\end{equation}
and thus we find that the anomaly coefficients are
\begin{equation}\label{eqn:e6egIR}
    \begin{gathered}
        \alpha^\text{IR} = 576N^3 - 622N + 256 - \frac{12}{N} \,, \quad \beta^\text{IR} = \frac{100 - 166N}{2} \,, \\[0.3em]
        \gamma^\text{IR} = \frac{30N+446}{240} \,, \quad \delta^\text{IR} = \frac{-30N - 38}{60} \,, \quad \kappa_{L}^\text{IR} = \frac{1}{8} \,, \quad \kappa_{R}^\text{IR} = \frac{3}{32} \,, \\[0.3em]
        \nu_{L}^\text{IR} = \frac{3 - 6N}{2} + \frac{1}{4N} \,, \quad \nu_{R}^\text{IR} = \frac{3 - 6N}{2} + \frac{3N - 1}{4N} \,, \quad \mu_{L}^\text{IR} = 0 \,, \quad \mu_{R}^\text{IR} = \frac{1}{8} \,, \\[0.3em] \rho_{LL}^\text{IR} = \frac{3N - 1}{32N} \,, \quad \rho_{RR}^\text{IR} = \frac{2N - 1}{32N} \,, \quad \rho_{LR, RL}^\text{IR} = \frac{1}{16N} \,,
    \end{gathered}
\end{equation}
where the subscript $L$ denotes the unbroken $\mathfrak{e}_6$ flavor symmetry, and $R$ is for the $\mathfrak{su}(6)$ flavor that right $\mathfrak{e}_6$ is broken to by the Higgsing. 

We would now like to reproduce these infrared anomalies from the Nambu--Goldstone analysis of Algorithm \ref{alg:I8IR}. The nilpotent orbit $A_1$ of $\mathfrak{e}_6$ is associated to the decomposition 
\begin{equation}
    \mathfrak{e}_6 \rightarrow \mathfrak{su}(2)_X \oplus \mathfrak{su}(6) \,,
\end{equation}
where both factors in the decomposition have embedding index one; this can be seen easily from the adjoint branching, which is
\begin{equation}\label{eqn:e6egadj}
    \bm{78} \rightarrow (\bm{3, 1}) \oplus (\bm{1, 35}) \oplus (\bm{2, 20}) \,.
\end{equation}
Thus, we can see that we must decompose the curvature of the right $\mathfrak{e}_6$ flavor bundle as
\begin{equation}
    \operatorname{Tr}F_R^2 \rightarrow \operatorname{Tr}F_X^2 + \operatorname{Tr}F_R^2 \,,
\end{equation}
where we have abused notation and used $F_R$ on the left to refer to the curvature of the UV $\mathfrak{e}_6$ bundle, and on the right to refer to the curvature of the IR $\mathfrak{su}(6)$ bundle. We do not need to discuss the decomposition of the $\operatorname{Tr}F_R^4$ as $\mathfrak{e}_6$ does not possess a quartic Casimir.

From the decomposition of the adjoint representation in equation \eqref{eqn:e6egadj}, the Nambu--Goldstone fermions transform in the following representations of the $\mathfrak{su}(2)_{R^\text{IR}} \oplus \mathfrak{su}(6)$ infrared global symmetry:
\begin{equation}
    (\bm{2, 1}) \,, \qquad 
    (\bm{1, 20}) \,.
\end{equation}
Thus, the contribution to the anomaly polynomial from the Nambu--Goldstone modes is
\begin{equation}
    I_8^\text{NG} = \frac{1}{2}\operatorname{ch}_{\bm{2}}(R^\text{IR})\widehat{A}(T) + \frac{1}{2}\operatorname{ch}_{\bm{20}}(F_{R})\widehat{A}(T) \bigg|_{\text{8-form}} \,,
\end{equation}
where we have just written $R^\text{IR}$ for the curvature of the $\mathfrak{su}(2)_{R^\text{IR}}$ bundle. To expand this, we must convert traces in the $\bm{20}$ representation of $\mathfrak{su}(6)$ to one-instanton normalized traces:
\begin{equation}
    \operatorname{ch}_{\bm{20}}(F_R) = 20 - \frac{3}{2}\operatorname{Tr} F_R^2 + \frac{1}{24}\bigg(-6\operatorname{Tr}F_R^4 + \frac{3}{2} \left(\operatorname{Tr}F_R^2\right)^2 \bigg) + \cdots \,.
\end{equation}
Altogether, then, we find that the Nambu--Goldstone modes contribute to the infrared anomaly in the following explicit way:
\begin{equation}
  \begin{aligned}
    I_8^\text{NG} &= 11 \times \frac{1}{5760}\left( 7p_1(T)^2 - 4p_2(T) \right) + \left( - \frac{3}{4}\operatorname{Tr}F_R^2 - \frac{1}{2}c_2(R^\text{IR}) \right) \times \left( - \frac{1}{24}p_1(T) \right) \\ &\qquad\qquad + \frac{1}{24}c_2(R^\text{IR})^2 + \frac{1}{32} \left(\operatorname{Tr}F_R^2\right)^2 - \frac{1}{8}\operatorname{Tr}F_R^4 \,.
  \end{aligned}
\end{equation}
Now that we have determined $I_8^\text{NG}$, we can combine it with the UV anomaly coefficients as in equation \eqref{eqn:e6egUV}, following Algorithm \ref{alg:I8IR}, and we see that the infrared anomalies that were given via the geometry in equation \eqref{eqn:e6egIR} appear directly. It is straightforward to apply this simple procedure to any exceptional parent theory by utilizing the nilpotent orbit data collated in Appendix \ref{app:nilps}.

\section{Discussion}\label{sec:discussion}

The main results of our work are the closed-form expressions for the anomaly
polynomials of any long quivers, or equivalently Algorithm \ref{alg:I8IR},
giving a prescription to find the complete anomaly polynomial of a 6d $(1, 0)$
SCFT obtained through nilpotent renormalization group flows of the parent theories. Beyond their usefulness
as tools to efficiently obtain the numerical values of its coefficients, they
also enable us to study the deformed theories purely in terms of
gauge-invariant quantities computed directly at the conformal fixed point,
without invoking the effective field theory on the generic point of the tensor
branch. That is, we can understand the complete anomaly polynomial of an SCFT in terms
of its conformal spectrum rather than its geometric description.

As we have already alluded to, a particularly important set of quantities of a
6d conformal field theory are its central charges. These are part of the
conformal data, and are in principle defined independently of any gauge
description or geometric engineering. For instance, the central charges $C_T$
and $C_J$ are obtained by computing the two-point correlators of the
energy-momentum tensor $T_{\mu\nu}$ and flavor currents $J_a^\mu$, respectively
\cite{Osborn:1993cr, Cordova:2019wns}:
\begin{equation}
\left<T^{\mu\nu}(x)T^{\rho\sigma}(0)\right> = \frac{C_T}{\text{Vol}(S^5)^2}\,\frac{P^{\mu\nu\rho\sigma}}{x^{12}}\,,\qquad
		\left<J^{\mu}_a(x)J_b^{\mu}(0)\right> = \frac{C_{J,a}}{\text{Vol}(S^5)^2}\,\frac{\delta^{ab}P^{\mu\nu}}{x^{10}}\,,
\end{equation}
where $P^{\mu\nu\rho\sigma}$ and $P^{\mu\nu}$ are the spin-two and spin-one
projectors, respectively. It is also well-known that in the presence of a
background metric, there is a Weyl anomaly and the tracelessness
condition of the energy-momentum tensor is broken \cite{Deser:1993yx, Duff:1993wm}:
\begin{equation}\label{Weyl-anomaly}
		\left<T^\mu_\mu\right> = \frac{a}{(4\pi)^3} E_6 + \cdots \,,
\end{equation}
where $E_6$ is the six-dimensional Euler density, while the $\cdots$ encode
both Weyl-invariant and scheme-dependent terms. In unitary theories, the central charges and the coefficient $a$ must
be positive. They are in particular also related to OPE coefficients, which makes
them particularly relevant in the modern incarnation -- numerical or analytical
-- of the conformal bootstrap \cite{Poland:2022qrs, Hartman:2022zik}.
These quantities are furthermore related to particular combinations of the
coefficients appearing in the anomaly polynomial of a given (1,0) SCFT
\cite{Cordova:2015fha, Cordova:2019wns}:
\begin{align}
		C_T &= 168 (2\alpha - 3\beta + 4\gamma + \delta)\,,\\
		C_{J,a} &= 240 (\kappa_a - \nu_a)\,,\\
		a &= \frac{16}{7}(\alpha-\beta+\gamma+\frac{3}{8}\delta)\,.\label{def-a-abcd}
\end{align}

Our results therefore make it particularly easy to find them directly without
going through geometric engineering.

In Section \ref{ana-cont}, we discuss how the anomaly polynomial of theories described by short quivers can be understood as limits of long quivers, and how one can define additional minimal building blocks encoding the nilpotent breakings that are well behaved under fusion, despite possibly having negative central charges. In Section \ref{sec:a-thm}, we come back to the Weyl anomaly coefficient $a$ defined in equation \eqref{Weyl-anomaly}, and give a proof of the $a$-theorem for nilpotent RG flows using only the group theory related to nilpotent orbits. We close by discussing possible future directions in Section \ref{sec:future}.

\subsection{Building Blocks and Analytic Continuation}\label{ana-cont}

As we go along the Higgs branch RG flows, there are fewer and fewer curves in the
tensor branch description, and we have assumed throughout that the quiver is
long enough so that either of its ends cannot influence the other. On the
other hand, even when one of its tails is kept undeformed, if the quiver is
short enough, as we go through nilpotent RG flows we could ultimately have no
curves left even though the end of the Hasse diagram of the associated flavor
symmetry has not been reached.

It is however possible to ``analytically continue'' the anomaly polynomial of
long quivers \cite{Heckman:2018pqx, Mekareeya:2017sqh} and consider values of $N$ that are smaller than the number of minimal conformal matter affected by the nilpotent orbits. When $\mathfrak{g}=\mathfrak{su}_K$ for instance, a nilpotent orbit can affect the
gauge symmetry of up to $K$ curves, but we can in principle set $N<K$ in the anomaly polynomial. This enables us to formally define a deformed version of minimal conformal matter:
\begin{equation}\label{deformed-mCM}
		\mathcal{A}_{0;1,f}^\mathfrak{g}(\varnothing, O):\qquad [\mathfrak{g}]\text{---}[O]\,,
\end{equation}
which we have depicted in a pictorial way as before. These putative theories, along with
the other types of building blocks we have encountered, allow us to construct
any long quiver, even when the nilpotent orbit corresponds to a curve
configuration where one or more conformal matter links are ``eaten'' by a
nilpotent deformation, and the building block therefore does not have an associated quiver. In fact, in many cases, when the nilpotent orbit $O$ is
located too deep in the Hasse diagram, the central charges, $C_T$ or $C_{J,a}$,
of $\mathcal{A}_{0;1,f}^{\mathfrak{g}}(\varnothing, O)$ are negative, in apparent
violation of unitarity.

In Section \ref{sec:anomaly-conformal-matter}, we have however seen that we can
compute the anomaly polynomial directly at the conformal fixed point, and
consider minimal conformal matter as a ``one-loop'' contribution. Even though
$\mathcal{A}_{0;1,f}^{\mathfrak{g}}(\varnothing, O)$ might have a seemingly inconsistent
anomaly polynomial, when it is fused with more minimal conformal matter
theories, we obtain an anomaly polynomial that matches exactly that of the
tensor branch computation. This is shown in the same way we have done around
equation \eqref{GS-CM} for higher-rank conformal matter, but now using
$\mathcal{A}^{\mathfrak{g}}_{0;1,f}(\varnothing,O)$ at both ends of the quiver.
This further has the advantage of completely bypassing the possible
propagation of the breaking throughout the quiver. 

Furthermore, the fact that the minimal building blocks defined in equation
\eqref{deformed-mCM} do not have a quiver that can be read out from a table or
a partition does not means that that they do not correspond to well-defined
theories. In fact, a number of theories with short bases -- that is, those with
a small number of curves on the tensor branch -- can be understood as an
analytical continuation of a long quiver \cite{Heckman:2018pqx, Mekareeya:2017sqh}.
For instance, one can find that the following non-Higgsable cluster:
\begin{equation}
		\mathcal{T}:\qquad \overset{\mathfrak{su}_2}{2}\overset{\mathfrak{g}_2}{3} \,,
\end{equation}
has the same anomaly polynomial as a fractional $(\mathfrak{e}_6,
\mathfrak{e}_6)$ conformal matter deformed with a nilpotent orbit $O=A_4+A_1$
when $N\to2$:
\begin{equation}
		I_8(\mathcal{T}) = I_8(\mathcal{A}_{1;1,\frac{2}{3}}^{\mathfrak{e}_6}(A_4+A_1, \varnothing))\,,\qquad \mathcal{T} ~\simeq~ [A_4+A_1]\text{---}\mathfrak{g}\text{---}[\mathfrak{g}_{\frac{2}{3}}=\varnothing]\,.
\end{equation}
These two theories have \emph{a priori} nothing in common. No $\mathfrak{e}_6$ node
appears in the quiver of $\mathcal{T}$, and as can be seen from Table
\ref{tbl:E6-nilp-1}, a complete minimal conformal matter has been ``eaten'' in
the quiver of
$\mathcal{A}_{N-1;1,\frac{2}{3}}^{\mathfrak{e}_6}(A_4+A_1,\varnothing)$.
Nonetheless, it is straightforward to check that both anomaly polynomials
precisely match. This can be extended to all non-Higgsable clusters \cite{Heckman:2018pqx}.
In this sense, as our results do not involve the tensor branch
description and we can work directly at the conformal fixed point; we can not
only deal with long quivers by fusing the building blocks defined in equation
\eqref{deformed-mCM} to other conformal matter, but they can describe short quivers
as well. 

In \cite{Heckman:2018pqx}, it was also shown that in some cases, short quivers
obtained by analytic continuation exhibit flavor enhancement. It would
therefore be interesting to see if more general types of short quivers can be
obtained using these deformed building blocks, and whether the enhancement can
be understood from the Jacobson--Morozov decomposition.

On the other hand, we stress the fact that using analytical continuation gives rise to
negative central charges does not necessarily mean that giving a vacuum
expectation value to the moment map with the corresponding nilpotent orbit is
forbidden. At the field theory level, there is \emph{a priori} nothing
preventing one to do so, and the apparent violation of unitarity should be
thought of more as a failure of Algorithm \ref{alg:I8IR} rather than an
obstruction to the corresponding deformation. Indeed we have assumed the quiver
to be long enough precisely to avoid these kinds of edge cases. It might well
be that there are additional modes decoupling, or that our prescription to find
the IR R-symmetry is not correct. Since the relations between the central
charges and the anomaly polynomial all involve R-symmetry terms, it might be
that a modification of the prescription in the case of short bases cures this apparent problem. Note that in
three dimensions something similar happens for ``ugly'' theories, where a naive
IR R-symmetry assignment for certain BPS monopoles lead to similar
``violations'' of unitarity \cite{Gaiotto:2008ak}. We leave a
systematic analysis of short quivers, including the question of these types of
breaking and analytic continuation for future work.

\subsection{The \texorpdfstring{$a$}{a}-theorem for Nilpotent Deformations}\label{sec:a-thm}

The $a$-anomaly has a preeminent role in the study of RG flows.
In two \cite{Zamolodchikov:1986gt} and four \cite{Komargodski:2011vj}
dimensions, an $a$-theorem has been shown: along an RG flow between two
(\emph{a priori} non-supersymmetric) CFTs, the coefficient $a$ decreases:
\begin{equation}\label{a-theorem}
		a^\text{UV} - a^\text{IR} > 0\,.
\end{equation}
This can be understood as a statement on the irreversibility of RG flows, as
$a$ is a measure of the number of degrees of freedom of the CFT. In six
dimensions, the question has been tackled using a background dilaton
\cite{Elvang:2012st, Baume:2013ika, Stergiou:2016uqq, Heckman:2021nwg}, but the
fate of a general $a$-theorem remains uncertain in even dimensions higher than
four. For 6d $(1,0)$ SCFTs, the $a$-theorem has been shown for tensor branch flows,
as well as for large classes of Higgs branch flows \cite{Heckman:2015axa,
Cordova:2015fha, Mekareeya:2016yal, Cordova:2020tij}.

Following the spirit of \cite{Heckman:2015axa, Mekareeya:2016yal}, we can use
our results to establish the $a$-theorem for nilpotent flows in a short and
concise way. Since the anomaly polynomial depends only on group-theoretical
quantities, we need only describe how the coefficients change as we go from one
theory to another. We will focus on conformal matter with a nilpotent deformation on one
side $A_{N-1}^{\mathfrak{g}}(O,\varnothing)$ for ease of exposition, but the
argument extends to all other theories. From equation
\eqref{shift-grav-coeffs}, the shifts in the relevant coefficients are given by:
\begin{equation}\label{shifts-CM-a-theorem}
		\begin{gathered}
		\alpha^\text{UV} - \alpha^\text{IR}  =\, 12\Gamma NI_{X} + \frac{12}{N}I_{X}^2 + 4(\beta^\text{UV}-\beta^\text{IR})
		 - \big(4\varphi_3(w) + \varphi_0(w) \big)\,,\\
		\beta^\text{UV} - \beta^\text{IR}  = \,  - \varphi_1(w) + \frac{1}{2}\varphi_0(w) \,,\\
		\gamma^\text{UV} - \gamma^\text{IR} =\, +\frac{7}{240}\text{dim}(O)\,,\qquad
		\delta^\text{UV} - \delta^\text{IR} =\, -\frac{1}{120}\text{dim}(O)\,.
		\end{gathered}
\end{equation}
Using equation \eqref{def-a-abcd}, the change in the quantity $a$ between the
two theories related by a nilpotent deformation is given by:
\begin{equation}
		a^\text{UV}-a^\text{IR}=\frac{16}{7}\left( (\alpha^\text{UV}-\alpha^\text{IR})-(\beta^\text{UV}-\beta^\text{IR})+(\gamma^\text{UV}-\gamma^\text{IR})+\frac{3}{8}(\delta^\text{UV}-\delta^\text{IR})\right)\,,
\end{equation}
The shifts of $\gamma,\delta$ depending only on the dimension of the nilpotent
orbit, their combined contribution is clearly positive. To establish the
$a$-theorem for nilpotent deformations, we therefore need only show that the
shift in $\alpha$ is always positive, while that of $\beta$ is always negative.
The signs of these shifts depends on the behavior of the function
$\varphi_n(w)$ defined in equation \eqref{def-varphi}, which we recall here for
convenience:
\begin{equation}
		\varphi_n(w) = \sum_{\alpha\in\Lambda^+}\left<\alpha,w\right>^n\,,
\end{equation}
As discussed around equation \eqref{def-wdd}, the entries of the vector $w$
defining the weighted Dynkin diagram labelling the nilpotent orbit correspond to the charges of the simple roots under the $\mathfrak{su}(2)_X$ Cartan element, which can only take
values $0, 1, 2$. One can then show that for a given nilpotent orbit with
weighted Dynkin diagram $w$, if $m\leq n$, then $\varphi_m(w)\leq\varphi_n(w)$.
The shift in $\beta$ is therefore always negative. On the other hand that of
$\alpha$ is more involved and depends on the embedding index $I_X$, which can
also be written as a function of $\varphi_2(w)$:
\begin{equation}
		I_X = \frac{1}{2}\left<w,w\right>=\frac{\varphi_2(w)}{2h^\vee_{\mathfrak{g}}}\,,
\end{equation}
as a consequence of the definition of the Killing form. As can be seen from Appendix \ref{app:nilp2TB}, long quivers must have $N\geq
r_\mathfrak{g}$.\footnote{For exceptional algebras, the long quiver condition
is $N\geq5$, but the bound on the embedding indices is also satisfied in that
case.} From the strange formula of Freudenthal and de Vries, one can then show that
$I_X<6r_\mathfrak{g}\Gamma$, and using $\varphi_3(w)\leq\varphi_2(w)^2$, it is
then straightforward to see that the shift in $\alpha$ is always positive for
long quivers. The contribution from $\gamma$ and $\delta$ depending only on the
dimension of the nilpotent orbits, this establishes the $a$-theorem for RG
flows between conformal matter and the SCFT associated with the nilpotent orbit
$O$.

We can however do better and consider flows between two theories in the Hasse
diagram of the corresponding flavor algebras. Under the partial ordering of
nilpotent orbits, see Section \ref{sec:nilpotent-deformations}, one can check
that if $O_1<O_2$ then $\varphi_n(w_1)\geq\varphi_n(w_2)$ for $n \leq 3$. This can be
understood as follows: as we go deeper in the Hasse diagram of a given algebra,
more and more roots are charged under the $\mathfrak{su}(2)$ subalgebra
defining the embedding $\rho_O$. The quantity $\varphi_n(w)$ being essentially
a positive weighted sum over those charges, this explains the relation. For
instance, in the case of $\varphi_0(w)$, this is immediate, and for
$\varphi_2(w)$ this can be understood as a consequence of the embedding index
$I_X$ growing along the Hasse diagram. 

Now, since the $a$-theorem is satisfied for any choice of nilpotent
deformations of the parent theory, these relations imply that if $O_1<O_2$, it
is also satisfied for an RG flow between
$A_{N-1}^{\mathfrak{g}}(O_1,\varnothing)$ and
$A_{N-1}^{\mathfrak{g}}(O_2,\varnothing)$.

This can be extended to any type of long quiver, including possible fractions,
generalizing  the results previously obtained in \cite{Mekareeya:2016yal}.
From these simple group-theoretical arguments, we have therefore shown that $a$
is monotonically decreasing as we go down in the Hasse diagram without needing
to ever refer to the tensor branch description or the F-theory construction.

We note that there can be theories $A_{N-1}^{\mathfrak{g}}(O,\varnothing)$ for
which $N<r_{\mathfrak{g}}$. From equation \eqref{shifts-CM-a-theorem}, one
will sometimes find that the value of $a^\text{UV}-a^\text{IR}$ becomes
negative. Those cases however correspond to analytically-continued theories in
the sense used in the previous subsection: the nilpotent orbit is too large and
there is no associated quiver describing the tensor branch of the theory.
Similar cases also occur when there is a deformation on both sides and one end
of the quiver is affecting the other.

\subsection{Beyond Nilpotent Orbits}\label{sec:future}

Algorithm \ref{alg:I8IR} heavily utilizes the properties of the
Jacobson--Morozov decomposition to find the IR anomaly polynomial, as well as
the fact that in the infrared, the R-symmetry is diagonal combination of the $\mathfrak{su}(2)_R$ UV
R-symmetry and the $\mathfrak{su}(2)_X$ subalgebra of the flavor symmetry. A
natural generalization would be to see if a similar prescription can be found
in the case of deformations of orbi-instantons related to embedding of ADE
discrete groups into $E_8$. In the case of
$\mathcal{O}^{\mathfrak{su}_K}_{N}(\sigma, \varnothing)$, closed-form
expressions for the gravitational and R-symmetry coefficients resembling those
appearing in equation \eqref{shift-grav-coeffs} are known
\cite{Mekareeya:2017jgc}. Indeed, in those cases the homomorphisms
$\sigma\in\operatorname{Hom}(\mathbb{Z}_K,E_8)$ are classified by
Kac labels, which can be interpreted as weighted Dynkin diagrams similar to
those labelling nilpotent orbits.

When the Kac labels are equivalent to the weighted Dynkin diagrams of nilpotent
orbits of $\mathfrak{e}_8$, applying Algorithm \ref{alg:I8IR} as if the
orbi-instanton deformation was nilpotent often, but not always, leads to the correct result. However,
this only occurs in a handful of cases given a choice of $K$, and Kac labels
have no generalizations to algebras of DE type. Moreover, expressing the IR
R-symmetry in terms of the UV data is more opaque in those cases, as we now
deal with embeddings of discrete groups rather than $\mathfrak{su}(2)$
subalgebras into the unbroken flavor symmetry. Finding an algorithm in terms of the Nambu--Goldstone modes decoupling from the
UV theory $\mathcal{O}_N^{\mathfrak{g}}$ would however help us better
understand the orbi-instanton theory. Indeed if there are modes beyond those
associated with the moment map becoming massive along the RG flow, this would
teach us about the other low-lying protected superconformal multiplets in its
gauge-invariant spectrum, and how they are related to the moment map. Having closed-form expressions for the anomaly polynomials of Higgsed orbi-instanton theories is particularly interesting as these 6d $(1,0)$ SCFTs are \emph{very Higgsable}, and thus their anomalies behave in a simple way under torus-compactification \cite{Ohmori:2015pua,Ohmori:2015pia,Mekareeya:2017jgc}. Such understanding would be especially useful to study torus-compactifications with non-trivial twists, such as Stiefel--Whitney twists, turned on along the torus \cite{Ohmori:2018ona,Giacomelli:2020jel,Heckman:2022suy}.  

Beyond superconformal theories, the techniques we have used throughout this
work can also be applied to Little String Theories (LSTs). These theories,
specific to six dimensions, describe strings decoupled from gravity and can
be realized in F-theory in a very similar way to SCFT, and also admit a
classification scheme \cite{Bhardwaj:2015oru}, see \cite{DelZotto:2023ahf}
for a concise review. In the case of heterotic LSTs, the structure of the Higgs branch flows are
similar to those of orbi-instantons, and has been under recent scrutiny, in
particular due to their connection to fiber-base duality
\cite{DelZotto:2022ohj, DelZotto:2022xrh, DelZotto:2023nrb, DelZotto:2023myd,
Ahmed:2023lhj, Lawrie:2023uiu}. For type-II LSTs however, the
flavor symmetries are severely constrained by both field- and string-theoretic
arguments \cite{Baume:2024oqn}, and therefore so are their Higgs branch flows. Since some of
the quantities relevant to the study of the duality are captured via anomalies \cite{Cordova:2020tij}, performing a similar analysis as in
this work for LSTs could shed additional light on how gauge-invariant
quantities are related under the duality from a bottom-up perspective. 

\subsection*{Acknowledgements}

We thank Hamza Ahmed, Ant\'onio Antunes, Chris Couzens, Jacques Distler, Jonathan Heckman, Monica Jinwoo Kang, Lorenzo Mansi, Paul-Konstantin Oehlmann, Fabian R\"uhle, and Matteo Sacchi for discussions. 
This work was partially
supported by the Deutsche Forschungsgemeinschaft under Germany’s Excellence Strategy --
EXC 2121 “Quantum Universe” -- 390833306, and the Collaborative Research Center - SFB
1624 “Higher Structures, Moduli Spaces, and Integrability” - 506632645. The work of F.B. was partly supported by the Swiss National Science Foundation (SNSF), grant
number P400P2\_194341, the German Research Foundation through a German-Israeli Project Cooperation (DIP) grant ``Holography and the Swampland''. The work of C.L.~is supported by DESY (Hamburg, Germany), a member of the Helmholtz Association HGF.

\appendix

\section{Characteristic Classes and Trace Relations}\label{app:representations}

In $d$ dimensions, the contribution of a left-handed Weyl fermion transforming
in a representation $\bm{R}$ to the anomaly polynomial is given by the
index of the Dirac operator \cite{AlvarezGaume:1983ig}, which can in turn
be written in terms of characteristic classes of the curvatures via the
Atiyah--Singer theorem:
\begin{equation}
		I_{d+2}^\text{fermion} = \left.\frac{1}{2}\widehat{A}(T)\text{ch}_{\bm{R}}(F)\right|_{d+2}\,.
\end{equation}
Note that we are using a convention giving only the contribution of the
representation $\bm{R}$, justifying the presence of the one-half factor. Other
conventions may not have this prefactor, as the counting includes the conjugate
representation: $\bm{R}\oplus\overline{\bm{R}}$. The A-roof genus $\widehat{A}(T)$
can be expanded in terms of Pontryagin classes $p_n(T)$ of the tangent bundle
of spacetime. Up to eighth order, one finds:
\begin{align}
		\widehat{A}(T) &= 1 - \frac{1}{24}p_1(T) + \frac{1}{5760}\big(7p_1(T)^2-4p_2(T)\big) + \,\cdots \,, \\
		L(T) &= 1 + \frac{1}{3}p_1(T) - \frac{1}{45}\big(p_1(T)^2-7p_2(T)\big) + \,\cdots \,,
\end{align}
where we also defined the Hirzebruch genus $L(T)$ appearing in the anomaly
polynomials of anti-symmetric chiral two-forms. Similarly, the Chern character
associated with the various gauge and flavor bundles can be expanded into
traces of their field strength $F$. As in \cite{Ohmori:2014kda}, we
follow a convention where $F$ is anti-Hermitian and rescaled to absorb the
usual factors of $(2\pi)$, so that the Chern character of non-Abelian algebras
is defined as:
\begin{equation}\label{def-chern-character}
		\text{ch}_{\bm{R}}(F) = \text{tr}_{\bm{R}}\, e^{iF} = \text{dim}(\bm{R}) - \frac{1}{2}\text{tr}_{\bm{R}}F^2 + \frac{1}{24}\text{tr}_{\bm{R}}F^4 + \cdots \,.
\end{equation}
For the special case of the R-symmetry $\mathfrak{su}(2)_R$ bundle, we have for the $d$-dimensional representation: 
\begin{equation}
		\text{ch}_{\bm{d}}(R) = d - \frac{d(d^2-1)}{6}c_2(R) + \frac{d(7-10d^2+3d^4)}{360}c_2(R)^2 + \cdots \,,\qquad c_2(R) = \frac{1}{4}\text{Tr}(R^2)\,,
\end{equation}
where $R$ is understood as the background field strength, and the traces are
one-instanton normalized. The prefactors are explained by the trace-relation
identities, see below. For instance, the anomaly polynomial of a free
left-handed fermion in six dimensions transforming as a doublet of the
$\mathfrak{su}(2)_R$ R-symmetry is given by 
\begin{equation}\label{I8-free-hyper}
		I_\text{free} = \frac{1}{2}\widehat{A}(T) \text{ch}_{\bm{2}}(R)\bigg|_\text{8-form} = \frac{1}{24}c_2(R)^2 + \frac{1}{48}c_2(R)p_1(T)+ \frac{1}{5760}(7p_1(T)^2 -4p_2(T)) \,.
\end{equation}
One of the advantages of writing the ``one-loop'' part of the anomaly in terms
of characteristic classes is that the Chern character satisfies a set of useful
properties under the tensor product and direct sum of representations, which
are utilized extensively in the main text:
\begin{gather}
		\text{ch}_{\bm{R}\otimes\bm{R}'}(F) = \text{ch}_{\bm{R}}(F)\,\text{ch}_{\bm{R}'}(F)\,,\qquad
		\text{ch}_{\bm{R}\oplus\bm{R}'}(F) = \text{ch}_{\bm{R}}(F)\,+\,\text{ch}_{\bm{R}'}(F)\,.
\end{gather}
Indeed, using these relations we find that under a given branching rule:
\begin{equation}
		\begin{aligned}
				\mathfrak{g}\, &\longrightarrow\,\,\mathfrak{g}'\,,\\
				\bm{R}  &\longrightarrow\,\bigoplus_\ell m_\ell\bm{R}_\ell\,,
		\end{aligned}
\end{equation}
where the representation $\bm{R}_\ell$ of $\mathfrak{g}'$ appears with a
possible non-trivial multiplicity, $m_\ell$, the character simply decomposes as
\begin{equation}
		\text{ch}_{\bm{R}}(F_{\mathfrak{g}}) = \sum_{\ell}m_\ell\, \text{ch}_{\bm{R}_\ell}(F_{\mathfrak{g}'})\,,
\end{equation}
where $F_{\mathfrak{g}}$, $F_{\mathfrak{g}'}$ are the field strengths
associated with $\mathfrak{g}$ and $\mathfrak{g}'$, respectively.

\subsection{Representation Indices and Trace Relations}\label{app:indices}

The various traces in the Chern character, as in equation
\eqref{def-chern-character}, must then be converted to one-instanton normalized
traces, $\text{Tr} F^n$. We again follow the conventions of
\cite{Ohmori:2014kda, Ohmori:2015pua}. For the adjoint representation, we
have:
\begin{equation}
		\text{tr}_\textbf{adj}F^2=h^\vee\,\text{Tr}F^2\,, \qquad 
		\text{Tr}F^2 = \text{Tr}(T_aT_b)\,F^a\wedge F^b\,,
\end{equation}
which fixes the overall normalization of the Killing form. For our purpose we
will only be interested in quadratic and quartic traces, and the trace
relations always take the generic form:
\begin{equation}\label{TrF2-app}
		\text{tr}_{\bm{R}} F^2 = A_{\bm{R}}\text{Tr}F^2\,,\qquad 
		\text{tr}_{\bm{R}} F^4 = B_{\bm{R}}\text{Tr}F^4 + C_{\bm{R}}(\text{Tr}F^2)^2\,.
\end{equation}
The quadratic index, $A_{\bm{R}}$, was first introduced by Dynkin
\cite{Dynkin:1957um}, and was generalized to higher order in
\cite{Patera:1976is}.

For specific cases, the trace relations defined in equation \eqref{TrF2-app}
are usually computed using algebra-specific relations or via so-called
\emph{Birdtrack} techniques \cite{Cvitanovic:2008zz}, and the results having
been tabulated for the most common representations, see, e.g.,
\cite{vanRitbergen:1998pn}, as well as \cite{Erler:1993zy, Heckman:2018jxk} for
applications in six dimensions specifically. While the literature is often
mostly concerned about the adjoint and fundamental representations, when
discussing nilpotent orbits we are led to deal with more exotic
representations. Furthermore as there are various normalizations for the
trace-relation coefficients in the literature, we now review how to obtain them
for arbitrary representations following the works pioneered by Okubo and Patera
\cite{Okubo:1981td, Okubo:1982dt, Okubo:1985qk}, enabling us to at the same
time set the conventions used in this work.

Representation indices are closely related to Casimir invariants, defined as
polynomial operators of a fixed degree commuting with all generators:
\begin{equation}
		C_p = g^{a_1\dots a_p}T_{a_1}\dots T_{a_p}\,,\qquad [C_p,T_a]=0\,,
\end{equation}
where $g_{a_1\dots a_p}$ is an invariant symmetric tensor and indices are
raised and lowered with the Killing metric, $g_{ab}= \text{Tr}(T_a T_b)$.

It is well-known that if the algebra $\mathfrak{g}$ is simple there are exactly
$r_\mathfrak{g}$ independent invariant tensors. It is then always possible to
choose those invariants so that they satisfy orthogonality relations, e.g.,
$g^{ab}g^{cd}g_{abcd}=0$. In that basis, the Casimir operators are unique up to
a normalization constant, and $C_p=0$ if there are no independent Casimir
operator of order $p$. There is then at most one Casimir operator for a given
$p$, except for $\mathfrak{g}=D_n$ which has two at order $n$. For $p=2,4$, the
quartic and quadratic invariants form a basis of the symmetrized traces over
the generator of any algebra $\mathfrak{g}\neq \mathfrak{so}(8)$ and we
therefore have the decomposition:
\begin{align}
		\text{tr}_{\bm{R}}F^2 &= \widetilde{\ell}_2(\bm{R}) g_{ab}F^a\wedge F^b\,,\\
		\text{tr}_{\bm{R}}F^4 &= \left(\widetilde{\ell}_4(\bm{R}) g_{abcd} + \widetilde{\ell}_{2,2}(\bm{R})g_{(ab}g_{cd)}\right)F^a\wedge F^b\wedge F^c\wedge F^d\,.\label{trF4-fund-indices}
\end{align}
The special case of $\mathfrak{g}=\mathfrak{so}(8)$ is treated below.  The
coefficients $\widetilde{\ell}_{n}$ were first studied in \cite{Okubo:1978qe} and
are called the fundamental indices of $\bm{R}$. They were furthermore
shown to be well behaved under branching rules and tensor products
\cite{Okubo:1978qe, Okubo:1981td, Okubo:1982dt}. They moreover can be obtained
directly from the weight system $W(\bm{R})$ of a representation. We first
define the quantities:
\begin{equation}\label{def-ell_2n}
		\begin{aligned}
		\ell_{2k}(\bm{R}) &= \sum_{\mu\in W(\bm{R})}\left<\mu,\mu\right>^k\,,\\
				K(\bm{R}) &= \frac{3}{2+\text{dim}(\mathfrak{g})}\left(\frac{\text{dim}(\mathfrak{g})}{\text{dim}(\bm{R})} - \frac{1}{6}\frac{\ell_2(\textbf{adj})}{\ell_2(\bm{R})}\right)\,,
		\end{aligned}
\end{equation}
with $\left<\cdot,\cdot\right>$ the pairing on the root space -- normalized such
that the longest root has length two. We note that $\ell_{2n}$ also defines
representation indices of order $2n$, but they are not fundamental in the sense that
they are not independent. The fundamental indices are instead given by
\cite{Okubo:1978qe, Okubo:1981td, Okubo:1982dt}: 
\begin{equation}\label{fund-indices}
    \begin{gathered}
		\widetilde{\ell}_2(\bm{R}) = \frac{\ell_2(\bm{R})}{2r_\mathfrak{g}}\,,\quad 
		\widetilde{\ell}_4(\bm{R}) = \ell_4(\bm{R}) - \frac{2 + r_\mathfrak{g}}{3r_\mathfrak{g}}K(\bm{R})(\ell_2(\bm{R}))^2\,,\\
		\widetilde{\ell}_{2,2}(\bm{R})=K(\bm{R})(\widetilde{\ell}_2(\bm{R}))^2\,.
    \end{gathered}
\end{equation}
From there we can compute the trace relations for any representation. For
quadratic traces, and using our normalization of the Killing form, one can
straightforwardly show that $\widetilde{\ell}_2(\bm{R})$ is equivalent to the usual definition of the Dynkin index:
\begin{equation}
		A_{\bm{R}} = \widetilde{\ell}_2(\bm{R}) = \frac{\ell_2(\bm{R})}{2r_\mathfrak{g}} 
		= \frac{\text{dim}(\bm{R})}{2\text{dim}(\mathfrak{g})} \left<\Lambda, \Lambda + 2\rho \right>\,,
\end{equation}
where $\Lambda$ is the highest weight of the representation, and satisfies
$A_{\bm{R}}=h^\vee_\mathfrak{g}$, as expected. At quartic order, we need to set a
reference representation, $\mathcal{F}$, such that $\text{tr}_\mathcal{F}(F^4)
= \text{Tr}F^4$, i.e., $B_\mathcal{F}=1$. The usual convention -- which we adopt
here -- is to choose the reference representation to be the so-called defining
representation for a classical algebra, namely the fundamental $\bm{n}$ of
$\mathfrak{su}(n)$, $\bm{2n}$ of $\mathfrak{sp}(n)$, and the vector $\bm{n}_v$
of $\mathfrak{so}(n)$.\footnote{In our convention, $\mathfrak{sp}(k)$ has rank
$k$, such that $\mathfrak{sp}(1)=\mathfrak{su}(2)$, and the fundamental
representation has dimension $2k$.} Comparing equation
\eqref{trF4-fund-indices} for both the desired and reference representations,
we finally find that when $\mathfrak{g}\neq\mathfrak{so}_8$:
\begin{equation}
		B_{\bm{R}} = \frac{\widetilde{\ell}_4(\bm{R})}{\widetilde{\ell}_4(\mathcal{F})}\,,\qquad
		C_{\bm{R}} = K(\bm{R}) (\widetilde{\ell}_2(\bm{R}))^2 - B_{\bm{R}} K(\mathcal{F}) (\widetilde{\ell}_2(\mathcal{F}))^2\,. 
\end{equation}
For exceptional algebras as well as $\mathfrak{su}(2)$ and $\mathfrak{su}(3)$,
there is no independent quartic Casimir, and the reference representation is
ill-defined. However, in those cases
$\widetilde{\ell}_4(\bm{R})=0=B_{\bm{R}}$ by construction and the formula
given above for $\mathcal{C}_{\bm{R}}$ also works for exceptional
algebras.

Given the machinery reviewed in this appendix, it is then straightforward to
compute the trace-relation indices for a given representation of any algebra,
as everything can be obtained from its weight lattice. It can then be achieved
in a programmatic way using dedicated software such as \texttt{LieART}
\cite{Feger:2012bs, Feger:2019tvk}, or by finding them in tables
\cite{mckay1977computation}. We note that for the latter, only the values of
$\ell_{2n}(\bm{R})$ defined in equation \eqref{def-ell_2n} are given, not
the fundamental indices.

\paragraph{The case of $\mathfrak{so}(8)$:} due to the presence of an additional
quartic Casimir, computing the fundamental indices is slightly more involved than in other cases.
However, in this work we only have to deal with the adjoint, vector, and
spinor representations of $\mathfrak{so}(8)$. For brevity, we have
simply collated the trace relations for those cases in Table \ref{tab:rep_so8}.
Additional details on the fourth-order Casimir invariants and their trace relations can be found in, e.g., \cite{Okubo:1981td}.

\begin{table}[H]
		\centering
		\begin{tabular}{cccrrc}
				\toprule
				$\bm{R}$ & $\bm{1}$ & $\bm{8}_v$  & $\bm{8}_s$ & $\bm{8}_c$& $\bm{28}$  \\\midrule
				$A_{\bm{R}}$ & $0$    & $1$ & $1$ & $1$                      & $6$       \\
				$B_{\bm{R}}$ & $0$    & $1$ & $-\frac{1}{2}$ & $-\frac{1}{2}$& $0$       \\
				$C_{\bm{R}}$ & $0$    & $0$ & $\frac{3}{8}$ & $\frac{3}{8}$  & $3$       \\
				\bottomrule
		\end{tabular}
		\caption{Trace-relation coefficients for representations of $\mathfrak{so}(8)$ relevant in this work.}
		\label{tab:rep_so8}
\end{table}

\section{Nilpotent Orbits and Their Branching Rules}\label{app:nilps}

As we have used repeatedly throughout this paper, each nilpotent orbit, $O$, of
a simple Lie algebra, $\mathfrak{g}$, is associated to a homomorphism $\rho_O:
\mathfrak{su}(2)_X \rightarrow \mathfrak{g}$ via the Jacobson--Morozov theorem.
Let $\mathfrak{f}$ be the centralizer in $\mathfrak{g}$ of the image of $\rho_O$.
Then each nilpotent orbit has an associated decomposition
\begin{equation}\label{eqn:JMdecomp}
  \mathfrak{g} \rightarrow \mathfrak{su}(2)_X \oplus \mathfrak{f} \,,
\end{equation}
where we choose to ignore the non-semi-simple part of $\mathfrak{f}$. In the
special case of the maximal nilpotent orbit, the homomorphism $\rho_O$ embeds
$\mathfrak{su}(2)_X$ trivially into $\mathfrak{g}$, and thus $\mathfrak{f} =
\mathfrak{g}$; despite this subtlety in this case, we keep the notation as in
equation \eqref{eqn:JMdecomp} for convenience. Under such a decomposition, we
are interested in the branching rule of the adjoint representation:
\begin{equation}
		\textbf{adj} \rightarrow \bigoplus_\ell \, (\bm{d}_\ell, \bm{R}_\ell) \,,
\end{equation}
where $\bm{d}_\ell$ are irreducible $\mathfrak{su}(2)_X$ representations, and
$\bm{R}_\ell$ are (not-necessarily-irreducible) representations of
$\mathfrak{f}$. 

First, we consider the case of $\mathfrak{g} = \mathfrak{su}(K)$.\footnote{When
$\mathfrak{g}$ is a classical Lie algebra, the adjoint branching rules induced
by nilpotent orbits is reviewed in detail in \cite{CKLL}.} Nilpotent orbits of
$\mathfrak{su}(K)$ are in one-to-one correspondence with integer partitions of
$K$. Given a nilpotent orbit $O$, we write the partition associated to $O$ in
the following way
\begin{equation}
    [1^{m_1}, 2^{m_2}, \cdots, K^{m_K}] \,, \quad \text{ where } \quad \sum_{i=1}^K i m_i = K \,.
\end{equation}
The branching rule of the fundamental representation under the decomposition as
in equation \eqref{eqn:JMdecomp} is given in terms of the partition as
\begin{equation}
    \bm{K} \rightarrow \bigoplus_{i=1}^K (\bm{i}, \bm{m_i}) \,.
\end{equation}
Here, we have used that the commutant of the image of $\mathfrak{su}(2)_X$ is 
\begin{equation}
    \mathfrak{f} = \,\bigoplus_{i=1}^K\, \mathfrak{su}(m_i) \,,
\end{equation}
and $\bm{m_i}$ is the representation of $\mathfrak{f}$ obtained by taking the
tensor product of the fundamental representation of $\mathfrak{su}(m_i)$ with
the singlet representation of all other factors. Once the branching rule of the
fundamental representation of $\mathfrak{su}(K)$ is known, the branching rules
of all other irreducible representations can be determined. In particular, the
branching rule of the adjoint representation can be straightforwardly derived
from the tensor product:
\begin{equation}
    \bm{K} \otimes \overline{\bm{K}} = \textbf{adj} \oplus \bm{1} \,.
\end{equation}

Similarly, we can consider the case of $\mathfrak{g} = \mathfrak{usp}(K)$, for
which nilpotent orbits are in one-to-one correspondence with C-partitions of
$K$.\footnote{To avoid the proliferation of half-integer quantities, in this
appendix we temporarily use the notation $\mathfrak{usp}(K)$ rather than
$\mathfrak{sp}(\frac{K}{2}$), recalling that we follow the convention where
$\mathfrak{usp}(2)=\mathfrak{sp}(1)=\mathfrak{su}(2)$. In the rest of the main
text, the notation $\mathfrak{sp}$ is used.} A C-partition of $K$ can be
written as
\begin{equation}\label{eqn:YYY}
    [1^{m_1}, 2^{m_2}, \cdots, K^{m_K}] \quad \text{ where } \quad \sum_{i=1}^K i m_i = K \quad \text{ such that } \quad i \text{ odd } \, \Rightarrow \, m_i \text{ even. }
\end{equation}
The commutant of the image of $\mathfrak{su}(2)_X$ is
\begin{equation}
    \mathfrak{f} = \bigoplus_{i=1}^K \mathfrak{j}(m_i) \qquad \text{ where } \qquad \mathfrak{j} = \begin{cases}
        \mathfrak{usp} \quad &\text{ if } i \text{ odd } \\
        \mathfrak{so} \quad &\text{ if } i \text{ even } \,,
    \end{cases}
\end{equation}
and the branching of the fundamental representation is
\begin{equation}\label{eqn:uspfunddecomp}
    \bm{K} \rightarrow \bigoplus_{i=1}^K (\bm{i}, \bm{m_i}) \,.
\end{equation}
Here, $\bm{m_i}$ is either the fundamental representation of
$\mathfrak{usp}(m_i)$ (if $i$ is odd) or the vector representation of
$\mathfrak{so}(m_i)$ (if $i$ is even), tensored with the trivial representation
of all other factors in $\mathfrak{f}$. The branching rule for the adjoint
representation then follows from
\begin{equation}
    \textbf{adj} = \operatorname{Sym}(\bm{K} \otimes \bm{K}) \,.
\end{equation}

Next, we can consider the case of $\mathfrak{g} = \mathfrak{so}(K)$. Each nilpotent orbit of $\mathfrak{so}(K)$ has an underlying BD-partition; a BD-partition of $K$ can be written as
\begin{equation}\label{eqn:AAA}
    [1^{m_1}, 2^{m_2}, \cdots, K^{m_K}] \quad \text{ where } \quad \sum_{i=1}^K i m_i = K \quad \text{ such that } \quad i \text{ even } \, \Rightarrow \, m_i \text{ even. }
\end{equation}
The commutant of the image of $\mathfrak{su}(2)_X$ is
\begin{equation}
    \mathfrak{f} = \bigoplus_{i=1}^K \mathfrak{j}(m_i) \qquad \text{ where } \qquad \mathfrak{j} = \begin{cases}
        \mathfrak{so} \quad &\text{ if } i \text{ odd } \\
        \mathfrak{usp} \quad &\text{ if } i \text{ even } \,,
    \end{cases}
\end{equation}
and the branching of the vector representation is
\begin{equation}
    \bm{K} \rightarrow \bigoplus_{i=1}^K (\bm{i}, \bm{m_i}) \,.
\end{equation}
Again, $\bm{m_i}$ is either the fundamental representation of
$\mathfrak{usp}(m_i)$ (if $i$ is even) or the vector representation of
$\mathfrak{so}(m_i)$ (if $i$ is odd), tensored with the trivial representation
of all other factors in $\mathfrak{f}$. The branching rule for the adjoint
representation then follows directly from
\begin{equation}
    \textbf{adj} = \operatorname{ASym}(\bm{K} \otimes \bm{K}) \,.
\end{equation}

Finally, we turn to the cases where $\mathfrak{g}$ is an exceptional Lie
algebra, where we use both the Bala--Carter notation \cite{bala1976classesI,
bala1976classesII} and the weighted Dynkin diagrams to label the nilpotent orbits. Recall that
a corollary of the Jacobson--Morozov theorem is that for each nilpotent orbit,
it is always possible to uniquely define an $\mathfrak{su}(2)$ triplet $(X,Y,H)$
of generators such that $H$ is in the Cartan subalgebra, and for which each of
the simple roots $E_i$ has eigenvalue $w_i=0,1,2$, see equation
\eqref{def-wdd}. Of course, this labelling then depends on the ordering of the
simple roots of the simple algebra $\mathfrak{g}$. We choose a basis where they
correspond to the rows of the Cartan matrix, $C$, and the weights defining the
nilpotent orbits are arranged as in Table \ref{tab:cartan}. Note that for a
classical algebra, one can obtain obtain the weighted Dynkin diagram of a
nilpotent orbit given directly from its partition \cite{collingwood2017nilpotent}.

\begin{table}[p]
	\centering
	\begin{tabular}{ccc}
		\toprule
		$\mathfrak{g}$ & $C$ & $w$ \\
		\midrule
		$\mathfrak{g}_2$ &  
		$\begin{pmatrix}
            2&  -3\\
            -1 & 2
		\end{pmatrix}$ & 
		$w_1\,w_2$\\
		\midrule
		$\mathfrak{f}_4$ &
		$\begin{pmatrix}
            2& -1& 0& 0\\
            -1& 2& -2& 0\\
            0& -1& 2& -1\\
            0& 0& -1& 2
		\end{pmatrix}$ &
		$w_1\,w_2\,w_3\,w_4$\\
		\midrule
		$\mathfrak{e}_6$ & 
		$\begin{pmatrix}
            2& -1& 0& 0& 0& 0\\
            -1& 2& -1& 0& 0& 0\\
            0& -1& 2& -1& 0& -1\\
            0& 0& -1& 2& -1& 0\\
            0& 0& 0& -1& 2& 0\\
            0& 0& -1& 0& 0& 2
		\end{pmatrix}$ &
		$w_1\,w_2\,\overset{\displaystyle{w}_6}{w_3}\,w_4\,w_5$\\
		\midrule
		$\mathfrak{e}_7$ & 
		$\begin{pmatrix}
            2& -1& 0& 0& 0& 0& 0\\
            -1& 2& -1& 0& 0& 0& 0\\
            0& -1& 2& -1& 0& 0& -1\\
            0& 0& -1& 2& -1& 0& 0\\
            0& 0& 0& -1& 2& -1& 0\\
            0& 0& 0& 0& -1& 2& 0\\
            0& 0& -1& 0& 0& 0& 2
		\end{pmatrix}$ &
		$w_1\,w_2\,\overset{\displaystyle{w}_7}{w_3}\,w_4\,w_5\,w_6$\\
		\midrule
		$\mathfrak{e}_8$ & 
		$\begin{pmatrix}
            2& -1& 0& 0& 0& 0& 0& 0\\
            -1& 2& -1& 0& 0& 0& 0& 0\\
            0& -1& 2& -1& 0& 0& 0& -1\\
            0& 0& -1& 2& -1& 0& 0& 0\\
            0& 0& 0& -1& 2& -1& 0& 0\\
            0& 0& 0& 0& -1& 2& -1& 0\\
            0& 0& 0& 0& 0& -1& 2& 0\\
            0& 0& -1& 0& 0& 0& 0& 2
		\end{pmatrix}$ &
		$w_1\,w_2\,\overset{\displaystyle{w}_8}{w_3}\,w_4\,w_5\,w_6\,w_7$\\
	\bottomrule
	\end{tabular}
	\caption{Cartan matrices for the exceptional algebras, with our choice of
	ordering for the weighted Dynkin diagrams uniquely labelling every
	nilpotent orbit.}
	\label{tab:cartan}
\end{table}

For exceptional algebras, the Jacobson--Morozov decomposition involves
representations that go beyond the usual adjoint, and defining representations.
The associated branching rules can be found in, e.g., \cite{Chacaltana:2014jba,
		Chacaltana:2015bna, Chacaltana:2016shw, Chacaltana:2017boe,
Chacaltana:2018vhp}, and have been reproduced in Tables
\ref{tbl:nilpDecompG2}--\ref{tbl:nilpDecompE8}. For convenience, we also
included the weighted Dynkin diagram of the nilpotent orbits following the above
ordering, as well as the Dynkin embedding index for each flavor factor. 

\newpage

\begin{longtable}{>{\small}c>{\small}c>{\small}l>{\small}c>{\small}l}
    \toprule
	$w$ & Bala--Carter & $\mathfrak{g}_2\to\mathfrak{su}_2^{(I_\mathcal{O})}\oplus \mathfrak{f}^{(I)}$ & $\text{dim}\mathcal{O}$ & $\textbf{adj}\to \bigoplus_i(\bm{d}_i,\bm{R}_i)$
    \\\midrule
        $[0, 0]$ & $0$ & $\mathfrak{su}_2^{(0)} \oplus\,\mathfrak{g}_{2}^{(1)}$ & $0$ & $(\bm{1}, \bm{14})$\\
        $[1, 0]$ & $A_1$ & $\mathfrak{su}_2^{(1)} \oplus\,\mathfrak{su}_{2}^{(3)}$ & $6$ & $(\bm{3}, \bm{1}) \oplus (\bm{2}, \bm{4}) \oplus (\bm{1}, \bm{3})$\\
        $[0, 1]$ & $\tilde{A}_1$ & $\mathfrak{su}_2^{(3)} \oplus\,\mathfrak{su}_{2}^{(1)}$ & $8$ & $(\bm{3}, \bm{1}) \oplus (\bm{4}, \bm{2}) \oplus (\bm{1}, \bm{3})$\\
        $[2, 0]$ & $G_2(a_1)$ & $\mathfrak{su}_2^{(4)} \oplus\,\varnothing$ & $10$ & $\bm{5} \oplus 3\cdot\bm{3}$\\
        $[2, 2]$ & $G_2$ & $\mathfrak{su}_2^{(28)} \oplus\,\varnothing$ & $12$ & $\bm{11} \oplus \bm{3}$\\
    \caption{Nilpotent orbits of $\mathfrak{g}_{2}$.}
    \label{tbl:nilpDecompG2}
\end{longtable}

\begin{landscape}
\begin{longtable}{>{\small}c>{\small}c>{\small}l>{\small}c>{\small}l}
    \toprule
    $w$ & Bala--Carter & $\mathfrak{f}_4\to\mathfrak{su}_2^{(I_\mathcal{O})}\oplus \mathfrak{f}^{(I)}$ & $\text{dim}\mathcal{O}$ & $\textbf{adj}\to \bigoplus_i(\bm{d}_i,\bm{R}_i)$
    \\\midrule
        $[0, 0, 0, 0]$ & $0$ & $\mathfrak{su}_2^{(0)} \oplus\,\mathfrak{f}_{4}^{(1)}$ & $0$ & $(\bm{1}, \bm{52})$\\
        $[1, 0, 0, 0]$ & $A_1$ & $\mathfrak{su}_2^{(1)} \oplus\,\mathfrak{sp}_{3}^{(1)}$ & $16$ & $(\bm{3}, \bm{1}) \oplus (\bm{2}, \bm{14'}) \oplus (\bm{1}, \bm{21})$\\
        $[0, 0, 0, 1]$ & $\tilde{A}_1$ & $\mathfrak{su}_2^{(2)} \oplus\,\mathfrak{su}_{4}^{(1)}$ & $22$ & $(\bm{3}, \bm{1}) \oplus (\bm{3}, \bm{6}) \oplus (\bm{2}, \bm{4}) \oplus (\bm{2}, \bm{\bar{4}}) \oplus (\bm{1}, \bm{15})$\\
        $[0, 1, 0, 0]$ & $A_1+\tilde{A}_1$ & $\mathfrak{su}_2^{(3)} \oplus\, \big(\mathfrak{su}_{2}^{(1)}\,\oplus\,\mathfrak{su}_{2}^{(2)} \big)$ & $28$ & $(\bm{1}, \bm{3}, \bm{1}) \oplus (\bm{1}, \bm{1}, \bm{3}) \oplus (\bm{2}, \bm{5}, \bm{2}) \oplus (\bm{3}, \bm{1}, \bm{1}) \oplus (\bm{3}, \bm{5}, \bm{1}) \oplus (\bm{4}, \bm{1}, \bm{2})$\\
        $[2, 0, 0, 0]$ & $A_2$ & $\mathfrak{su}_2^{(4)} \oplus\,\mathfrak{su}_{3}^{(2)}$ & $30$ & $(\bm{3}, \bm{1}) \oplus (\bm{3}, \bm{6}) \oplus (\bm{3}, \bm{\bar{6}}) \oplus (\bm{5}, \bm{1}) \oplus (\bm{1}, \bm{8})$\\
        $[0, 0, 0, 2]$ & $\tilde{A}_2$ & $\mathfrak{su}_2^{(8)} \oplus\,\mathfrak{g}_{2}^{(1)}$ & $30$ & $(\bm{3}, \bm{1}) \oplus (\bm{5}, \bm{7}) \oplus (\bm{1}, \bm{14})$\\
        $[0, 0, 1, 0]$ & $A_2+\tilde{A}_1$ & $\mathfrak{su}_2^{(6)} \oplus\,\mathfrak{su}_{2}^{(6)}$ & $34$ & $(\bm{3}, \bm{1}) \oplus (\bm{3}, \bm{5}) \oplus (\bm{5}, \bm{3}) \oplus (\bm{4}, \bm{2}) \oplus (\bm{2}, \bm{4}) \oplus (\bm{1}, \bm{3})$\\
\end{longtable}

\newpage
\begin{longtable}{>{\small}c>{\small}c>{\small}l>{\small}c>{\small}l}
    \toprule
    $w$ & Bala--Carter & $\mathfrak{e}_6\to\mathfrak{su}_2^{(I_\mathcal{O})}\oplus \mathfrak{f}^{(I)}$ & $\text{dim}\mathcal{O}$ & $\textbf{adj}\to \bigoplus_i(\bm{d}_i,\bm{R}_i)$
    \\\midrule
        $[0, 0, 0, 0, 0, 0]$ & $0$ & $\mathfrak{su}_2^{(0)} \oplus\,\mathfrak{e}_{6}^{(1)}$ & $0$ & $(\bm{1}, \bm{78})$\\
        $[0, 0, 0, 0, 0, 1]$ & $A_1$ & $\mathfrak{su}_2^{(1)} \oplus\,\mathfrak{su}_{6}^{(1)}$ & $22$ & $(\bm{1}, \bm{35}) \oplus (\bm{2}, \bm{20}) \oplus (\bm{3}, \bm{1})$\\
        $[1, 0, 0, 0, 1, 0]$ & $2A_1$ & $\mathfrak{su}_2^{(2)} \oplus\,\mathfrak{so}_{7}^{(1)}$ & $32$ & $(\bm{1}, \bm{21}) \oplus (\bm{1}, \bm{1}) \oplus 2\cdot(\bm{2}, \bm{8}) \oplus (\bm{3}, \bm{7}) \oplus (\bm{3}, \bm{1})$\\
        $[0, 0, 1, 0, 0, 0]$ & $3A_1$ & $\mathfrak{su}_2^{(3)} \oplus\, \big(\mathfrak{su}_{2}^{(1)}\,\oplus\,\mathfrak{su}_{3}^{(2)} \big)$ & $40$ & $(\bm{1}, \bm{1}, \bm{8}) \oplus (\bm{1}, \bm{3}, \bm{1}) \oplus (\bm{2}, \bm{2}, \bm{8}) \oplus (\bm{3}, \bm{1}, \bm{1}) \oplus (\bm{3}, \bm{1}, \bm{8}) \oplus (\bm{4}, \bm{2}, \bm{1})$\\
        $[0, 0, 0, 0, 0, 2]$ & $A_2$ & $\mathfrak{su}_2^{(4)} \oplus\, \big(\mathfrak{su}_{3}^{(1)}\,\oplus\,\mathfrak{su}_{3}^{(1)} \big)$ & $42$ & $(\bm{1}, \bm{8}, \bm{1}) \oplus (\bm{1}, \bm{1}, \bm{8}) \oplus (\bm{3}, \bm{1}, \bm{1}) \oplus (\bm{3}, \bm{3}, \bm{3}) \oplus (\bm{3}, \bm{\bar{3}}, \bm{\bar{3}}) \oplus (\bm{5}, \bm{1}, \bm{1})$\\
        $[1, 0, 0, 0, 1, 1]$ & $A_2+A_1$ & $\mathfrak{su}_2^{(5)} \oplus\,\mathfrak{su}_{3}^{(1)}$ & $46$ & $(\bm{1}, \bm{1}) \oplus (\bm{1}, \bm{8}) \oplus (\bm{2}, \bm{3}) \oplus (\bm{2}, \bm{\bar{3}}) \oplus 2\cdot(\bm{2}, \bm{1}) \oplus (\bm{3}, \bm{3}) \oplus (\bm{3}, \bm{\bar{3}}) \oplus 2\cdot(\bm{3}, \bm{1}) $\\
		& & & & $\quad\oplus\, (\bm{4}, \bm{3}) \oplus (\bm{4}, \bm{\bar{3}}) \oplus (\bm{5}, \bm{1})$\\
\end{longtable}

\newpage
\begin{longtable}{>{\small}c>{\small}c>{\small}l>{\small}c>{\small}l}
    \toprule
	$w$ & Bala--Carter & $\mathfrak{e}_7\to\mathfrak{su}_2^{(I_\mathcal{O})}\oplus \mathfrak{f}^{(I)}$ & $\text{dim}\mathcal{O}$ & $\textbf{adj}\to \bigoplus_i(\bm{d}_i,\bm{R}_i)$
    \\\midrule
        $[0, 0, 0, 0, 0, 0, 0]$ & $0$ & $\mathfrak{su}_2^{(0)} \oplus\,\mathfrak{e}_{7}^{(1)}$ & $0$ & $(\bm{1}, \bm{133})$\\
        $[1, 0, 0, 0, 0, 0, 0]$ & $A_1$ & $\mathfrak{su}_2^{(1)} \oplus\,\mathfrak{so}_{12}^{(1)}$ & $34$ & $(\bm{1}, \bm{66}) \oplus (\bm{2}, \bm{\bar{32}}) \oplus (\bm{3}, \bm{1})$\\
        $[0, 0, 0, 0, 1, 0, 0]$ & $2A_1$ & $\mathfrak{su}_2^{(2)} \oplus\, \big(\mathfrak{so}_{9}^{(1)}\,\oplus\,\mathfrak{su}_{2}^{(1)} \big)$ & $52$ & $(\bm{1}, \bm{1}, \bm{3}) \oplus (\bm{1}, \bm{36}, \bm{1}) \oplus (\bm{2}, \bm{16}, \bm{2}) \oplus (\bm{3}, \bm{1}, \bm{1}) \oplus (\bm{3}, \bm{9}, \bm{1})$\\
        $[0, 1, 0, 0, 0, 0, 0]$ & $(3A_1)'$ & $\mathfrak{su}_2^{(3)} \oplus\, \big(\mathfrak{sp}_{3}^{(1)}\,\oplus\,\mathfrak{su}_{2}^{(1)} \big)$ & $64$ & $(\bm{1}, \bm{1}, \bm{3}) \oplus (\bm{1}, \bm{21}, \bm{1}) \oplus (\bm{2}, \bm{14}, \bm{2}) \oplus (\bm{3}, \bm{1}, \bm{1}) \oplus (\bm{3}, \bm{14}, \bm{1}) \oplus (\bm{4}, \bm{1}, \bm{2})$\\
        $[0, 0, 0, 0, 0, 2, 0]$ & $(3A_1)''$ & $\mathfrak{su}_2^{(3)} \oplus\,\mathfrak{f}_{4}^{(1)}$ & $54$ & $(\bm{1}, \bm{52}) \oplus (\bm{3}, \bm{1}) \oplus (\bm{3}, \bm{26})$\\
        $[0, 0, 0, 0, 0, 1, 1]$ & $4A_1$ & $\mathfrak{su}_2^{(4)} \oplus\,\mathfrak{sp}_{3}^{(1)}$ & $70$ & $(\bm{1}, \bm{21}) \oplus (\bm{2}, \bm{6}) \oplus (\bm{2}, \bm{14'}) \oplus 2\cdot(\bm{3}, \bm{1}) \oplus (\bm{3}, \bm{14}) \oplus (\bm{4}, \bm{6})$\\
        $[2, 0, 0, 0, 0, 0, 0]$ & $A_2$ & $\mathfrak{su}_2^{(4)} \oplus\,\mathfrak{su}_{6}^{(1)}$ & $66$ & $(\bm{1}, \bm{35}) \oplus (\bm{3}, \bm{1}) \oplus (\bm{3}, \bm{15}) \oplus (\bm{3}, \bm{\bar{15}}) \oplus (\bm{5}, \bm{1})$\\
\end{longtable}

\newpage
\begin{longtable}{>{\small}c>{\small}c>{\small}l>{\small}c>{\small}l}
    \toprule
	$w$ & Bala--Carter & $\mathfrak{e}_8\to\mathfrak{su}_2^{(I_\mathcal{O})}\oplus \mathfrak{f}^{(I)}$ & $\text{dim}\mathcal{O}$ & $\textbf{adj}\to \bigoplus_i(\bm{d}_i,\bm{R}_i)$
    \\\midrule
        $[0, 0, 0, 0, 0, 0, 0, 0]$ & $0$ & $\mathfrak{su}_2^{(0)} \oplus\,\mathfrak{e}_{8}^{(1)}$ & $0$ & $(\bm{1}, \bm{248})$\\
        $[0, 0, 0, 0, 0, 0, 1, 0]$ & $A_1$ & $\mathfrak{su}_2^{(1)} \oplus\,\mathfrak{e}_{7}^{(1)}$ & $58$ & $(\bm{3}, \bm{1}) \oplus (\bm{2}, \bm{56}) \oplus (\bm{1}, \bm{133})$\\
        $[1, 0, 0, 0, 0, 0, 0, 0]$ & $2A_1$ & $\mathfrak{su}_2^{(2)} \oplus\,\mathfrak{so}_{13}^{(1)}$ & $92$ & $(\bm{1}, \bm{78}) \oplus (\bm{2}, \bm{64}) \oplus (\bm{3}, \bm{1}) \oplus (\bm{3}, \bm{13})$\\
        $[0, 0, 0, 0, 0, 1, 0, 0]$ & $3A_1$ & $\mathfrak{su}_2^{(3)} \oplus\, \big(\mathfrak{f}_{4}^{(1)}\,\oplus\,\mathfrak{su}_{2}^{(1)} \big)$ & $112$ & $(\bm{1}, \bm{1}, \bm{3}) \oplus (\bm{1}, \bm{52}, \bm{1}) \oplus (\bm{2}, \bm{26}, \bm{2}) \oplus (\bm{3}, \bm{1}, \bm{1}) \oplus (\bm{3}, \bm{26}, \bm{1}) \oplus (\bm{4}, \bm{1}, \bm{2})$\\
        $[0, 0, 0, 0, 0, 0, 2, 0]$ & $A_2$ & $\mathfrak{su}_2^{(4)} \oplus\,\mathfrak{e}_{6}^{(1)}$ & $114$ & $(\bm{1}, \bm{78}) \oplus (\bm{3}, \bm{1}) \oplus (\bm{3}, \bm{27}) \oplus (\bm{3}, \bm{\bar{27}}) \oplus (\bm{5}, \bm{1})$\\
        $[0, 0, 0, 0, 0, 0, 0, 1]$ & $4A_1$ & $\mathfrak{su}_2^{(4)} \oplus\,\mathfrak{sp}_{4}^{(1)}$ & $128$ & $(\bm{1}, \bm{36}) \oplus (\bm{2}, \bm{48}) \oplus (\bm{3}, \bm{1}) \oplus (\bm{3}, \bm{27}) \oplus (\bm{4}, \bm{8})$\\
        $[1, 0, 0, 0, 0, 0, 1, 0]$ & $A_2+A_1$ & $\mathfrak{su}_2^{(5)} \oplus\,\mathfrak{su}_{6}^{(1)}$ & $136$ & $(\bm{1}, \bm{35}) \oplus (\bm{2}, \bm{6}) \oplus (\bm{2}, \bm{\bar{6}}) \oplus (\bm{2}, \bm{20}) \oplus 2\cdot(\bm{3}, \bm{1}) \oplus (\bm{3}, \bm{15}) \oplus (\bm{3}, \bm{\bar{15}}) $\\
\end{longtable}

\end{landscape}

\section{Hasse Diagrams}\label{app:hasse}

For completeness, in this appendix we collate the Hasse diagrams of the nilpotent
orbits associated with the Lie algebras appearing in long quivers of
exceptional types. For each orbit, we give the weighted Dynkin diagram and the
Jacobson--Morozov decomposition
$\mathfrak{g}\to\mathfrak{su}(2)_X\oplus\mathfrak{f}$. The superscripts indicate
the Dynkin embedding index of each factor.  For the reader's convenience, individual
versions of all the Hasse diagrams can be found as ancillary files in the arXiv
version of this work.

For completeness, we have also included the name of the transverse Slodowy
slices between nilpotent orbits, following the notation introduced by Kraft and
Procesi \cite{kraft1981minimal}, see also \cite{fu2017generic} for exceptional
cases. Transitions colored in blue indicate than in the associated quiver,
another minimal conformal matter is affected, and the transition is not allowed
if the quiver is too short.

\begin{figure}[H]
		\centering
        \resizebox{!}{.45\paperheight}{
		\begin{subfigure}[b]{.2\textwidth}
		\centering
			\begin{tikzpicture}[>=latex,line join=bevel,]
\node (0) at (31.239bp,126.74bp) [draw,rectangle] {\begin{tabular}{c}$[1^2]$\\[1mm] $\scalebox{.8}{[0]}$\\[.5mm]\scalebox{.8}{$\mathfrak{su}_2^{(0)} \oplus\,\mathfrak{su}_{2}^{(1)}$}\end{tabular}};
  \node (1) at (31.239bp,23.579bp) [draw,rectangle] {\begin{tabular}{c}$[2]$\\[1mm] $\scalebox{.8}{[2]}$\\[.5mm]\scalebox{.8}{$\mathfrak{su}_2^{(1)} \oplus\,\varnothing$}\end{tabular}};
  \draw [->] (0) ..controls (31.239bp,89.589bp) and (31.239bp,72.78bp)  .. (1);
  \definecolor{strokecol}{rgb}{0.0,0.0,0.0};
  \pgfsetstrokecolor{strokecol}
  \draw (40.24bp,69.86bp) node[right] {$A_1$};
\end{tikzpicture}
			\caption{$\mathfrak{g}=\mathfrak{su}_2$}
		\end{subfigure}
		\begin{subfigure}[b]{.25\textwidth}
		\centering
			\begin{tikzpicture}[>=latex,line join=bevel,]
\node (0) at (31.239bp,226.89bp) [draw,rectangle] {\begin{tabular}{c}$[1^3]$\\[1mm] $\scalebox{.8}{[0, 0]}$\\[.5mm]\scalebox{.8}{$\mathfrak{su}_2^{(0)} \oplus\,\mathfrak{su}_{3}^{(1)}$}\end{tabular}};
  \node (1) at (31.239bp,126.74bp) [draw,rectangle] {\begin{tabular}{c}$[2,1]$\\[1mm] $\scalebox{.8}{[1, 1]}$\\[.5mm]\scalebox{.8}{$\mathfrak{su}_2^{(1)} \oplus\,\varnothing$}\end{tabular}};
  \node (2) at (31.239bp,23.579bp) [draw,rectangle] {\begin{tabular}{c}$[3]$\\[1mm] $\scalebox{.8}{[2, 2]}$\\[.5mm]\scalebox{.8}{$\mathfrak{su}_2^{(4)} \oplus\,\varnothing$}\end{tabular}};
  \draw [->] (0) ..controls (31.239bp,190.83bp) and (31.239bp,175.4bp)  .. (1);
  \definecolor{strokecol}{rgb}{0.0,0.0,0.0};
  \pgfsetstrokecolor{strokecol}
  \draw (38.74bp,170.01bp) node[right] {$a_2$};
  \draw [->] (1) ..controls (31.239bp,89.589bp) and (31.239bp,72.78bp)  .. (2);
  \draw (40.24bp,69.86bp) node[right] {$A_2$};
\end{tikzpicture}
			\caption{$\mathfrak{g}=\mathfrak{su}_3$}
		\end{subfigure}
		\begin{subfigure}[b]{.25\textwidth}
		\centering
			\begin{tikzpicture}[>=latex,line join=bevel,]
\node (0) at (49.195bp,638.52bp) [draw,rectangle] {\begin{tabular}{c}$[1^7]$\\[1mm] $\scalebox{.8}{[0, 0, 0]}$\\[.5mm]\scalebox{.8}{$\mathfrak{su}_2^{(0)} \oplus\,\mathfrak{so}_{7}^{(1)}$}\end{tabular}};
  \node (1) at (49.195bp,535.36bp) [draw,rectangle] {\begin{tabular}{c}$[2^2,1^3]$\\[1mm] $\scalebox{.8}{[0, 1, 0]}$\\[.5mm]\scalebox{.8}{$\mathfrak{su}_2^{(1)} \oplus\, \big(\mathfrak{su}_{2}^{(1)}\,\oplus\,\mathfrak{su}_{2}^{(2)} \big)$}\end{tabular}};
  \node (2) at (49.195bp,435.21bp) [draw,rectangle] {\begin{tabular}{c}$[3, 1^4]$\\[1mm] $\scalebox{.8}{[2, 0, 0]}$\\[.5mm]\scalebox{.8}{$\mathfrak{su}_2^{(2)} \oplus\, \big(\mathfrak{su}_{2}^{(1)}\,\oplus\,\mathfrak{su}_{2}^{(1)} \big)$}\end{tabular}};
  \node (3) at (49.195bp,333.05bp) [draw,rectangle] {\begin{tabular}{c}$[3, 2^2]$\\[1mm] $\scalebox{.8}{[1, 0, 1]}$\\[.5mm]\scalebox{.8}{$\mathfrak{su}_2^{(3)} \oplus\,\mathfrak{su}_{2}^{(1)}$}\end{tabular}};
  \node (4) at (49.195bp,229.89bp) [draw,rectangle] {\begin{tabular}{c}$[3^2, 1]$\\[1mm] $\scalebox{.8}{[0, 2, 0]}$\\[.5mm]\scalebox{.8}{$\mathfrak{su}_2^{(4)} \oplus\,\varnothing$}\end{tabular}};
  \node (5) at (49.195bp,126.74bp) [draw,rectangle] {\begin{tabular}{c}$[5, 1^2]$\\[1mm] $\scalebox{.8}{[2, 2, 0]}$\\[.5mm]\scalebox{.8}{$\mathfrak{su}_2^{(10)} \oplus\,\varnothing$}\end{tabular}};
  \node (6) at (49.195bp,23.579bp) [draw,rectangle] {\begin{tabular}{c}$[7]$\\[1mm] $\scalebox{.8}{[2, 2, 2]}$\\[.5mm]\scalebox{.8}{$\mathfrak{su}_2^{(28)} \oplus\,\varnothing$}\end{tabular}};
  \draw [->] (0) ..controls (49.195bp,601.37bp) and (49.195bp,584.57bp)  .. (1);
  \definecolor{strokecol}{rgb}{0.0,0.0,0.0};
  \pgfsetstrokecolor{strokecol}
  \draw (56.19bp,581.64bp) node[right] {$b_3$};
  \draw [->] (1) ..controls (49.195bp,499.3bp) and (49.195bp,483.87bp)  .. (2);
  \draw (56.19bp,478.49bp) node[right] {$c_1$};
  \draw [->] (2) ..controls (49.195bp,398.64bp) and (49.195bp,382.33bp)  .. (3);
  \draw (59.19bp,378.33bp) node[right] {$2a_1$};
  \draw [->] (3) ..controls (49.195bp,295.9bp) and (49.195bp,279.09bp)  .. (4);
  \draw (58.19bp,276.17bp) node[right] {$A_1$};
  \draw [->] (4) ..controls (49.195bp,192.75bp) and (49.195bp,175.94bp)  .. (5);
  \draw (58.19bp,173.01bp) node[right] {$A_3$};
  \draw [blue,->] (5) ..controls (49.195bp,89.589bp) and (49.195bp,72.78bp)  .. (6);
  \draw (58.19bp,69.86bp) node[right] {$A_5$};
\end{tikzpicture}
			\caption{$\mathfrak{g}=\mathfrak{so}_7$}
		\end{subfigure}
		\begin{subfigure}[b]{.25\textwidth}
		\centering
			\begin{tikzpicture}[>=latex,line join=bevel,]
\node (0) at (31.239bp,431.55bp) [draw,rectangle] {\begin{tabular}{c}$0$\\[1mm] $\scalebox{.8}{[0, 0]}$\\[.5mm]\scalebox{.8}{$\mathfrak{su}_2^{(0)} \oplus\,\mathfrak{g}_{2}^{(1)}$}\end{tabular}};
  \node (1) at (31.239bp,330.39bp) [draw,rectangle] {\begin{tabular}{c}$A_1$\\[1mm] $\scalebox{.8}{[1, 0]}$\\[.5mm]\scalebox{.8}{$\mathfrak{su}_2^{(1)} \oplus\,\mathfrak{su}_{2}^{(3)}$}\end{tabular}};
  \node (2) at (31.239bp,230.56bp) [draw,rectangle] {\begin{tabular}{c}$\widetilde{A}_1$\\[1mm] $\scalebox{.8}{[0, 1]}$\\[.5mm]\scalebox{.8}{$\mathfrak{su}_2^{(3)} \oplus\,\mathfrak{su}_{2}^{(1)}$}\end{tabular}};
  \node (3) at (31.239bp,126.74bp) [draw,rectangle] {\begin{tabular}{c}$G_2(a_1)$\\[1mm] $\scalebox{.8}{[2, 0]}$\\[.5mm]\scalebox{.8}{$\mathfrak{su}_2^{(4)} \oplus\,\varnothing$}\end{tabular}};
  \node (4) at (31.239bp,23.579bp) [draw,rectangle] {\begin{tabular}{c}$G_2$\\[1mm] $\scalebox{.8}{[2, 2]}$\\[.5mm]\scalebox{.8}{$\mathfrak{su}_2^{(28)} \oplus\,\varnothing$}\end{tabular}};
  \draw [->] (0) ..controls (31.239bp,394.91bp) and (31.239bp,379.0bp)  .. (1);
  \definecolor{strokecol}{rgb}{0.0,0.0,0.0};
  \pgfsetstrokecolor{strokecol}
  \draw (38.74bp,374.67bp) node[right] {$g_2$};
  \draw [->] (1) ..controls (31.239bp,294.62bp) and (31.239bp,279.51bp)  .. (2);
  \draw (38.24bp,273.52bp) node[right] {$m$};
  \draw [->] (2) ..controls (31.239bp,192.65bp) and (31.239bp,175.83bp)  .. (3);
  \draw (40.24bp,173.01bp) node[right] {$A_1$};
  \draw [blue,->] (3) ..controls (31.239bp,89.589bp) and (31.239bp,72.78bp)  .. (4);
  \draw (40.24bp,69.86bp) node[right] {$G_2$};
\end{tikzpicture}
			\caption{$\mathfrak{g}=\mathfrak{g}_2$}
		\end{subfigure}
        }
		\caption{Hasse diagram for the nilpotent orbit of various low-rank algebras. Each orbit is labelled by its partition or Bala--Carter label for classical or exceptional algebras, respectively.}
		\label{fig:Hasse-1}
\end{figure}

\begin{figure}[H]
		\centering
		\resizebox{!}{.7\paperheight}{
			\begin{tikzpicture}[>=latex,line join=bevel,]
\node (0) at (119.78bp,1297.3bp) [draw,rectangle] {\begin{tabular}{c}$0$\\[1mm] $\scalebox{.8}{[0, 0, 0, 0]}$\\[.5mm]\scalebox{.8}{$\mathfrak{su}_2^{(0)} \oplus\,\mathfrak{f}_{4}^{(1)}$}\end{tabular}};
  \node (1) at (119.78bp,1194.2bp) [draw,rectangle] {\begin{tabular}{c}$A_1$\\[1mm] $\scalebox{.8}{[1, 0, 0, 0]}$\\[.5mm]\scalebox{.8}{$\mathfrak{su}_2^{(1)} \oplus\,\mathfrak{sp}_{3}^{(1)}$}\end{tabular}};
  \node (2) at (119.78bp,1093.4bp) [draw,rectangle] {\begin{tabular}{c}$\widetilde{A}_1$\\[1mm] $\scalebox{.8}{[0, 0, 0, 1]}$\\[.5mm]\scalebox{.8}{$\mathfrak{su}_2^{(2)} \oplus\,\mathfrak{su}_{4}^{(1)}$}\end{tabular}};
  \node (3) at (119.78bp,985.86bp) [draw,rectangle] {\begin{tabular}{c}$A_1 + \widetilde{A}_1$\\[1mm] $\scalebox{.8}{[0, 1, 0, 0]}$\\[.5mm]\scalebox{.8}{$\mathfrak{su}_2^{(3)} \oplus\, \big(\mathfrak{su}_{2}^{(8)}\,\oplus\,\mathfrak{su}_{2}^{(1)} \big)$}\end{tabular}};
  \node (4) at (89.783bp,882.03bp) [draw,rectangle] {\begin{tabular}{c}$A_2$\\[1mm] $\scalebox{.8}{[2, 0, 0, 0]}$\\[.5mm]\scalebox{.8}{$\mathfrak{su}_2^{(4)} \oplus\,\mathfrak{su}_{3}^{(2)}$}\end{tabular}};
  \node (5) at (149.78bp,816.2bp) [draw,rectangle] {\begin{tabular}{c}$\widetilde{A}_2$\\[1mm] $\scalebox{.8}{[0, 0, 0, 2]}$\\[.5mm]\scalebox{.8}{$\mathfrak{su}_2^{(8)} \oplus\,\mathfrak{g}_{2}^{(1)}$}\end{tabular}};
  \node (6) at (83.783bp,749.69bp) [draw,rectangle] {\begin{tabular}{c}$A_2 + \widetilde{A}_1$\\[1mm] $\scalebox{.8}{[0, 0, 1, 0]}$\\[.5mm]\scalebox{.8}{$\mathfrak{su}_2^{(6)} \oplus\,\mathfrak{su}_{2}^{(6)}$}\end{tabular}};
  \node (7) at (150.78bp,645.19bp) [draw,rectangle] {\begin{tabular}{c}$\widetilde{A}_2 + A_1$\\[1mm] $\scalebox{.8}{[0, 1, 0, 1]}$\\[.5mm]\scalebox{.8}{$\mathfrak{su}_2^{(9)} \oplus\,\mathfrak{su}_{2}^{(3)}$}\end{tabular}};
  \node (8) at (50.783bp,645.19bp) [draw,rectangle] {\begin{tabular}{c}$B_2$\\[1mm] $\scalebox{.8}{[2, 0, 0, 1]}$\\[.5mm]\scalebox{.8}{$\mathfrak{su}_2^{(10)} \oplus\, \big(\mathfrak{su}_{2}^{(1)}\,\oplus\,\mathfrak{su}_{2}^{(1)} \big)$}\end{tabular}};
  \node (9) at (100.78bp,539.36bp) [draw,rectangle] {\begin{tabular}{c}$C_3(a_1)$\\[1mm] $\scalebox{.8}{[1, 0, 1, 0]}$\\[.5mm]\scalebox{.8}{$\mathfrak{su}_2^{(11)} \oplus\,\mathfrak{su}_{2}^{(1)}$}\end{tabular}};
  \node (10) at (100.78bp,436.21bp) [draw,rectangle] {\begin{tabular}{c}$F_4(a_3)$\\[1mm] $\scalebox{.8}{[0, 2, 0, 0]}$\\[.5mm]\scalebox{.8}{$\mathfrak{su}_2^{(12)} \oplus\,\varnothing$}\end{tabular}};
  \node (11) at (59.783bp,333.05bp) [draw,rectangle] {\begin{tabular}{c}$B_3$\\[1mm] $\scalebox{.8}{[2, 2, 0, 0]}$\\[.5mm]\scalebox{.8}{$\mathfrak{su}_2^{(28)} \oplus\,\mathfrak{su}_{2}^{(8)}$}\end{tabular}};
  \node (12) at (143.78bp,333.05bp) [draw,rectangle] {\begin{tabular}{c}$C_3$\\[1mm] $\scalebox{.8}{[1, 0, 1, 2]}$\\[.5mm]\scalebox{.8}{$\mathfrak{su}_2^{(35)} \oplus\,\mathfrak{su}_{2}^{(1)}$}\end{tabular}};
  \node (13) at (100.78bp,229.89bp) [draw,rectangle] {\begin{tabular}{c}$F_4(a_2)$\\[1mm] $\scalebox{.8}{[0, 2, 0, 2]}$\\[.5mm]\scalebox{.8}{$\mathfrak{su}_2^{(36)} \oplus\,\varnothing$}\end{tabular}};
  \node (14) at (100.78bp,126.74bp) [draw,rectangle] {\begin{tabular}{c}$F_4(a_1)$\\[1mm] $\scalebox{.8}{[2, 2, 0, 2]}$\\[.5mm]\scalebox{.8}{$\mathfrak{su}_2^{(60)} \oplus\,\varnothing$}\end{tabular}};
  \node (15) at (100.78bp,23.579bp) [draw,rectangle] {\begin{tabular}{c}$F_4$\\[1mm] $\scalebox{.8}{[2, 2, 2, 2]}$\\[.5mm]\scalebox{.8}{$\mathfrak{su}_2^{(156)} \oplus\,\varnothing$}\end{tabular}};
  \draw [->] (0) ..controls (119.78bp,1260.2bp) and (119.78bp,1243.4bp)  .. (1);
  \definecolor{strokecol}{rgb}{0.0,0.0,0.0};
  \pgfsetstrokecolor{strokecol}
  \draw (127.28bp,1240.47bp) node[right] {$f_4$};
  \draw [->] (1) ..controls (119.78bp,1157.8bp) and (119.78bp,1142.3bp)  .. (2);
  \draw (126.78bp,1137.31bp) node[right] {$c_3$};
  \draw [->] (2) ..controls (119.78bp,1054.7bp) and (119.78bp,1036.8bp)  .. (3);
  \draw (128.28bp,1034.56bp) node[right] {$a_3^+$};
  \draw [->] (3) ..controls (107.83bp,955.71bp) and (105.59bp,949.48bp)  .. (103.78bp,943.6bp) .. controls (101.13bp,934.97bp) and (98.711bp,925.51bp)  .. (4);
  \draw (112.78bp,928.3bp) node[right] {$A_1$};
  \draw [->] (3) ..controls (125.55bp,955.53bp) and (126.73bp,949.34bp)  .. (127.78bp,943.6bp) .. controls (133.54bp,912.31bp) and (139.72bp,876.58bp)  .. (5);
  \draw (140.78bp,928.3bp) node[right] {$A_1$};
  \draw [->] (4) ..controls (87.793bp,837.8bp) and (86.426bp,808.09bp)  .. (6);
  \draw (97.28bp,811.15bp) node[right] {$a_2^+$};
  \draw [->] (5) ..controls (150.09bp,762.9bp) and (150.38bp,713.85bp)  .. (7);
  \draw (158.28bp,743.39bp) node[right] {$g_2$};
  \draw [->] (6) ..controls (101.07bp,713.45bp) and (108.77bp,699.37bp)  .. (116.78bp,687.44bp) .. controls (118.87bp,684.35bp) and (121.13bp,681.2bp)  .. (7);
  \draw (123.78bp,690.14bp) node[right] {$m$};
  \draw [->] (6) ..controls (70.167bp,719.3bp) and (67.734bp,713.22bp)  .. (65.783bp,707.44bp) .. controls (62.794bp,698.6bp) and (60.131bp,688.84bp)  .. (8);
  \draw (74.78bp,692.14bp) node[right] {$A_1$};
  \draw [->] (7) ..controls (132.63bp,606.5bp) and (123.88bp,588.32bp)  .. (9);
  \draw (137.78bp,584.64bp) node[right] {$m$};
  \draw [->] (8) ..controls (64.017bp,609.13bp) and (70.273bp,594.01bp)  .. (76.783bp,580.94bp) .. controls (78.152bp,578.19bp) and (79.634bp,575.38bp)  .. (9);
  \draw (94.28bp,586.89bp) node[right] {$[2A_1]^+$};
  \draw [->] (9) ..controls (100.78bp,502.22bp) and (100.78bp,485.41bp)  .. (10);
  \draw (109.78bp,482.49bp) node[right] {$A_1$};
  \draw [->] (10) ..controls (86.028bp,398.8bp) and (79.052bp,381.59bp)  .. (11);
  \draw (93.78bp,379.33bp) node[right] {$G_2$};
  \draw [->] (10) ..controls (116.26bp,398.8bp) and (123.57bp,381.59bp)  .. (12);
  \draw (138.28bp,379.33bp) node[right] {$4G_2$};
  \draw [->] (11) ..controls (74.539bp,295.64bp) and (81.514bp,278.43bp)  .. (13);
  \draw (93.78bp,276.17bp) node[right] {$A_1$};
  \draw [->] (12) ..controls (128.31bp,295.64bp) and (120.99bp,278.43bp)  .. (13);
  \draw (135.78bp,276.17bp) node[right] {$A_1$};
  \draw [->] (13) ..controls (100.78bp,192.75bp) and (100.78bp,175.94bp)  .. (14);
  \draw (109.28bp,173.01bp) node[right] {$C_3$};
  \draw [->] (14) ..controls (100.78bp,89.589bp) and (100.78bp,72.78bp)  .. (15);
  \draw (108.78bp,69.86bp) node[right] {$F_4$};
\end{tikzpicture}
		}
		\caption{Hasse diagram for the nilpotent orbits of $\mathfrak{f}_4$. Each orbit is labelled by its Bala--Carter label.}
		\label{fig:Hasse-F4}
\end{figure}

\begin{figure}[H]
		\centering
		\resizebox{!}{.7\paperheight}{
			\begin{tikzpicture}[>=latex,line join=bevel,]
\node (0) at (80.843bp,1697.2bp) [draw,rectangle] {\begin{tabular}{c}$0$\\[1mm] $\scalebox{.8}{[0, 0, 0, 0, 0, 0]}$\\[.5mm]\scalebox{.8}{$\mathfrak{su}_2^{(0)} \oplus\,\mathfrak{e}_{6}^{(1)}$}\end{tabular}};
  \node (1) at (80.843bp,1597.1bp) [draw,rectangle] {\begin{tabular}{c}$A_1$\\[1mm] $\scalebox{.8}{[0, 0, 0, 0, 0, 1]}$\\[.5mm]\scalebox{.8}{$\mathfrak{su}_2^{(1)} \oplus\,\mathfrak{su}_{6}^{(1)}$}\end{tabular}};
  \node (2) at (80.843bp,1496.9bp) [draw,rectangle] {\begin{tabular}{c}$2A_1$\\[1mm] $\scalebox{.8}{[1, 0, 0, 0, 1, 0]}$\\[.5mm]\scalebox{.8}{$\mathfrak{su}_2^{(2)} \oplus\,\mathfrak{so}_{7}^{(1)}$}\end{tabular}};
  \node (3) at (80.843bp,1393.8bp) [draw,rectangle] {\begin{tabular}{c}$3A_1$\\[1mm] $\scalebox{.8}{[0, 0, 1, 0, 0, 0]}$\\[.5mm]\scalebox{.8}{$\mathfrak{su}_2^{(3)} \oplus\, \big(\mathfrak{su}_{2}^{(1)}\,\oplus\,\mathfrak{su}_{3}^{(2)} \big)$}\end{tabular}};
  \node (4) at (80.843bp,1290.6bp) [draw,rectangle] {\begin{tabular}{c}$A_2$\\[1mm] $\scalebox{.8}{[0, 0, 0, 0, 0, 2]}$\\[.5mm]\scalebox{.8}{$\mathfrak{su}_2^{(4)} \oplus\, \big(\mathfrak{su}_{3}^{(1)}\,\oplus\,\mathfrak{su}_{3}^{(1)} \big)$}\end{tabular}};
  \node (5) at (80.843bp,1185.5bp) [draw,rectangle] {\begin{tabular}{c}$A_2 + A_1$\\[1mm] $\scalebox{.8}{[1, 0, 0, 0, 1, 1]}$\\[.5mm]\scalebox{.8}{$\mathfrak{su}_2^{(5)} \oplus\,\mathfrak{su}_{3}^{(1)}$}\end{tabular}};
  \node (6) at (35.843bp,1082.3bp) [draw,rectangle] {\begin{tabular}{c}$A_2 + 2A_1$\\[1mm] $\scalebox{.8}{[0, 1, 0, 1, 0, 0]}$\\[.5mm]\scalebox{.8}{$\mathfrak{su}_2^{(6)} \oplus\,\mathfrak{su}_{2}^{(6)}$}\end{tabular}};
  \node (7) at (125.84bp,1082.3bp) [draw,rectangle] {\begin{tabular}{c}$2A_2$\\[1mm] $\scalebox{.8}{[2, 0, 0, 0, 2, 0]}$\\[.5mm]\scalebox{.8}{$\mathfrak{su}_2^{(8)} \oplus\,\mathfrak{g}_{2}^{(1)}$}\end{tabular}};
  \node (8) at (35.843bp,979.15bp) [draw,rectangle] {\begin{tabular}{c}$A_3$\\[1mm] $\scalebox{.8}{[1, 0, 0, 0, 1, 2]}$\\[.5mm]\scalebox{.8}{$\mathfrak{su}_2^{(10)} \oplus\,\mathfrak{sp}_{2}^{(1)}$}\end{tabular}};
  \node (9) at (125.84bp,979.15bp) [draw,rectangle] {\begin{tabular}{c}$2A_2 + A_1$\\[1mm] $\scalebox{.8}{[1, 0, 1, 0, 1, 0]}$\\[.5mm]\scalebox{.8}{$\mathfrak{su}_2^{(9)} \oplus\,\mathfrak{su}_{2}^{(3)}$}\end{tabular}};
  \node (10) at (79.843bp,875.99bp) [draw,rectangle] {\begin{tabular}{c}$A_3 + A_1$\\[1mm] $\scalebox{.8}{[0, 1, 0, 1, 0, 1]}$\\[.5mm]\scalebox{.8}{$\mathfrak{su}_2^{(11)} \oplus\,\mathfrak{su}_{2}^{(1)}$}\end{tabular}};
  \node (11) at (79.843bp,772.83bp) [draw,rectangle] {\begin{tabular}{c}$D_4(a_1)$\\[1mm] $\scalebox{.8}{[0, 0, 2, 0, 0, 0]}$\\[.5mm]\scalebox{.8}{$\mathfrak{su}_2^{(12)} \oplus\,\varnothing$}\end{tabular}};
  \node (12) at (51.843bp,604.52bp) [draw,rectangle] {\begin{tabular}{c}$D_4$\\[1mm] $\scalebox{.8}{[0, 0, 2, 0, 0, 2]}$\\[.5mm]\scalebox{.8}{$\mathfrak{su}_2^{(28)} \oplus\,\mathfrak{su}_{3}^{(2)}$}\end{tabular}};
  \node (13) at (121.84bp,669.68bp) [draw,rectangle] {\begin{tabular}{c}$A_4$\\[1mm] $\scalebox{.8}{[2, 0, 0, 0, 2, 2]}$\\[.5mm]\scalebox{.8}{$\mathfrak{su}_2^{(20)} \oplus\,\mathfrak{su}_{2}^{(1)}$}\end{tabular}};
  \node (14) at (48.843bp,436.21bp) [draw,rectangle] {\begin{tabular}{c}$D_5(a_1)$\\[1mm] $\scalebox{.8}{[1, 1, 0, 1, 1, 2]}$\\[.5mm]\scalebox{.8}{$\mathfrak{su}_2^{(30)} \oplus\,\varnothing$}\end{tabular}};
  \node (15) at (123.84bp,539.36bp) [draw,rectangle] {\begin{tabular}{c}$A_4 + A_1$\\[1mm] $\scalebox{.8}{[1, 1, 0, 1, 1, 1]}$\\[.5mm]\scalebox{.8}{$\mathfrak{su}_2^{(21)} \oplus\,\varnothing$}\end{tabular}};
  \node (16) at (92.843bp,333.05bp) [draw,rectangle] {\begin{tabular}{c}$E_6(a_3)$\\[1mm] $\scalebox{.8}{[2, 0, 2, 0, 2, 0]}$\\[.5mm]\scalebox{.8}{$\mathfrak{su}_2^{(36)} \oplus\,\varnothing$}\end{tabular}};
  \node (17) at (138.84bp,436.21bp) [draw,rectangle] {\begin{tabular}{c}$A_5$\\[1mm] $\scalebox{.8}{[2, 1, 0, 1, 2, 1]}$\\[.5mm]\scalebox{.8}{$\mathfrak{su}_2^{(35)} \oplus\,\mathfrak{su}_{2}^{(1)}$}\end{tabular}};
  \node (18) at (92.843bp,229.89bp) [draw,rectangle] {\begin{tabular}{c}$D_5$\\[1mm] $\scalebox{.8}{[2, 0, 2, 0, 2, 2]}$\\[.5mm]\scalebox{.8}{$\mathfrak{su}_2^{(60)} \oplus\,\varnothing$}\end{tabular}};
  \node (19) at (92.843bp,126.74bp) [draw,rectangle] {\begin{tabular}{c}$E_6(a_1)$\\[1mm] $\scalebox{.8}{[2, 2, 0, 2, 2, 2]}$\\[.5mm]\scalebox{.8}{$\mathfrak{su}_2^{(84)} \oplus\,\varnothing$}\end{tabular}};
  \node (20) at (92.843bp,23.579bp) [draw,rectangle] {\begin{tabular}{c}$E_6$\\[1mm] $\scalebox{.8}{[2, 2, 2, 2, 2, 2]}$\\[.5mm]\scalebox{.8}{$\mathfrak{su}_2^{(156)} \oplus\,\varnothing$}\end{tabular}};
  \draw [->] (0) ..controls (80.843bp,1661.2bp) and (80.843bp,1645.8bp)  .. (1);
  \definecolor{strokecol}{rgb}{0.0,0.0,0.0};
  \pgfsetstrokecolor{strokecol}
  \draw (88.34bp,1640.37bp) node[right] {$e_6$};
  \draw [->] (1) ..controls (80.843bp,1561.0bp) and (80.843bp,1545.6bp)  .. (2);
  \draw (88.34bp,1540.21bp) node[right] {$a_5$};
  \draw [->] (2) ..controls (80.843bp,1459.8bp) and (80.843bp,1443.0bp)  .. (3);
  \draw (87.84bp,1440.06bp) node[right] {$b_3$};
  \draw [->] (3) ..controls (80.843bp,1356.6bp) and (80.843bp,1339.8bp)  .. (4);
  \draw (89.84bp,1336.9bp) node[right] {$A_1$};
  \draw [->] (4) ..controls (80.843bp,1253.0bp) and (80.843bp,1235.2bp)  .. (5);
  \draw (96.84bp,1232.99bp) node[right] {$[2a_2]^+$};
  \draw [->] (5) ..controls (64.648bp,1148.1bp) and (56.992bp,1130.8bp)  .. (6);
  \draw (70.34bp,1127.08bp) node[right] {$a_2$};
  \draw [->] (5) ..controls (97.038bp,1148.1bp) and (104.69bp,1130.8bp)  .. (7);
  \draw (116.84bp,1128.58bp) node[right] {$A_2$};
  \draw [->] (6) ..controls (35.843bp,1045.2bp) and (35.843bp,1028.4bp)  .. (8);
  \draw (44.84bp,1025.43bp) node[right] {$A_1$};
  \draw [->] (6) ..controls (68.693bp,1044.4bp) and (84.728bp,1026.4bp)  .. (9);
  \draw (95.34bp,1023.43bp) node[right] {$\tau$};
  \draw [->] (7) ..controls (125.84bp,1045.2bp) and (125.84bp,1028.4bp)  .. (9);
  \draw (133.34bp,1024.43bp) node[right] {$g_2$};
  \draw [->] (8) ..controls (51.678bp,941.74bp) and (59.164bp,924.53bp)  .. (10);
  \draw (68.84bp,922.27bp) node[right] {$b_2$};
  \draw [->] (9) ..controls (109.29bp,941.74bp) and (101.46bp,924.53bp)  .. (10);
  \draw (114.84bp,920.27bp) node[right] {$m$};
  \draw [->] (10) ..controls (79.843bp,838.85bp) and (79.843bp,822.04bp)  .. (11);
  \draw (88.84bp,819.11bp) node[right] {$A_1$};
  \draw [blue,->] (11) ..controls (72.412bp,743.36bp) and (70.978bp,737.1bp)  .. (69.843bp,731.26bp) .. controls (63.825bp,700.28bp) and (58.911bp,664.65bp)  .. (12);
  \draw (78.84bp,715.96bp) node[right] {$G_2$};
  \draw [blue,->] (11) ..controls (94.958bp,735.43bp) and (102.1bp,718.22bp)  .. (13);
  \draw (115.84bp,715.96bp) node[right] {$3C_2$};
  \draw [->] (12) ..controls (50.921bp,552.41bp) and (50.044bp,503.8bp)  .. (14);
  \draw (59.34bp,532.56bp) node[right] {$a_2$};
  \draw [->] (13) ..controls (122.51bp,625.76bp) and (122.97bp,596.53bp)  .. (15);
  \draw (132.84bp,599.22bp) node[right] {$A_1$};
  \draw [->] (14) ..controls (64.678bp,398.8bp) and (72.164bp,381.59bp)  .. (16);
  \draw (83.84bp,379.33bp) node[right] {$A_2$};
  \draw [->] (15) ..controls (96.564bp,501.57bp) and (83.353bp,483.75bp)  .. (14);
  \draw (102.84bp,482.49bp) node[right] {$A_2$};
  \draw [->] (15) ..controls (129.2bp,502.22bp) and (131.7bp,485.41bp)  .. (17);
  \draw (141.84bp,482.49bp) node[right] {$A_2$};
  \draw [blue,->] (16) ..controls (92.843bp,295.9bp) and (92.843bp,279.09bp)  .. (18);
  \draw (101.34bp,276.17bp) node[right] {$C_3$};
  \draw [->] (17) ..controls (122.29bp,398.8bp) and (114.46bp,381.59bp)  .. (16);
  \draw (129.84bp,379.33bp) node[right] {$A_1$};
  \draw [->] (18) ..controls (92.843bp,192.75bp) and (92.843bp,175.94bp)  .. (19);
  \draw (101.84bp,173.01bp) node[right] {$A_5$};
  \draw [blue,->] (19) ..controls (92.843bp,89.589bp) and (92.843bp,72.78bp)  .. (20);
  \draw (101.34bp,69.86bp) node[right] {$E_6$};
\end{tikzpicture}
		}
		\caption{Hasse diagram for the nilpotent orbits of $\mathfrak{e}_6$. Each orbit is labelled by its Bala--Carter label.}
		\label{fig:Hasse-E6}
\end{figure}

\begin{figure}[H]
		\centering
		\resizebox{!}{.7\paperheight}{
			\input{figures/hasse_E7.tex}
		}
		\caption{Hasse diagram for the nilpotent orbits of $\mathfrak{e}_7$. Each orbit is labelled by its Bala--Carter label.}
		\label{fig:Hasse-E7}
\end{figure}

\begin{figure}[H]
		\centering
		\resizebox{!}{.7\paperheight}{
			\input{figures/hasse_E8.tex}
		}
		\caption{Hasse diagram for the nilpotent orbits of $\mathfrak{e}_8$. Each orbit is labelled by its Bala--Carter label.}
		\label{fig:Hasse-E8}
\end{figure}

\section{From a Nilpotent Orbit to a Tensor Branch}\label{app:nilp2TB}

While our main result is that the anomaly polynomial of long quivers rely only
on a few parameters that do not depend on the details of the geometric
engineering, the tensor branch effective field theory of a 6d $(1,0)$ SCFT
associated to a particular non-compact elliptically-fibered Calabi--Yau
threefold is a pivotal component underlying our analysis. As was noted
already in \cite{Heckman:2015bfa} (see also \cite{Heckman:2018pqx}), the tensor
branch configurations appear to organize themselves into families associated to
nilpotent orbits of simple Lie algebras. In this appendix, we provide a
comprehensive mapping between the nilpotent orbits and the tensor branch
descriptions that they are purported to be associated to. 

In the main body of this paper, we show that the difference in the anomaly
polynomials for the theories with the tensor branch configurations associated
to the maximal nilpotent orbit and any other nilpotent orbit is precisely what
we would expect from a bottom-up Higgsing by giving a vacuum expectation value
valued in the nilpotent orbit to the moment map of the flavor symmetry
associated to the maximal nilpotent orbit.

It is important to note that the correspondence described in the previous
literature does not directly relate a nilpotent orbit to a tensor branch.
First, one needs to pick a parent theory of the family with a $\mathfrak{g}$
flavor symmetry factor, and then, for certain choices of parent theory, one
notices that there is a family of descendant theories that are in one-to-one
correspondence with the nilpotent orbits of $\mathfrak{g}$.\footnote{In fact,
one defines the notion of a parent theory in this way.} The parent theories for
such families were enumerated in \cite{Heckman:2018pqx}, and in this appendix
we give the correspondence between nilpotent orbits and descendant tensor
branch configurations for each family.

\subsection{A-series}

We start by considering parent SCFTs with an
$\mathfrak{su}(K)$ flavor algebra and a tensor branch configuration that takes
the form
\begin{equation}\label{eqn:appAgen}
    \overset{\mathfrak{su}_K}{2}\cdots\overset{\mathfrak{su}_K}{2} \cdots \,,
\end{equation}
where the $\cdots$ on the right indicates any attached collection of curves and
algebras. To satisfy the long quiver condition, which we assume throughout this
paper, it is sufficient to take the number of $\overset{\mathfrak{su}_K}{2}$ on
the left to be at least $K+1$. Let $O$ be a nilpotent orbit of
$\mathfrak{su}(K)$; $O$ can be written uniquely as an integer partition of $K$:
\begin{equation}\label{eqn:appAnilp}
    [1^{m_1}, 2^{m_2}, \cdots, K^{m_K}] \,, \qquad \text{ such that } \qquad \sum_{i=1}^K i m_i = K \,.
\end{equation}
Then, the tensor branch effective field theory associated to the nilpotent
orbit $O$ can be described as \cite{Mekareeya:2016yal}
\begin{equation}
    \overset{\mathfrak{su}_{k_1}}{2}
    \overset{\mathfrak{su}_{k_2}}{2}
    \overset{\mathfrak{su}_{k_3}}{2}\cdots\overset{\mathfrak{su}_{k_K}}{2} 
	\overset{\mathfrak{su}_{K_{\phantom{K}}}}{2}\cdots \,.
\end{equation}
The rightmost-written $\mathfrak{su}(K)$ gauge algebra is always present when we
consider long quivers. The $k_i$ are fixed in terms of the exponents of the
partition by the anomaly cancellation conditions:
\begin{equation}\label{eqn:appAktom}
    2k_i - k_{i-1} - k_{i+1} = m_i \,, 
\end{equation}
where, for convenience, we have defined $k_0 = 0$ and $k_{K+1} = K$. This
system of equations can be solved to yield an exact expression for $k_i$.
Taking into account that the coefficients $m_i$ define a partition of $K$, it is
straightforward to show that:
\begin{equation}\label{A-k-to-m}
		k_{i} = K - \sum_{j=1}^{K-i} j\,m_{i+j}\,.
\end{equation}

Thus, we have determined the putative tensor branch description of the 6d
$(1,0)$ SCFT associated to the parent SCFT in equation \eqref{eqn:appAgen} and
the nilpotent orbit in equation \eqref{eqn:appAnilp}.

\subsection{C-series}

Next, we consider the tensor branch descriptions associated to parent theories which take the form:
\begin{equation}\label{eqn:appCgen}
    \overset{\mathfrak{so}_{2K+8}}{4}
    \overset{\mathfrak{sp}_{K}}{1}
    \overset{\mathfrak{so}_{2K+8}}{4}\cdots\overset{\mathfrak{sp}_{K}}{1}
    \overset{\mathfrak{so}_{2K+8}}{4} \cdots \,,
\end{equation}
and a choice of nilpotent orbit of $\mathfrak{sp}(K)$. The $\cdots$ on the
right in equation \eqref{eqn:appCgen} represents any combinations of curves and
algebras that can be consistently attached. In this case, the long quiver
condition is satisfied if we have at least $K+1$ copies of
$\overset{\mathfrak{so}_{2K+8}}{4}$ on the left in equation
\eqref{eqn:appCgen}. A nilpotent orbit of $\mathfrak{sp}(K)$ is given by a C-partition of $2K$, as described around equation \eqref{eqn:YYY}.

It was proposed that the descendant of the parent theory in equation \eqref{eqn:appCgen} associated to the nilpotent orbit in equation \eqref{eqn:YYY} has the tensor branch description:
\begin{equation}
    \overset{\mathfrak{so}_{k_1}}{4}
    \overset{\mathfrak{sp}_{k_2}}{1}
    \overset{\mathfrak{so}_{k_3}}{4}\cdots\overset{\mathfrak{sp}_{k_{2K}}}{1}
    \overset{\mathfrak{so}_{2K+8}}{4} \cdots \,.
\end{equation}
The $k_i$ are again fixed in terms of the C-partition via the anomaly cancellation conditions. We have
\begin{equation}\label{eqn:appCmtok}
    \begin{aligned}
        k_i - 8 - k_{i-1} - k_{i+1} = \frac{m_i}{2} &\qquad \text{ if $i$ odd } \,, \\
        4k_i + 16 - k_{i-1} - k_{i+1} = m_i &\qquad \text{ if $i$ even } \,.
    \end{aligned}
\end{equation}
Note that the RHS is always guaranteed to be an integer from the C-partition
condition, and we have defined $k_0 = 0$ and $k_{2K+1} = 2K + 8$. Again, as for
the A-series, we can solve for $k_i$ and find a compact expression:
\begin{equation}\label{eqn:solCmtok}
    \begin{aligned}
        k_i =& 2(K+4)- \sum_{j=1}^{2K-i}j\, m_{i+j}  &\qquad \text{ if $i$ odd } \,, \\
		k_i =& K - \sum_{j=1}^{2K-i}\frac{j}{2}\, m_{i+j} &\qquad \text{ if $i$ even } \,.
    \end{aligned}
\end{equation}

\subsection{D-series}

The next set of families of 6d $(1,0)$ SCFTs that we consider are those
progenated from a parent theory of the form:
\begin{equation}\label{eqn:appDgen}
    \overset{\mathfrak{sp}_{K-4}}{1}
    \overset{\mathfrak{so}_{2K}}{4}
    \overset{\mathfrak{sp}_{K-4}}{1}
    \cdots
    \overset{\mathfrak{so}_{2K}}{4}
    \overset{\mathfrak{sp}_{K-4}}{1} \cdots \,.
\end{equation}
Such a parent theory has an $\mathfrak{so}(2K)$ flavor symmetry. Each nilpotent orbit $O$ has an underlying D-partition, as in equation \eqref{eqn:AAA}.

As we have already discussed, the association of a nilpotent orbit to be
D-partition is not unique. In this paper, we determine the anomaly polynomials
of the relevant 6d SCFTs from both a top-down tensor branch perspective, and
from a bottom-up nilpotent Higgsing perspective, where the latter depends only
on the parent theory and the choice of nilpotent orbit. For parent theories
with an $\mathfrak{so}(2K)$ flavor algebra, the anomaly polynomial is actually
agnostic to the precise nilpotent orbit, and instead only depends on the
underlying D-partition. The dependence of the tensor branch description on the
choice of nilpotent orbit itself has been discussed in detail in
\cite{Distler:2022yse}, however, for our purposes, we only need to know the
tensor branch as depending on the D-partition; while this would appear to give
two distinct SCFTs with the same tensor branch, there are in fact choices of
discrete $\theta$-angles by which the tensor branch effective field theories
differ. We do not write the $\theta$-angles here and refer the reader to
\cite{Distler:2022yse} for the full details.

Given a parent theory of the form in equation \eqref{eqn:appDgen} and a
nilpotent orbit associated to a D-partition as in equation
\eqref{eqn:AAA}, the descendant theory is proposed to have the tensor
branch effective descriptions:
\begin{equation}\label{eqn:appDprescription}
    \overset{\mathfrak{sp}_{k_1}}{1}
    \overset{\mathfrak{so}_{k_2}}{4}
    \overset{\mathfrak{sp}_{k_3}}{1}
    \cdots
    \overset{\mathfrak{so}_{k_{2K}}}{4}
    \overset{\mathfrak{sp}_{K-4}}{1} \cdots \,.
\end{equation}
As usual, the anomaly cancellation conditions fix the ranks of the various
gauge algebras in terms of the multiplicities specifying the D-partition. We
must have
\begin{equation}
    \begin{aligned}
        k_i - 8 - k_{i-1} - k_{i+1} = \frac{m_i}{2} &\qquad \text{ if $i$ even } \,, \\
        4k_i + 16 - k_{i-1} - k_{i+1} = m_i &\qquad \text{ if $i$ odd } \,.
    \end{aligned}
\end{equation}
The D-partition condition now guarantees that $m_i$ is an even integer for $i$
even, and thus the RHS is always an integer. We have defined $k_0 = 0$ and
$k_{2K+1} = K-4$ for convenience. As for the C-series, a closed expression can
be found:
\begin{equation}\label{eqn:solDmtok}
    \begin{aligned}
		k_i =&\, 2K - \sum_{j=1}^{2K-i}j\, m_{i+j}  &\qquad \text{ if $i$ even } \,, \\
		k_i =&\, (K-4) - \sum_{j=1}^{2K-i}\frac{j}{2}\, m_{i+j} &\qquad \text{ if $i$ odd } \,.
    \end{aligned}
\end{equation}

Interestingly, the
$k_i$ that one obtains in this way are sometimes inconsistent with the allowed
F-theory configurations. For example, one may obtain a $k_i = 7$ for $i$ even,
which would indicate that the tensor description is of the form $\cdots
\overset{\mathfrak{so}_7}{4}\cdots$, which cannot be engineered from a
non-compact elliptically-fibered Calabi--Yau threefold in the F-theory
construction of 6d $(1,0)$ SCFTs. In fact, when the ranks of the gauge algebras
are ``too small'', the prescription for the tensor branch geometry associated
to the nilpotent orbit, which we gave in equation \eqref{eqn:appDprescription}
needs to be modified. The modification can be summarized in the following short
list of replacement rules:
\begin{equation}
    \begin{aligned}
        \overset{\mathfrak{sp}_{-3}}{1}\overset{\mathfrak{so}_3}{4}\overset{\mathfrak{sp}_{-2}}{1}\overset{\mathfrak{so}_5}{4}\overset{\mathfrak{sp}_{-1}}{1}\overset{\mathfrak{so}_7}{4} \cdots &\quad\longrightarrow\quad & &2\overset{\mathfrak{su}_2}{2}\overset{\mathfrak{g}_2}{3} \cdots \,, \\
        \overset{\mathfrak{sp}_{-3}}{1}\overset{\mathfrak{so}_4}{4}\overset{\mathfrak{sp}_{-1}}{1}\overset{\mathfrak{so}_7}{4} \cdots &\quad\longrightarrow\quad & &\overset{\mathfrak{su}_2}{2}\overset{\mathfrak{g}_2}{3} \cdots \,,\\
        \overset{\mathfrak{sp}_{-3}}{1}\overset{\mathfrak{so}_4}{4}\overset{\mathfrak{sp}_{-1}}{1}\overset{\mathfrak{so}_8}{4} \cdots &\quad\longrightarrow\quad & &\overset{\mathfrak{su}_2}{2}\overset{\mathfrak{so}_7}{3} \cdots \,,\\
        \overset{\mathfrak{sp}_{-2}}{1}\overset{\mathfrak{so}_5}{4}\overset{\mathfrak{sp}_{-1}}{1}\overset{\mathfrak{so}_7}{4} \cdots &\quad\longrightarrow\quad & &\overset{\mathfrak{su}_2}{2}\overset{\mathfrak{so}_7}{3} \cdots \,,\\
        \overset{\mathfrak{sp}_{-2}}{1}\overset{\mathfrak{so}_6}{4} \cdots &\quad\longrightarrow\quad & &\overset{\mathfrak{su}_3}{3} \cdots \,,\\
        \overset{\mathfrak{sp}_{-2}}{1}\overset{\mathfrak{so}_7}{4} \cdots &\quad\longrightarrow\quad & &\overset{\mathfrak{g}_2}{3} \cdots \,, \\
        \overset{\mathfrak{sp}_{-2}}{1}\overset{\mathfrak{so}_8}{4} \cdots &\quad\longrightarrow\quad & &\overset{\mathfrak{so}_7}{3} \cdots \,, \\
        \overset{\mathfrak{sp}_{-1}}{1}\overset{\mathfrak{g}}{4} \cdots &\quad\longrightarrow\quad & &\overset{\mathfrak{g}}{3} \cdots \,.
    \end{aligned}
\end{equation}

In this way, we have specified a way of assigning to a parent theory of the
form in equation \eqref{eqn:appDgen} and a choice of nilpotent orbit of
$\mathfrak{so}(2K)$ as in equation \eqref{eqn:AAA} a 6d $(1,0)$ tensor
branch. 

\subsection{E-series}

Finally, we turn to what we call the ``E-series'' of tensor branch descriptions. These will be long quivers whose spine contains a repeating pattern of exceptional conformal matter. Note that when the
fractions $f\neq1$, the tensor branch translations that we describe herein do
\emph{not} all involve a nilpotent orbit of an exceptional Lie algebra as
$\mathfrak{g}_f\subseteq\mathfrak{g}$. 

Due to the sporadic nature of this series, we present the dictionary between
parent theory with flavor symmetry factor $\mathfrak{g}$ plus each nilpotent
orbit of $\mathfrak{g}$ and the tensor branch description in the form of
tables, which were first described explicitly in \cite{Heckman:2016ssk} when $f=1$. These are Tables \ref{tbl:E6-nilp-1-2}, \ref{tbl:E6-nilp-1} for
$\mathfrak{e}_6$, Tables \ref{tbl:E7-nilp-1-3}, \ref{tbl:E7-nilp-1-2},
\ref{tbl:E7-nilp-2-3}, \ref{tbl:E7-nilp-1} for $\mathfrak{e}_7$, and Tables
\ref{tbl:E8-nilp-1-4}, \ref{tbl:E8-nilp-1-3}, \ref{tbl:E8-nilp-1-2},
\ref{tbl:E8-nilp-2-3}, \ref{tbl:E8-nilp-3-4}, and \ref{tbl:E8-nilp-1} for $\mathfrak{e}_8$. In each
table, the nilpotent orbit labelled as either $[1^K]$ or $0$ -- and
appearing on the first row -- provides the form of the tensor branch description of the
parent theory. Subsequent rows provide the tensor branch effective field
theory corresponding to that parent theory plus each nilpotent of the given
flavor algebra of the parent theory.

\begin{longtable}{ccr}
	\toprule
	Nilpotent Orbit & $\mathfrak{f}$ & \multicolumn{1}{c}{Tensor Branch}\\
	\midrule
		$[1^3]$ & $\mathfrak{su}_{3}$ & $1\overset{\mathfrak{e}_6}{6}\cdots$ \\
		$[2,1]$ & $\varnothing$ & $\overset{\mathfrak{e}_6}{5}\cdots$ \\
		$[3]$ & $\varnothing$ & $\overset{\mathfrak{f}_4}{5}\cdots$ \\
	\bottomrule
	\caption{The tensor branch descriptions associated to nilpotent Higgsing of the $\mathfrak{su}_{3}$ flavor symmetry for a long quiver associated to $\mathfrak{e}_{6}$ conformal matter with fraction number $f=\frac{1}{2}$.}
\label{tbl:E6-nilp-1-2}
\end{longtable}

\begin{longtable}{ccr}
	\toprule
	Nilpotent Orbit & $\mathfrak{f}$ & \multicolumn{1}{c}{Tensor Branch}\\
	\midrule
		$0$ & $\mathfrak{e}_{6}$ & $1\overset{\mathfrak{su}_3}{3}1\overset{\mathfrak{e}_6}{6}1\overset{\mathfrak{su}_3}{3}1\overset{\mathfrak{e}_6}{6}1\overset{\mathfrak{su}_3}{3}1\overset{\mathfrak{e}_6}{6}1\overset{\mathfrak{su}_3}{3}1\overset{\mathfrak{e}_6}{6}\cdots$ \\
		$A_1$ & $\mathfrak{su}_{6}$ & $\overset{\mathfrak{su}_3}{2}1\overset{\mathfrak{e}_6}{6}1\overset{\mathfrak{su}_3}{3}1\overset{\mathfrak{e}_6}{6}1\overset{\mathfrak{su}_3}{3}1\overset{\mathfrak{e}_6}{6}1\overset{\mathfrak{su}_3}{3}1\overset{\mathfrak{e}_6}{6}\cdots$ \\
		$2A_1$ & $\mathfrak{so}_{7}$ & $\overset{\mathfrak{su}_2}{2}1\overset{\mathfrak{e}_6}{6}1\overset{\mathfrak{su}_3}{3}1\overset{\mathfrak{e}_6}{6}1\overset{\mathfrak{su}_3}{3}1\overset{\mathfrak{e}_6}{6}1\overset{\mathfrak{su}_3}{3}1\overset{\mathfrak{e}_6}{6}\cdots$ \\
		$3A_1$ & $\mathfrak{su}_{2}\oplus\mathfrak{su}_{3}$ & $21\overset{\mathfrak{e}_6}{6}1\overset{\mathfrak{su}_3}{3}1\overset{\mathfrak{e}_6}{6}1\overset{\mathfrak{su}_3}{3}1\overset{\mathfrak{e}_6}{6}1\overset{\mathfrak{su}_3}{3}1\overset{\mathfrak{e}_6}{6}\cdots$ \\
		$A_2$ & $\mathfrak{su}_{3}\oplus\mathfrak{su}_{3}$ & $1\overset{\displaystyle 1}{\overset{\mathfrak{e}_6}{6}}1\overset{\mathfrak{su}_3}{3}1\overset{\mathfrak{e}_6}{6}1\overset{\mathfrak{su}_3}{3}1\overset{\mathfrak{e}_6}{6}1\overset{\mathfrak{su}_3}{3}1\overset{\mathfrak{e}_6}{6}\cdots$ \\
		$A_2 + A_1$ & $\mathfrak{su}_{3}$ & $1\overset{\mathfrak{e}_6}{5}1\overset{\mathfrak{su}_3}{3}1\overset{\mathfrak{e}_6}{6}1\overset{\mathfrak{su}_3}{3}1\overset{\mathfrak{e}_6}{6}1\overset{\mathfrak{su}_3}{3}1\overset{\mathfrak{e}_6}{6}\cdots$ \\
		$2A_2$ & $\mathfrak{g}_{2}$ & $1\overset{\mathfrak{f}_4}{5}1\overset{\mathfrak{su}_3}{3}1\overset{\mathfrak{e}_6}{6}1\overset{\mathfrak{su}_3}{3}1\overset{\mathfrak{e}_6}{6}1\overset{\mathfrak{su}_3}{3}1\overset{\mathfrak{e}_6}{6}\cdots$ \\
\end{longtable}

%
%
%
\begin{longtable}{ccr}
	\toprule
	Nilpotent Orbit & $\mathfrak{f}$ & \multicolumn{1}{c}{Tensor Branch}\\
	\midrule
		$[1^2]$ & $\mathfrak{su}_{2}$ & $1\overset{\mathfrak{e}_7}{8}\cdots$ \\
		$[2]$ & $\varnothing$ & $\overset{\mathfrak{e}_7}{7}\cdots$ \\
	\bottomrule
	\caption{The tensor branch descriptions associated to nilpotent Higgsing of the $\mathfrak{su}_{2}$ flavor symmetry for a long quiver associated to $\mathfrak{e}_{7}$ conformal matter with fraction number $f=\frac{1}{3}$.}
\label{tbl:E7-nilp-1-3}
\end{longtable}

\begin{longtable}{ccr}
	\toprule
	Nilpotent Orbit & $\mathfrak{f}$ & \multicolumn{1}{c}{Tensor Branch}\\
	\midrule
		$[1^7]$ & $\mathfrak{so}_{7}$ & $\overset{\mathfrak{su}_2}{2} \, 1 \overset{\mathfrak{e}_7}{8} 1 \overset{\mathfrak{su}_2}{2} \overset{\mathfrak{so}_7}{3} \overset{\mathfrak{su}_2}{2} 1 \overset{\mathfrak{e}_7}{8} \cdots$ \\
		$[2^2,1^3]$ & $\mathfrak{su}_{2}\oplus\mathfrak{su}_{2}$ & $2 \, 1 \overset{\mathfrak{e}_7}{8} 1 \overset{\mathfrak{su}_2}{2} \overset{\mathfrak{so}_7}{3} \overset{\mathfrak{su}_2}{2} 1 \overset{\mathfrak{e}_7}{8} \cdots$ \\
		$[3, 1^4]$ & $\mathfrak{su}_{2}\oplus\mathfrak{su}_{2}$ & $ 1 \overset{\displaystyle 1}{\overset{\mathfrak{e}_7}{8}} 1 \overset{\mathfrak{su}_2}{2} \overset{\mathfrak{so}_7}{3} \overset{\mathfrak{su}_2}{2} 1 \overset{\mathfrak{e}_7}{8}  \cdots$ \\
		$[3, 2^2]$ & $\mathfrak{su}_{2}$ & $ 1 \overset{\mathfrak{e}_7}{7} 1 \overset{\mathfrak{su}_2}{2} \overset{\mathfrak{so}_7}{3} \overset{\mathfrak{su}_2}{2} 1 \overset{\mathfrak{e}_7}{8}  \cdots$ \\
		$[3^2, 1]$ & $\varnothing$ & $  \overset{\mathfrak{e}_7}{6} 1 \overset{\mathfrak{su}_2}{2} \overset{\mathfrak{so}_7}{3} \overset{\mathfrak{su}_2}{2} 1 \overset{\mathfrak{e}_7}{8}  \cdots$ \\
		$[5, 1^2]$ & $\varnothing$ & $  \overset{\mathfrak{e}_6}{6} 1 \overset{\mathfrak{su}_2}{2} \overset{\mathfrak{so}_7}{3} \overset{\mathfrak{su}_2}{2} 1 \overset{\mathfrak{e}_7}{8}  \cdots$ \\
		$[7]$ & $\varnothing$ & $ \overset{\mathfrak{f}_4}{5} 1 \overset{\mathfrak{g}_2}{3} \overset{\mathfrak{su}_2}{2} 1 \overset{\mathfrak{e}_7}{8}  \cdots$ \\
	\bottomrule
	\caption{The tensor branch descriptions associated to nilpotent Higgsing of the $\mathfrak{so}_{7}$ flavor symmetry for a long quiver associated to $\mathfrak{e}_{7}$ conformal matter with fraction number $f=\frac{1}{2}$.}
\label{tbl:E7-nilp-1-2}
\end{longtable}

\begin{longtable}{ccr}
	\toprule
	Nilpotent Orbit & $\mathfrak{f}$ & \multicolumn{1}{c}{Tensor Branch}\\
	\midrule
		$[1^2]$ & $\mathfrak{su}_{2}$ & $\overset{\mathfrak{so}_7}{3} \overset{\mathfrak{su}_2}{2} 1\overset{\mathfrak{e}_7}{8} \cdots$ \\
		$[2]$ & $\varnothing$ & $\overset{\mathfrak{g}_2}{3} \overset{\mathfrak{su}_2}{2} 1 \overset{\mathfrak{e}_7}{8} \cdots$ \\
	\bottomrule
	\caption{The tensor branch descriptions associated to nilpotent Higgsing of the $\mathfrak{su}_{2}$ flavor symmetry for a long quiver associated to $\mathfrak{e}_{7}$ conformal matter with fraction number $f=\frac{2}{3}$.}
\label{tbl:E7-nilp-2-3}
\end{longtable}

\begin{longtable}{ccr}
	\toprule
	Nilpotent Orbit & $\mathfrak{f}$ & \multicolumn{1}{c}{Tensor Branch}\\
	\midrule
		$0$ & $\mathfrak{e}_{7}$ & $1\overset{\mathfrak{su}_2}{2}\overset{\mathfrak{so}_7}{3}\overset{\mathfrak{su}_2}{2}1\overset{\mathfrak{e}_7}{8}1\overset{\mathfrak{su}_2}{2}\overset{\mathfrak{so}_7}{3}\overset{\mathfrak{su}_2}{2}1\overset{\mathfrak{e}_7}{8}1\overset{\mathfrak{su}_2}{2}\overset{\mathfrak{so}_7}{3}\overset{\mathfrak{su}_2}{2}1\overset{\mathfrak{e}_7}{8}1\overset{\mathfrak{su}_2}{2}\overset{\mathfrak{so}_7}{3}\overset{\mathfrak{su}_2}{2}1\overset{\mathfrak{e}_7}{8}1\overset{\mathfrak{su}_2}{2}\overset{\mathfrak{so}_7}{3}\overset{\mathfrak{su}_2}{2}1\overset{\mathfrak{e}_7}{8}\cdots$ \\
		$A_1$ & $\mathfrak{so}_{12}$ & $\overset{\mathfrak{su}_2}{1}\overset{\mathfrak{so}_7}{3}\overset{\mathfrak{su}_2}{2}1\overset{\mathfrak{e}_7}{8}1\overset{\mathfrak{su}_2}{2}\overset{\mathfrak{so}_7}{3}\overset{\mathfrak{su}_2}{2}1\overset{\mathfrak{e}_7}{8}1\overset{\mathfrak{su}_2}{2}\overset{\mathfrak{so}_7}{3}\overset{\mathfrak{su}_2}{2}1\overset{\mathfrak{e}_7}{8}1\overset{\mathfrak{su}_2}{2}\overset{\mathfrak{so}_7}{3}\overset{\mathfrak{su}_2}{2}1\overset{\mathfrak{e}_7}{8}1\overset{\mathfrak{su}_2}{2}\overset{\mathfrak{so}_7}{3}\overset{\mathfrak{su}_2}{2}1\overset{\mathfrak{e}_7}{8}\cdots$ \\
		$2A_1$ & $\mathfrak{so}_{9}\oplus\mathfrak{su}_{2}$ & $1\overset{\mathfrak{so}_7}{3}\overset{\mathfrak{su}_2}{2}1\overset{\mathfrak{e}_7}{8}1\overset{\mathfrak{su}_2}{2}\overset{\mathfrak{so}_7}{3}\overset{\mathfrak{su}_2}{2}1\overset{\mathfrak{e}_7}{8}1\overset{\mathfrak{su}_2}{2}\overset{\mathfrak{so}_7}{3}\overset{\mathfrak{su}_2}{2}1\overset{\mathfrak{e}_7}{8}1\overset{\mathfrak{su}_2}{2}\overset{\mathfrak{so}_7}{3}\overset{\mathfrak{su}_2}{2}1\overset{\mathfrak{e}_7}{8}1\overset{\mathfrak{su}_2}{2}\overset{\mathfrak{so}_7}{3}\overset{\mathfrak{su}_2}{2}1\overset{\mathfrak{e}_7}{8}\cdots$ \\
\end{longtable}

%
%
%
\begin{longtable}{ccr}
	\toprule
	Nilpotent Orbit & $\mathfrak{f}$ & \multicolumn{1}{c}{Tensor Branch}\\
	\midrule
		$[1^2]$ & $\mathfrak{su}_{2}$ & $2\,1\overset{\mathfrak{e}_8}{(12)} \cdots$ \\
		$[2]$ & $\varnothing$ & $1\overset{\displaystyle 1}{\overset{\mathfrak{e}_8}{(12)}} \cdots$ \\
	\bottomrule
	\caption{The tensor branch descriptions associated to nilpotent Higgsing of the $\mathfrak{su}_{2}$ flavor symmetry for a long quiver associated to $\mathfrak{e}_{8}$ conformal matter with fraction number $f=\frac{1}{4}$.}
\label{tbl:E8-nilp-1-4}
\end{longtable}

\begin{longtable}{ccr}
	\toprule
	Nilpotent Orbit & $\mathfrak{f}$ & \multicolumn{1}{c}{Tensor Branch}\\
	\midrule
		$0$ & $\mathfrak{g}_{2}$ & $\overset{\mathfrak{su}_2}{2} 2\,1\,\overset{\mathfrak{e}_8}{(12)} 1 \, 2 \overset{\mathfrak{su}_2}{2} \overset{\mathfrak{g}_2}{3} 1 \overset{\mathfrak{f}_4}{5} 1 \overset{\mathfrak{g}_2}{3} \overset{\mathfrak{su}_2}{2} 2 1 \overset{\mathfrak{e}_8}{(12)} \cdots$ \\
		$A_1$ & $\mathfrak{su}_{2}$ & $2\, 2\,1\,\overset{\mathfrak{e}_8}{(12)} 1 \, 2 \overset{\mathfrak{su}_2}{2} \overset{\mathfrak{g}_2}{3} 1 \overset{\mathfrak{f}_4}{5} 1 \overset{\mathfrak{g}_2}{3} \overset{\mathfrak{su}_2}{2} 2 1 \overset{\mathfrak{e}_8}{(12)} \cdots$ \\
		$\widetilde{A}_1$ & $\mathfrak{su}_{2}$ & $2\,1\,\overset{\displaystyle 1}{\overset{\mathfrak{e}_8}{(12)}} 1 \, 2 \overset{\mathfrak{su}_2}{2} \overset{\mathfrak{g}_2}{3} 1 \overset{\mathfrak{f}_4}{5} 1 \overset{\mathfrak{g}_2}{3} \overset{\mathfrak{su}_2}{2} 2 1 \overset{\mathfrak{e}_8}{(12)} \cdots$ \\
		$G_2(a_1)$ & $\varnothing$ & $1\,\underset{\displaystyle 1}{\overset{\displaystyle 1}{\overset{\mathfrak{e}_8}{(12)}}} 1 \, 2 \overset{\mathfrak{su}_2}{2} \overset{\mathfrak{g}_2}{3} 1 \overset{\mathfrak{f}_4}{5} 1 \overset{\mathfrak{g}_2}{3} \overset{\mathfrak{su}_2}{2} 2 1 \overset{\mathfrak{e}_8}{(12)} \cdots$ \\
		$G_2$ & $\varnothing$ & $\overset{\mathfrak{e}_7}{8} 1 \overset{\mathfrak{su}_2}{2} \overset{\mathfrak{g}_2}{3} 1 \overset{\mathfrak{f}_4}{5} 1 \overset{\mathfrak{g}_2}{3} \overset{\mathfrak{su}_2}{2} 2 1 \overset{\mathfrak{e}_8}{(12)} \cdots$ \\
	\bottomrule
	\caption{The tensor branch descriptions associated to nilpotent Higgsing of the $\mathfrak{g}_{2}$ flavor symmetry for a long quiver associated to $\mathfrak{e}_{8}$ conformal matter with fraction number $f=\frac{1}{3}$.}
\label{tbl:E8-nilp-1-3}
\end{longtable}

\begin{longtable}{ccr}
	\toprule
	Nilpotent Orbit & $\mathfrak{f}$ & \multicolumn{1}{c}{Tensor Branch}\\
	\midrule
		$0$ & $\mathfrak{f}_{4}$ & $1\overset{\mathfrak{g}_2}{3} \overset{\mathfrak{su}_2}{2} 2 1 \overset{\mathfrak{e}_8}{(12)} 1 \, 2 \overset{\mathfrak{su}_2}{2} \overset{\mathfrak{g}_2}{3} 1 \overset{\mathfrak{f}_4}{5} 1 \overset{\mathfrak{g}_2}{3} \overset{\mathfrak{su}_2}{2} 2 1 \overset{\mathfrak{e}_8}{(12)} \cdots$ \\
		$A_1$ & $\mathfrak{sp}_{3}$ & $\overset{\mathfrak{g}_2}{2} \overset{\mathfrak{su}_2}{2} 2 1 \overset{\mathfrak{e}_8}{(12)} 1 \, 2 \overset{\mathfrak{su}_2}{2} \overset{\mathfrak{g}_2}{3} 1 \overset{\mathfrak{f}_4}{5} 1 \overset{\mathfrak{g}_2}{3} \overset{\mathfrak{su}_2}{2} 2 1 \overset{\mathfrak{e}_8}{(12)} \cdots$ \\
		$\widetilde{A}_1$ & $\mathfrak{su}_{4}$ & $\overset{\mathfrak{su}_3}{2} \overset{\mathfrak{su}_2}{2} 2 1 \overset{\mathfrak{e}_8}{(12)} 1 \, 2 \overset{\mathfrak{su}_2}{2} \overset{\mathfrak{g}_2}{3} 1 \overset{\mathfrak{f}_4}{5} 1 \overset{\mathfrak{g}_2}{3} \overset{\mathfrak{su}_2}{2} 2 1 \overset{\mathfrak{e}_8}{(12)} \cdots$ \\
		$A_1 + \widetilde{A}_1$ & $\mathfrak{su}_{2}\oplus\mathfrak{su}_{2}$ & $\overset{\mathfrak{su}_2}{2} \overset{\mathfrak{su}_2}{2} 2 1 \overset{\mathfrak{e}_8}{(12)} 1 \, 2 \overset{\mathfrak{su}_2}{2} \overset{\mathfrak{g}_2}{3} 1 \overset{\mathfrak{f}_4}{5} 1 \overset{\mathfrak{g}_2}{3} \overset{\mathfrak{su}_2}{2} 2 1 \overset{\mathfrak{e}_8}{(12)} \cdots$ \\
		$A_2$ & $\mathfrak{su}_{3}$ & $2 \overset{\mathfrak{su}_2}{2} 2 1 \overset{\mathfrak{e}_8}{(12)} 1 \, 2 \overset{\mathfrak{su}_2}{2} \overset{\mathfrak{g}_2}{3} 1 \overset{\mathfrak{f}_4}{5} 1 \overset{\mathfrak{g}_2}{3} \overset{\mathfrak{su}_2}{2} 2 1 \overset{\mathfrak{e}_8}{(12)} \cdots$ \\
\end{longtable}

\begin{longtable}{ccr}
	\toprule
	Nilpotent Orbit & $\mathfrak{f}$ & \multicolumn{1}{c}{Tensor Branch}\\
	\midrule
		$0$ & $\mathfrak{g}_{2}$ & $1 \overset{\mathfrak{f}_4}{5} 1 \overset{\mathfrak{g}_2}{3} \overset{\mathfrak{su}_2}{2} 2\,1\,\overset{\mathfrak{e}_8}{(12)} \cdots$ \\
		$A_1$ & $\mathfrak{su}_{2}$ & $\overset{\mathfrak{f}_4}{4} 1 \overset{\mathfrak{g}_2}{3} \overset{\mathfrak{su}_2}{2} 2\,1\,\overset{\mathfrak{e}_8}{(12)}  \cdots$ \\
		$\widetilde{A}_1$ & $\mathfrak{su}_{2}$ & $\overset{\mathfrak{so}_9}{4} 1 \overset{\mathfrak{g}_2}{3} \overset{\mathfrak{su}_2}{2} 2\,1\,\overset{\mathfrak{e}_8}{(12)}  \cdots$ \\
		$G_2(a_1)$ & $\varnothing$ & $\overset{\mathfrak{so}_8}{4} 1 \overset{\mathfrak{g}_2}{3} \overset{\mathfrak{su}_2}{2} 2\,1\,\overset{\mathfrak{e}_8}{(12)}  \cdots$ \\
		$G_2$ & $\varnothing$ & $\overset{\mathfrak{g}_2}{3} \overset{\mathfrak{su}_2}{2} 2\,1\,\overset{\displaystyle 1}{\overset{\mathfrak{e}_8}{(12)}}  \cdots$ \\
	\bottomrule
	\caption{The tensor branch descriptions associated to nilpotent Higgsing of the $\mathfrak{g}_{2}$ flavor symmetry for a long quiver associated to $\mathfrak{e}_{8}$ conformal matter with fraction number $f=\frac{2}{3}$.}
\label{tbl:E8-nilp-2-3}
\end{longtable}

\newpage

\begin{longtable}{ccr}
	\toprule
	Nilpotent Orbit & $\mathfrak{f}$ & \multicolumn{1}{c}{Tensor Branch}\\
	\midrule
		$[1^2]$ & $\mathfrak{su}_{2}$ & $\overset{\mathfrak{g}_2}{3}1\overset{\mathfrak{f}_4}{5}1\overset{\mathfrak{g}_2}{3}\overset{\mathfrak{su}_2}{2}2\,1\overset{\mathfrak{e}_8}{(12)}\cdots$ \\
		$[2]$ & $\varnothing$ & $\overset{\mathfrak{su}_3}{3}1\overset{\mathfrak{f}_4}{5}1\overset{\mathfrak{g}_2}{3}\overset{\mathfrak{su}_2}{2}2\,1\overset{\mathfrak{e}_8}{(12)}\cdots$ \\
	\bottomrule
	\caption{The tensor branch descriptions associated to nilpotent Higgsing of the $\mathfrak{su}_{2}$ flavor symmetry for a long quiver associated to $\mathfrak{e}_{8}$ conformal matter with fraction number $f=\frac{3}{4}$.}
\label{tbl:E8-nilp-3-4}
\end{longtable}

\begin{landscape}
\begin{longtable}{>{\footnotesize}c>{\footnotesize}c>{\footnotesize}r}\\
	\toprule   
	Nilpotent Orbit & $\mathfrak{f}$ & \multicolumn{1}{c}{Tensor Branch}\\
	\midrule
		$0$ & $\mathfrak{e}_{8}$ & $1\,2\overset{\mathfrak{su}_2}{2}\overset{\mathfrak{g}_2}{3}1\overset{\mathfrak{f}_4}{5}1\overset{\mathfrak{g}_2}{3}\overset{\mathfrak{su}_2}{2}2\,1\overset{\mathfrak{e}_8}{(12)}1\,2\overset{\mathfrak{su}_2}{2}\overset{\mathfrak{g}_2}{3}1\overset{\mathfrak{f}_4}{5}1\overset{\mathfrak{g}_2}{3}\overset{\mathfrak{su}_2}{2}2\,1\overset{\mathfrak{e}_8}{(12)}1\,2\overset{\mathfrak{su}_2}{2}\overset{\mathfrak{g}_2}{3}1\overset{\mathfrak{f}_4}{5}1\overset{\mathfrak{g}_2}{3}\overset{\mathfrak{su}_2}{2}2\,1\overset{\mathfrak{e}_8}{(12)}1\,2\overset{\mathfrak{su}_2}{2}\overset{\mathfrak{g}_2}{3}1\overset{\mathfrak{f}_4}{5}1\overset{\mathfrak{g}_2}{3}\overset{\mathfrak{su}_2}{2}2\,1\overset{\mathfrak{e}_8}{(12)}1\,2\overset{\mathfrak{su}_2}{2}\overset{\mathfrak{g}_2}{3}1\overset{\mathfrak{f}_4}{5}1\overset{\mathfrak{g}_2}{3}\overset{\mathfrak{su}_2}{2}2\,1\overset{\mathfrak{e}_8}{(12)}\cdots$ \\
\end{longtable}

\end{landscape}

\bibliography{references}{}
\bibliographystyle{utphys}

\end{document}